\documentclass[letterpaper,11pt]{article}
\usepackage{iftex}
\ifPDFTeX
  \usepackage[T1]{fontenc}
  \usepackage[utf8]{inputenc}
\fi
\usepackage{graphicx}
\usepackage{placeins}
\usepackage{afterpage}
\usepackage{geometry}
\usepackage{caption}
\usepackage[
  backend=biber,
  style=nature,
  sorting=none,
  sortcites=true,
  autocite=superscript,
  giveninits=true,
  maxbibnames=5,
  minbibnames=1,
  doi=true,
  isbn=false,
  url=true
]{biblatex}
\usepackage{amsmath, amssymb, bm}
\usepackage{booktabs}
\usepackage{threeparttable}
\usepackage{siunitx}
\usepackage{multirow}
\usepackage{appendix}
\usepackage{xcolor}\usepackage[normalem]{ulem}
\usepackage{tcolorbox}
\usepackage{fancyhdr}
\usepackage{algorithm}
\usepackage[noend]{algpseudocode}
\usepackage{iftex}
\usepackage[version=4]{mhchem}
\DeclareSIUnit\angstrom{\text{\AA}}
\ifPDFTeX
  \newcommand{\reportfont}{\sffamily}
\else
  \usepackage{fontspec}
  \IfFontExistsTF{Verdana}{%
    \newfontfamily\reportfont{Verdana}
  }{%
    \newcommand{\reportfont}{\sffamily}
  }
\fi
\usepackage{hyperref}\usepackage{soul}\sethlcolor{yellow}\soulregister\ref7
\tcbuselibrary{skins,breakable}
\definecolor{MetaPanelBlue}{HTML}{EAF4FF}
\colorlet{MetaPanel}{MetaPanelBlue!61!white}
\definecolor{ReportBlue}{HTML}{0064E0}
\hypersetup{
  colorlinks=true,
  citecolor=black,
  linkcolor=ReportBlue,
  urlcolor=ReportBlue
}
\AtEveryBibitem{%
  \iffieldequalstr{entrykey}{maziarz2026chemist}{\clearfield{doi}}{}%
  \iffieldundef{eprint}{}{\clearfield{doi}\clearfield{url}}%
  \ifentrytype{article}{%
    \iffieldundef{doi}{}{\clearfield{url}\clearfield{urldate}}%
    \iffieldundef{volume}{}{\iffieldundef{pages}{}{\clearfield{doi}}}%
  }{}%
}
\newcommand{\nsampler}{Sampler}

\newcommand{\rnx}{ReactNexus}

\newcommand{\reporttitle}{Generative discovery of chemical reaction networks}
\newcommand{\reportauthors}{Yunhong Han\textsuperscript{1,*}, Gaopeng Ren\textsuperscript{1,*}, Yan Liu\textsuperscript{1,*}, Ronghua Zhang\textsuperscript{1}, Huangjiang Huang\textsuperscript{1}, Ruiqi Zhang\textsuperscript{1,2}, Zhaoxi Yu\textsuperscript{1,3}, Chuncheng Liu\textsuperscript{1}, Zhao Li\textsuperscript{1}, Xiaohan Lin\textsuperscript{1}, Jiaxu Wang\textsuperscript{1}, Yuxuan Qin\textsuperscript{1}, Meijie Wang\textsuperscript{1}, Guoshun Mei\textsuperscript{1}, Doudou Wang\textsuperscript{1}, Taoyong Cui\textsuperscript{1}, Haojun Jia\textsuperscript{1,\ensuremath{\dagger}}, Chenru Duan\textsuperscript{1,\ensuremath{\dagger}}, Qiyuan Zhao\textsuperscript{1,\ensuremath{\dagger}}}
\newcommand{\reportaffiliations}{%
\textsuperscript{1}Deep Principle, Hangzhou, Zhejiang, China.\\
\textsuperscript{2}Peking University Shenzhen Graduate School, Shenzhen, Guangdong, China.\\
\textsuperscript{3}Key Laboratory of Theoretical and Computational Photochemistry of Ministry of Education, College of Chemistry, Beijing Normal University, Beijing 100875, China.\\\textsuperscript{*}These authors contributed equally to this work.}
\newcommand{\reportdate}{November 19, 2026}
\newcommand{\reportcorrespondence}{Haojun Jia: \href{mailto:haojunjia@deepprinciple.com}{haojunjia@deepprinciple.com}\\
Chenru Duan: \href{mailto:duanchenru@deepprinciple.com}{duanchenru@deepprinciple.com}\\
Qiyuan Zhao: \href{mailto:zhaoqiyuan@deepprinciple.com}{zhaoqiyuan@deepprinciple.com}}
\newcommand{\reportcodeurl}{https://mira.deepprinciple.com}
\newcommand{\reportlogo}{figures/deep_principle_logo_cropped.png}
\newcommand{\footerlogo}{figures/deep_principle_logo_cropped.png}
\newenvironment{reportabstract}{\par\vspace{1.35em}\small}{\par}
\newcommand{\reportmetadata}{%
  {\reportfont\small\textbf{Date:} \reportdate\par}%
  {\reportfont\small\textsuperscript{\ensuremath{\dagger}}\textbf{Correspondence:}\par \reportcorrespondence\par}%
  {\reportfont\small\textbf{Available soon:} \href{\reportcodeurl}{\reportcodeurl}\par}%
}
\newcommand{\SIref}[1]{\csname RNXSIref@#1\endcsname}
\expandafter\def\csname RNXSIref@fig:si_multisite_l1_pes\endcsname{S8}
\expandafter\def\csname RNXSIref@si:snar_nmr\endcsname{S11}
\expandafter\def\csname RNXSIref@fig:si_nmr_332b\endcsname{S15}
\expandafter\def\csname RNXSIref@fig:si_nmr_332a\endcsname{S16}
\expandafter\def\csname RNXSIref@fig:si_nmr_332c\endcsname{S17}
\expandafter\def\csname RNXSIref@fig:si_nmr_p6\endcsname{S18}
\expandafter\def\csname RNXSIref@si:benchmark_details\endcsname{S5}
\expandafter\def\csname RNXSIref@si:carina\endcsname{S4}
\expandafter\def\csname RNXSIref@si:ic_conformers\endcsname{S7.1}
\expandafter\def\csname RNXSIref@si:ic_dft_comparison\endcsname{S7.2}
\expandafter\def\csname RNXSIref@si:identity_details\endcsname{S6.4}\expandafter\def\csname RNXSIref@si:energy_details\endcsname{S6.3}
\expandafter\def\csname RNXSIref@si:literature_baselines\endcsname{S5.4}
\expandafter\def\csname RNXSIref@si:model_evaluation\endcsname{S5}
\expandafter\def\csname RNXSIref@si:navigation_details\endcsname{S6.5}
\expandafter\def\csname RNXSIref@si:rd11e\endcsname{S1}
\expandafter\def\csname RNXSIref@si:sampler\endcsname{S2}
\expandafter\def\csname RNXSIref@si:stationary_verification\endcsname{S6.2}
\expandafter\def\csname RNXSIref@si:tracer\endcsname{S3}
\expandafter\def\csname RNXSIref@si:workflow_screening\endcsname{S6.1}

\begin{document}
\begingroup
\setlength{\parindent}{0pt}
\setlength{\parskip}{0.55em}
\begin{tcolorbox}[
  enhanced,
  breakable,
  colback=MetaPanel,
  colframe=MetaPanel,
  arc=10pt,
  boxrule=0pt,
  left=18pt,
  right=18pt,
  top=16pt,
  bottom=16pt,
  width=\textwidth
]
{\centering\reportfont\bfseries\Large \reporttitle\par}
\vspace{0.7em}
{\centering\reportfont\bfseries \reportauthors\par}
\vspace{0.35em}
{\reportfont\footnotesize\raggedright \reportaffiliations\par}
\vspace{0.9em}
\noindent
\begin{reportabstract}
Understanding chemical reactivity requires uncovering which pathways are accessible from a given set of reactants, how they compete, and how successive reactions reshape the mixture. Automated saddle-point searches can explore from reactants alone, but their computational cost rises sharply as competing pathways accumulate across successive steps. Machine-learning approaches to transition-state generation typically require predefined products, yet identifying those products is a central goal of exploration. Here we introduce ReactNexus, a generative framework that generates transition-state candidates directly from reactant graphs without specifying products, predicts the species they connect, and refines pathways with a reactive machine-learned potential to guide network exploration. Its models were trained on a quantum-chemical database of 130 million structures from 1.1 million reaction records, spanning 11 elements and multiple charge and spin states. On held-out reactions, 93.6\% of generated candidates refined to valid transition states at the density functional theory level, and 83.3\% connected the supplied reactants to distinct products. Across decomposition, complex rearrangements, and organocatalysis, ReactNexus reveals how competing mechanisms shape product formation, uncovering pathways missed in previous studies. In a multisite nucleophilic aromatic substitution, ReactNexus identified diarylation products that explained initially unassigned LC--MS signals, with structural assignments supported by NMR after isolation. ReactNexus also predicted an additional product family not detected under the initial conditions; subsequent higher-temperature experiments provided LC--MS evidence supporting its formation. This work highlights the potential of generative models to extend mechanistic analysis to increasingly complex reaction mixtures.
\end{reportabstract}
\par\vspace{0.2em}
\noindent
\raisebox{0.9em}{%
\begin{minipage}[b]{0.68\textwidth}
\reportmetadata
\end{minipage}%
}\hfill
\begin{minipage}[b]{0.22\textwidth}
\raggedleft
\includegraphics[width=1.25in]{\reportlogo}
\end{minipage}
\end{tcolorbox}
\endgroup

\section{Introduction}

Reaction networks provide a mechanistic basis for understanding molecular stability, catalytic selectivity, and the evolution of reacting mixtures.\autocite{dewyer2018methods,steiner2022autonomous,seumer2026practical,oliveira2017inverse,abdelgaid2026identification} Starting from a set of reactants, their construction requires discovering accessible species, identifying the elementary pathways connecting them, and evaluating the barriers that govern competition.\autocite{unsleber2020exploration,wen2023chemical,zhao2021simultaneously,maeda2018implementation,unsleber2022chemoton,wang2014discovering} Each discovered species can open further reactions, so pathway discovery and transition-state (TS) location must be repeated as the network grows.\autocite{zhao2023deep,bensberg2023concentration} Increasing molecular size, conformational flexibility, and the number of interacting components multiplies these possibilities, while an omitted pathway can redirect the predicted evolution of the mixture.\autocite{zhao2023thermally} The central challenge is therefore to sustain network exploration at a tractable computational cost without compromising mechanistic reliability.

Template-based generators, including Reaction Mechanism Generator (RMG) and HopCat, construct networks within encoded reaction families, while graph-based methods such as ZStruct and Yet Another Reaction Program (YARP) propose candidates through generic bond edits.\autocite{broadbelt1994computer,gao2016reaction,liu2021rmg,klucznik2024cationic,zimmerman2013automated,zhao2021simultaneously,zhao2022algorithmic} The former can omit reactions outside the encoded families, whereas the latter generate large candidate sets that require extensive TS searches to establish viable connections. Reactive dynamics and direct potential energy surface searches can discover pathways without predefined products, as demonstrated by the \emph{ab initio} nanoreactor, artificial force induced reaction (AFIR), and stochastic surface walking (SSW).\autocite{wang2014discovering,grimme2019exploration,wang2016automated,maeda2018implementation,shang2013stochastic} Extending these searches to larger networks remains costly because competing channels and conformations must be examined repeatedly.\autocite{hayashi2022silico,sumiya2020rate,staub2026accurate,liu2023optimal}

Machine learning accelerates these calculations through reactive machine-learned interatomic potentials (MLIPs) and direct generation of TS structures.\autocite{unke2021machine,anstine2026aimnet2rxn,zhao2025harnessing,yuan2024analytical,cui2025horm,burger2026hip,zhang2024exploring,ichino2024systematic} Reactive MLIPs lower the cost of energy and force evaluations but do not, by themselves, change the underlying search strategies or their input requirements. Diffusion and flow-based models generate three-dimensional TS candidates from molecular structures or reaction graphs.\autocite{duan2023accurate,kim2024diffusion,duan2025optimal,galustian2025goflow,xu2026automated} Models such as React-OT and GoFlow condition TS generation on specified reactants and products.\autocite{duan2025optimal,galustian2025goflow,sun2026unified} MolGEN extends this approach to network exploration by first generating candidate products and then generating TSs for the resulting reactant--product pairs.\autocite{tuo2026flow} In double-ended searches and reactant--product-conditioned TS generation, candidate products must still be supplied or predicted before TS determination.

Here we introduce \rnx{}, a generative framework that reverses this order by sampling possible TS geometries directly from reactant graphs (Fig.~\ref{fig:overview}). Trajectory predictions from these geometries identify the connected species, while a reactive MLIP evaluates energetics and supports optional pathway refinement. Accepted connections are assembled into a network, and product sets selected by barrier-guided ranking become inputs for subsequent rounds. To train models across diverse chemical environments, we assembled the Reaction Database with 11 Elements (RD11E), a quantum-chemical database containing 1.1 million reaction records and 130 million structures. The database spans 11 elements, systems with up to 25 heavy atoms, and multiple charge and spin states. Across decomposition, complex rearrangements, organocatalysis, and multisite substitution, the generated networks reveal previously unrecognized pathways and competing mechanisms of product formation. In proline-catalysed aldol addition, the network identifies a three-component pathway to enamine formation in which benzaldehyde participates before serving as the C--C coupling partner. In a multisite nucleophilic aromatic substitution, the network identified three diarylation structures consistent with initially unexplained LC--MS signals, and their assignments were supported by $^1$H and $^{13}$C NMR after isolation. The network additionally predicted a further-substitution product family not detected at 70~$^\circ$C. Subsequent experiments at 100~$^\circ$C provided LC--MS evidence for the predicted product family.

\section{Results}
\subsection{Reactant-conditioned generation yields productive network connections}\label{sec:framework}

\begin{figure}[!htbp]
\centering
\captionsetup{font=footnotesize,labelformat=empty,justification=justified,singlelinecheck=false,skip=4pt}
\includegraphics[width=\textwidth]{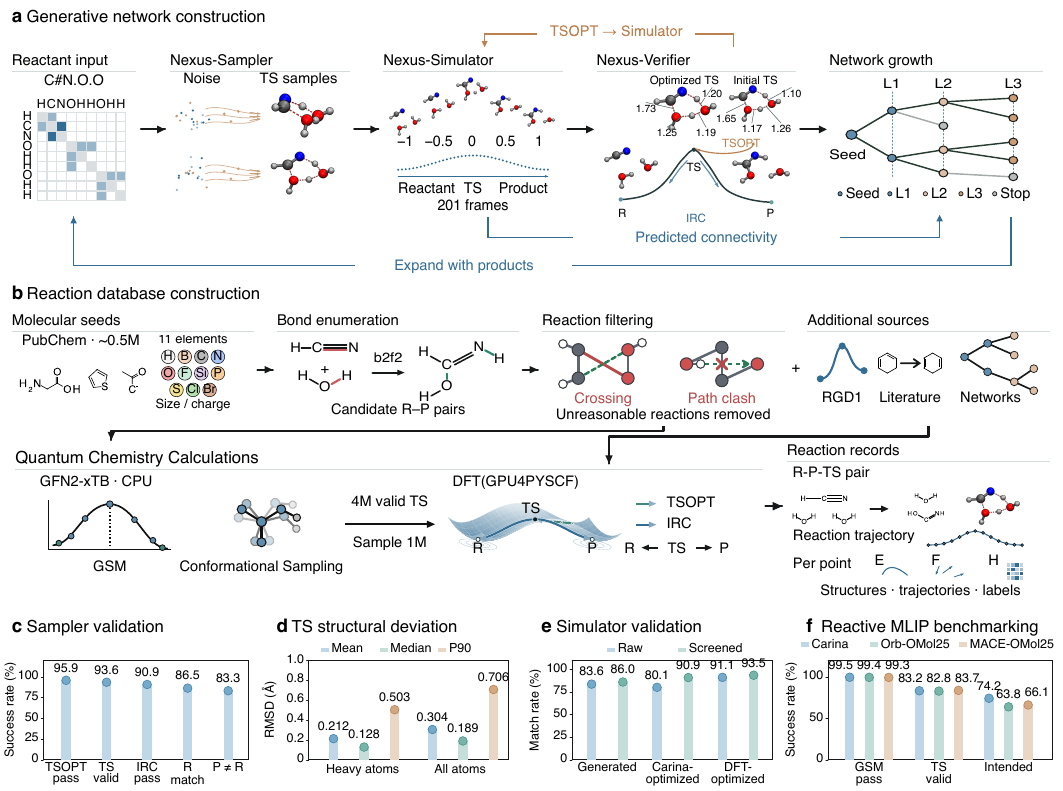}
\caption{\textbf{Fig. 1 \textbar{} Generative reaction-network construction and single-step validation.} (a) Reactant graphs condition Nexus-Sampler to generate candidate transition-state (TS) geometries; Nexus-Simulator predicts full reaction trajectories and their endpoints from sampled TSs. Nexus-Verifier uses Carina, a reactive MLIP, for single-point energy (SPE)-based screening, TS refinement, and intrinsic reaction coordinate (IRC) calculations. The modules support different levels of refinement, with accepted product sets serving as inputs for further generation. A 201-frame trajectory prediction and an example path obtained from an IRC calculation are shown. Distances are in \AA. (b) Database construction from molecular seeds and generic bond edits via conformational sampling, GFN2-xTB pre-explorations and DFT refinement. Additional reaction sources enter directly at DFT TS reoptimization and IRC calculations. Records retain structures and trajectories together with energies (E), forces (F), and Hessians (H). (c) Outcomes of DFT TS optimization, frequency analysis, and IRC calculations for 10,000 Sampler-generated TS structures. R match denotes recovery of the input reactant. P $\ne$ R denotes a match to the input reactant R at one endpoint and a distinct product at the other. All percentages use the initial 10,000 TS structures. (d) Displacements from generated TS structures to corresponding DFT-optimized geometries for 9,360 valid TSs; P90 denotes the 90th percentile. (e) Agreement of both Simulator-predicted endpoints with the corresponding DFT IRC endpoints for a common set of 8,622 TS structures, using Sampler-generated, Carina-optimized, and DFT-optimized TS inputs. \emph{Raw} includes all 8,622 input structures; \emph{Screened} retains those satisfying R match and P $\ne$ R based on Simulator predictions, with agreement calculated over the retained predictions. (f) GSM convergence, valid-TS formation, and intended-reaction recovery using Carina, Orb-OMol25, and MACE-OMol25 for the same 1,000 reference reactions, with optimized reactant and product endpoints supplied.}
\label{fig:overview}
\end{figure}

\paragraph{Method overview.}\label{subsec:rgraph2ts}\label{subsec:t2traj}\label{subsec:mlip}\label{subsec:network_construction}
From an input reactant set, Nexus-Sampler generates TS geometries without specifying products, and Nexus-Simulator predicts the reaction paths and connected species. Nexus-Verifier uses Carina, an EquiformerV3-based reactive MLIP,\autocite{liao2026equiformerv3} to evaluate energetics and refine pathways (Fig.~\ref{fig:overview}a).

Geometrical clustering and endpoint screening retain candidates connecting the input reactants to distinct products. Carina single-point energies (SPEs) and geometrical diversity then guide representative selection. Simulator predictions and SPEs support rapid exploration without TS optimization or intrinsic reaction coordinate (IRC) calculations. Selected pathways can undergo further TS refinement using Carina, followed by Simulator prediction or IRC calculations and endpoint verification.

After pathway evaluation for each input, accepted product sets are added to a persistent global candidate pool for further expansion. A barrier-guided policy ranks candidates using local electronic TS heights along their best available complete routes from the starting reactants, with the largest height contributing most.\autocite{zhao2023deep} The configurable branching factor $n$ specifies how many highest-priority eligible candidates are selected per round. Their complete product sets, including co-products and spectators, become new Sampler inputs. The pool is updated and reranked only after calculations for all inputs selected in the current round are complete. Newly found routes can change priorities, while previously discovered unexpanded candidates remain eligible. Screening, identity assignment, and network navigation are detailed in Sections~\ref{meth:screening}, \SIref{si:identity_details}, and~\ref{subsec:policy}, respectively.

\paragraph{Reaction dataset.}
RD11E contains optimized TS geometries, reactant and product structures, and trajectories obtained from IRC calculations, with electronic energies available for all structures and forces and Hessians for subsets (Fig.~\ref{fig:overview}b). The models were trained using only the training partition: TS geometries for Sampler, reaction trajectories for Simulator, and energy--force labels for Carina.

The database comprises approximately 1.1 million reaction records and 130 million structures obtained from TS optimization, IRC calculations, and endpoint optimization. Its chemical scope spans 11 elements (H, B, C, N, O, F, Si, P, S, Cl, and Br) and reaction systems containing up to 25 heavy atoms. Neutral, cationic, and anionic species are represented in open- or closed-shell states. The data include unimolecular and bimolecular reactions, as well as systems containing explicit catalysts or solvent molecules. Database construction, cleaning, and partitioning are described in Section~\ref{subsec:database}. Detailed dataset statistics and chemical-space analyses are provided in Section~\SIref{si:rd11e}.

\paragraph{Single-step generation and verification.}\label{sec:models}
We first evaluated whether reactant-conditioned TS generation produces valid transition states connecting the supplied reactants to distinct products. The benchmark used 1,000 reference reaction records held out from training and validation (Section~S5.1). Using only the corresponding reactant graphs, \nsampler{} generated ten TS candidates per reference reaction, yielding 10,000 candidates in total. Of these candidates, 93.6\% yield a valid first-order TS after DFT optimization at the same level of theory used for database construction; IRC calculations recover the input reactant for 86.5\% and connect it to a distinct product for 83.3\% of the initial candidates, allowing resonance-equivalent endpoint representations (Fig.~\ref{fig:overview}c). These are the \emph{valid-TS}, \emph{input-reactant match}, and \emph{productive-connection} rates, respectively. The latter directly measures the fraction of proposals that provide a potential new network connection. Although the reference reaction records were held out, some reactants were represented in the Sampler training data. Productive-connection rates were similar for candidates with and without reactant overlap (83.07\% and 83.62\%, respectively; Section~\SIref{si:benchmark_details}).

For the 9,360 valid TSs, median RMSDs between generated and corresponding DFT-optimized geometries are 0.128~\AA{} for heavy atoms and 0.189~\AA{} for all atoms (Fig.~\ref{fig:overview}d). Differences in task formulation and required inputs limit direct comparison with related generative models; contextual comparisons with MolGEN, React-OT, and GoFlow\autocite{tuo2026flow,duan2025optimal,galustian2025goflow} are provided in Section~\SIref{si:literature_baselines}.

Of the 10,000 \nsampler{}-generated TS candidates, 8,622 had completed DFT IRC calculations and converged Carina optimizations. This common subset was used to compare Simulator predictions from the \nsampler{}-generated, Carina-optimized, and DFT-optimized geometries. The Carina-optimized input set included 609 geometries (7.06\%) that converged in optimization but failed the strict TS criterion. \emph{Raw} agreement measures the fraction of predictions matching both endpoints obtained from DFT IRC calculations across all 8,622 candidates: 83.6\%, 80.1\%, and 91.1\%, respectively (Fig.~\ref{fig:overview}e). During network growth, only TS candidates whose simulated endpoints connect the input reactants to a distinct product are retained for expansion. To reflect this selection in the benchmark, \emph{Screened} agreement is calculated over the predictions retained in each scenario. Agreement with both DFT IRC endpoints is higher within these subsets, reaching 86.0\%, 90.9\%, and 93.5\%, respectively. Agreement with DFT IRC endpoints reflects both input geometry quality and differences between the Carina and DFT potential energy surfaces. To assess Simulator agreement within the Carina potential, a separate comparison used 7,967 candidates with strict Carina TSs and accepted IRC calculations on both potentials. Starting from the same Carina TS geometries, Simulator predictions matched both Carina IRC endpoints in 90.7\% of cases. Population definitions and complete results are provided in Section~\SIref{si:benchmark_details}.

Carina provides the energies and forces used for barrier estimation and pathway refinement. On the evaluated subset of held-out RD11E structures, it achieved mean absolute errors of 0.0183~eV for energies and 0.0054~eV/\AA{} for Cartesian force components (Fig.~S5e,f). Its performance in complete pathway searches was then assessed using the same 1,000 reference reactions following ReactBench.\autocite{zhao2025harnessing} Reactant and product inputs were taken from geometries obtained by optimizing the corresponding IRC endpoints. Carina recovers both specified endpoints by IRC for 74.2\% of reference reactions, compared with 63.8\% for Orb-OMol25 and 66.1\% for MACE-OMol25. GSM convergence and valid-TS formation rates are similar across the three MLIPs (Fig.~\ref{fig:overview}f).\autocite{rhodes2025orbv3,batatia2022mace}

\FloatBarrier

\enlargethispage{\baselineskip}
\subsection{Exploration captures competing pathways and low-energy TS conformations}\label{sec:validation}
\begin{figure}[!htbp]
\centering
\captionsetup{font=footnotesize,labelformat=empty,justification=justified,singlelinecheck=false,skip=4pt}
\includegraphics[width=\textwidth]{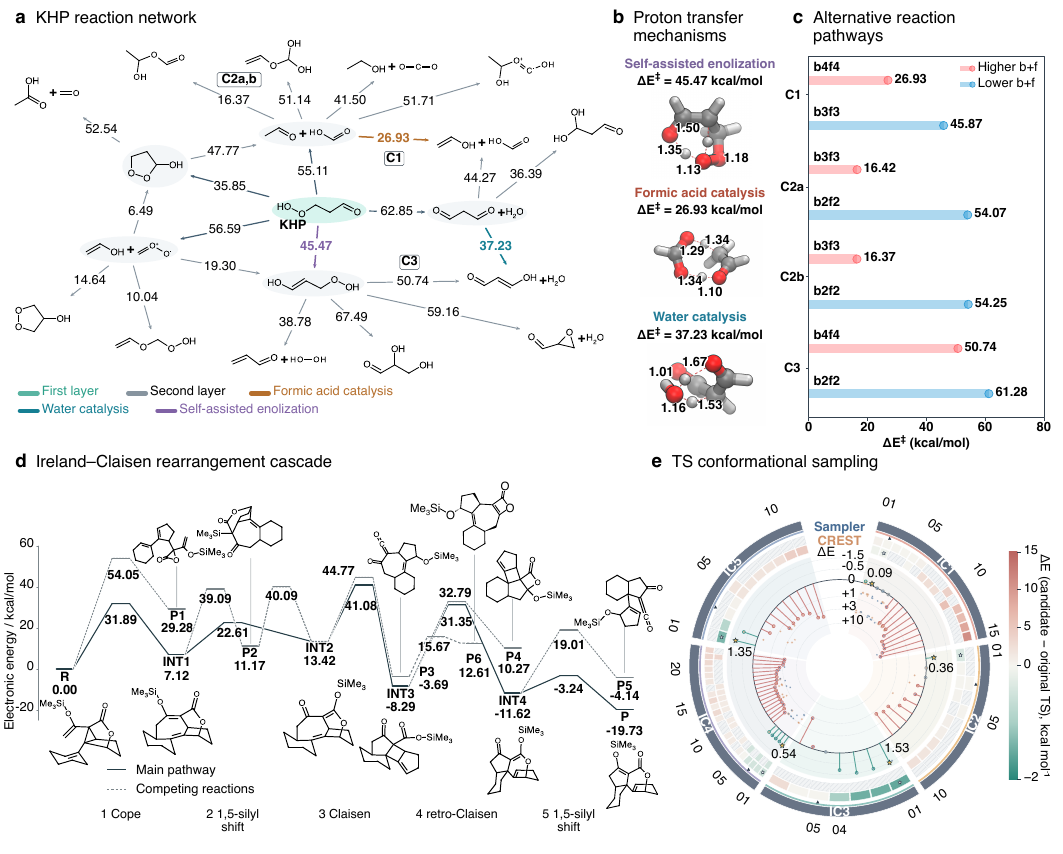}
\caption{\textbf{Fig. 2 \textbar{} Mechanism discovery and conformational refinement in reaction-network exploration.} (a) Two-layer $\gamma$-ketohydroperoxide (KHP) network, expanding the five first-layer product nodes reached through the lowest-barrier pathways ($n=5$). Co-products remain available for subsequent reactions. (b) Representative TSs for self-assisted, formic-acid-assisted, and water-assisted proton transfer. Colours correspond to highlighted pathways in a; distances are in \AA. (c) Alternative pathways connecting the same reactants and products. Labels $b_n f_n$ count broken and formed bond-order units through a single TS, including explicit hydrogen transfers. Labels C1--C3 link the network in a to the pathway comparisons in c. C2a and C2b distinguish the two enantiomeric product outcomes. (d) Five-step Ireland--Claisen (IC) reference sequence and selected competing connections. One closed-shell input is selected per global-pool expansion ($n=1$); the eleventh expanded input, INT4, supplies the connection to P. Dark solid and grey dashed profiles denote the reference sequence and competing reactions. Platforms show TS and product reference energies relative to the initial R. (e) TS conformational coverage from Sampler candidates and constrained sampling after Carina refinement (226 records in 66 geometrical groups). Tracks show source-specific group minima and points individual records; triangles and stars mark initial reference and selected TSs. $\Delta E$ is relative to the lowest TS retained for each reaction in the initial exploration. Central values give barrier reductions with the corresponding reactant reference fixed. All energies are Carina-computed electronic energies in kcal/mol.}
\label{fig:r1}
\end{figure}

The decomposition of $\gamma$-ketohydroperoxide (KHP) is an established benchmark for automated reaction discovery.\autocite{grambow2018unimolecular,zhao2021simultaneously,zhao2022algorithmic,tuo2026flow} For this system, ReactNexus generated a two-layer network by expanding the five first-layer product nodes reached through the lowest-barrier pathways (branching factor $n=5$; Fig.~\ref{fig:r1}a). The recovered chemistry includes the classical Korcek sequence through a five-membered cyclic peroxide, dehydration, and direct fragmentation. The relative kinetic ordering of previously reported channels agrees with earlier YARP-based exploration.\autocite{zhao2022algorithmic} Enolization is the exception: a hydroxyl-assisted intramolecular hydrogen-transfer pathway makes it more competitive (purple in Fig.~\ref{fig:r1}a; Fig.~\ref{fig:r1}b). By retaining decomposition co-products in subsequent exploration, ReactNexus automatically identifies water- and acid-catalysed proton-transfer pathways.\autocite{jalan2013new}

These pathways illustrate how cooperative proton transfer can provide lower-barrier mechanisms connecting the same reactants and products (Fig.~\ref{fig:r1}c). For acetaldehyde enolization, formic acid uses different oxygen atoms to donate and accept protons, coupling these transfers to C--O bond-order reorganization before being regenerated. This b4f4 pathway, involving the cleavage and formation of four bond-order units through a single TS, has a Carina-computed electronic barrier of 26.9~kcal/mol, compared with 45.9~kcal/mol for the b3f3 alternative, in which a single acid oxygen performs both roles. Similarly, formic acid addition to acetaldehyde couples proton transfer with C--O bond formation and bond-order reorganization through a b3f3 pathway. Its barrier is 16.4~kcal/mol, compared with 54.1~kcal/mol for the b2f2 bond-exchange pathway to the same product. Each comparison uses the same optimized reactant reference. Such cooperative mechanisms involve several coupled reactive coordinates and can fall outside restricted bond-edit enumeration or be difficult to locate through single-ended searches.

The Ireland--Claisen (IC) cascade tests multistep exploration in a conformationally flexible system. Lee and colleagues elucidated this unexpected cascade using trial-and-error calculations and AFIR searches.\autocite{lee2019unexpected} Here, ReactNexus recovered a five-step route to the final product P (INT5) after 11 expansion rounds that included competing branches (Fig.~\ref{fig:r1}d). The main sequence comprises Cope and Claisen/retro-Claisen rearrangements and 1,5-silyl migrations. Each round selected one closed-shell input ($n=1$) from the persistent global candidate pool, ranked by the barrier-based score of its best complete route from the starting reactant. Competing branches remained eligible for later exploration. The competing pathway from INT2 couples O-to-O silyl transfer with C--O cleavage to form a ketene-containing product (P3). From INT3, silyl transfer coupled to intramolecular C--O bond formation yields a $\beta$-lactone (P4). Figure~\ref{fig:r1}d shows competing connections at the level of molecular connectivity, including ring closure from P3 to the four-membered lactone P6 and a branch from P2 that rejoins the main sequence at INT2. Each local barrier is referenced to the corresponding optimized reactant. For comparison with earlier work, B3LYP-D3 single-point calculations with continuum n-hexane solvation were performed on the Carina-optimized structures (Section~\SIref{si:ic_dft_comparison}).

Because distinct TS conformations can yield different barriers,\autocite{young2021autode,zhao2022conformational} supplementary refinement of Sampler candidates and constrained conformational sampling were used to test the conformational coverage of the initial exploration (Fig.~\ref{fig:r1}e). With each reactant reference fixed, these searches lowered the Carina-computed electronic barriers by 0.09--1.53~kcal/mol relative to the lowest TSs retained in the initial exploration. The two approaches recovered both shared and distinct TS conformations for the five IC transformations. Supplementary Carina TS optimization covered 540 Sampler candidates assigned to the five reference transformations. Constrained conformational sampling generated 364 conformers around the five initially verified TSs with their reactive cores restrained.\autocite{pracht2020automated,zhao2025graphically} All restraints were released before Carina TS refinement and subsequent IRC calculations. Three of the five lowest retained TSs were obtained from constrained sampling, and two from additional refinement of Sampler candidates. These modest additional barrier reductions support effective coverage of the low-energy TS conformational space by Sampler and the initial screening workflow in this cascade. Sampling settings, geometrical grouping, and numerical diagnostics are detailed in Section~\SIref{si:ic_conformers}.

\subsection{Reaction networks reveal catalyst regeneration, and competing transformations}\label{sec:catalysis}
\refstepcounter{figure}\label{fig:cat}

\begin{figure}[!htbp]
\centering
\captionsetup{font=footnotesize,labelformat=empty,justification=justified,singlelinecheck=false,skip=4pt}
\includegraphics[width=\textwidth]{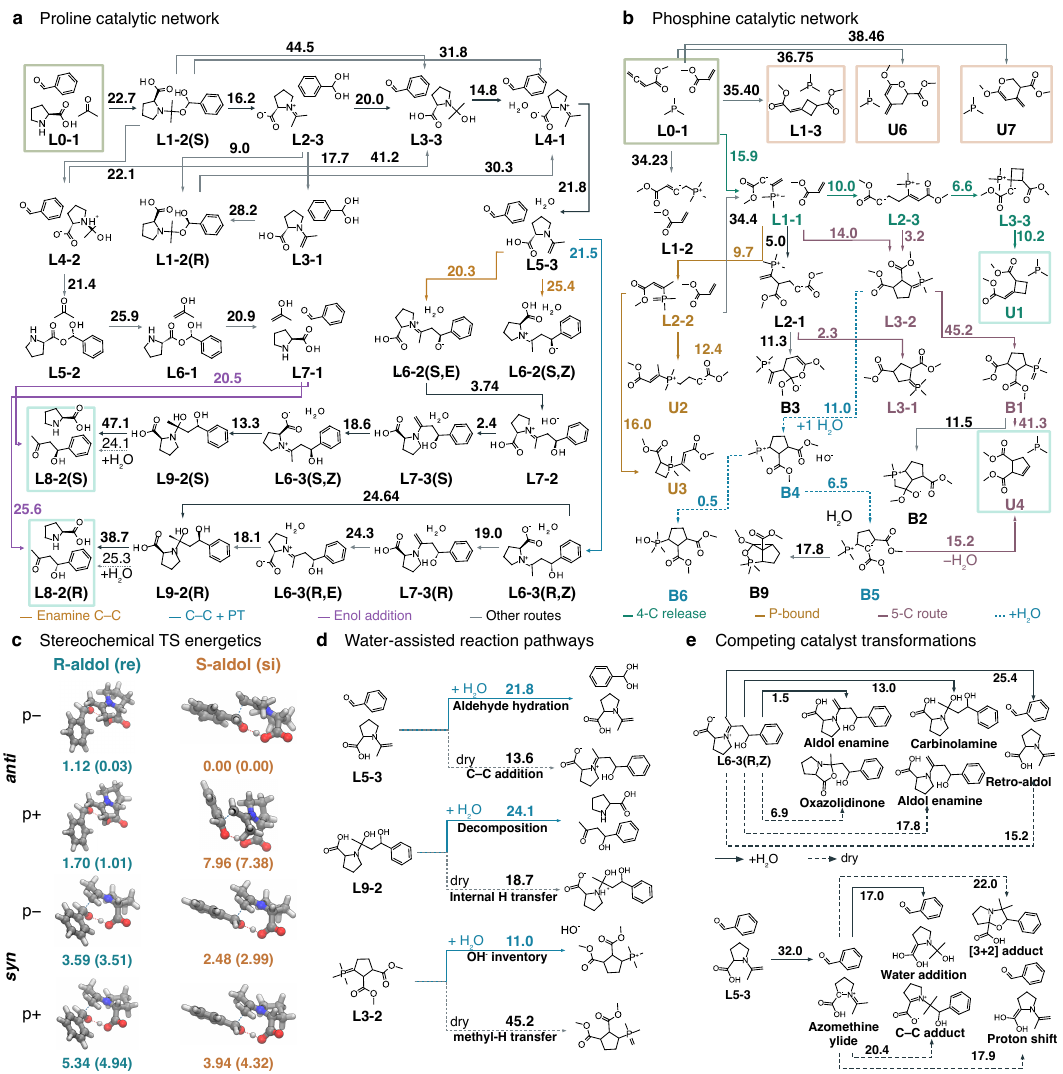}
\caption*{\textbf{Fig. 3 \textbar{} Connecting multicomponent reactants to product formation and catalyst regeneration.} (a) Proline-catalysed aldol network from acetone, benzaldehyde, and proline. Orange marks enamine C--C addition, blue marks C--C formation coupled to proton transfer, and purple marks free-enol addition to benzaldehyde assisted by the carboxylic acid group of free proline. Grey denotes other connections. (b) Reaction network generated from methyl allenoate, methyl acrylate, and trimethylphosphine (PMe$_3$). Green and muted purple mark four- and five-membered-ring pathways, respectively. Ochre marks reactions involving catalyst methyl groups or phosphorus-containing rings. Blue marks the added-water proton shuttle preceding PMe$_3$ release. (c) Eight R-proline C--C TS geometries classified by syn/anti enamine--carboxylic-acid arrangement, aldehyde re/si face, and ring conformation ($p-$: endo; $p+$: exo). Numbers are computed relative free energies, with relative DFT electronic energies in parentheses, each referenced to the lowest value among the eight TSs. Calculations used $\omega$B97X-D4/def2-TZVPD/SMD(DMSO) on Carina-optimized geometries. (d) Pathways with and without added water. For the phosphonium ylide, proton transfer from water (11.0) and hydrogen transfer from a phosphine methyl group (45.2) yield different products. (e) Competing reactions of proline-derived condensation intermediates and a COOH-retaining azomethine ylide. Solid and dashed arrows indicate reactions with and without water, respectively. Unless otherwise stated, arrow values are Carina-computed electronic barriers referenced to a common optimized reactant complex for each input, including all explicit molecules. Water-containing and water-free calculations use separate references. All energies are in kcal/mol. L labels denote exploration batch/rank; B, supplementary intermediates; U, unexpanded products.}
\end{figure}

ReactNexus was used to explore product formation, catalyst regeneration, and competing catalyst transformations in proline-catalysed aldol addition and phosphine-catalysed allenoate–acrylate coupling (Fig.~\ref{fig:cat}). In the proline system, benzaldehyde participates in catalyst activation before serving as the C--C coupling partner. Starting from acetone, benzaldehyde, and proline (L0-1), the network identifies a three-component adduct [L1-2(S)] that leads to the acetone--proline carbinolamine L3-3 via benzaldehyde hydrate. Benzaldehyde is regenerated during this sequence, and dehydration followed by proton transfer produces the enamine L5-3. The network also contains the conventional acetone--proline carbinolamine-forming pathway to L3-3, with benzaldehyde remaining a spectator. The conventional carbinolamine-forming pathway has an electronic TS height of 39.0~kcal/mol, compared with 22.7~kcal/mol for formation of L1-2(S) and 18.1~kcal/mol for a distinct three-component adduct. All three TS heights use the same initial mixture of acetone, benzaldehyde, and proline as the reference.

Both covalent enamine catalysis and free-proline-assisted enol addition were recovered during network exploration. In the enamine route, C--C addition followed by proton transfers and hydration leads to carbinolamines L9-2(S/R). Their C--N cleavage releases the aldol products and regenerates proline [L8-2(S/R)].\autocite{list2000proline,bahmanyar2001transition,bahmanyar2003quantum} With one added water molecule, product release from L9-2(S/R) has electronic barriers of 24.1 and 25.3~kcal/mol, respectively (Fig.~\ref{fig:cat}a,d). In the alternative route, the free enol of acetone couples with benzaldehyde while free proline serves as a proton shuttle (L7-1). Its carboxylic acid group accepts a proton from the enol and donates one to the aldehyde oxygen during C--C addition, yielding the aldol products and regenerating free proline [L8-2(S/R)].

Beyond recovering reaction connections, Sampler captures three-dimensional TS diversity from reactant graphs alone. For the water-free enamine--benzaldehyde system of the classical Houk model,\autocite{bahmanyar2001transition,bahmanyar2003quantum} 10,000 generated TS candidates covered eight reference conformational families for R-proline and seven for S-proline, spanning enamine--carboxylic-acid arrangements, aldehyde approach faces, and ring conformations (Fig.~\ref{fig:cat}c). DFT electronic energies and computed free energies evaluated on the Carina-optimized geometries place the lowest anti TSs below the syn alternatives, consistent with earlier calculations by Houk and co-workers.\autocite{allemann2010computational}

The phosphine network reveals different catalyst-regeneration requirements for four- and five-membered-ring formation (Fig.~\ref{fig:cat}b). Exploration started from methyl allenoate, methyl acrylate, and trimethyl\-phosphine (PMe$_3$). The four-membered-ring branch releases intact neutral PMe$_3$ to give U1 with a final-step barrier of 10.2~kcal/mol. In the five-membered-ring branch, however, the cyclic phosphonium ylide L3-2 requires proton transfer before phosphine elimination. Direct substrate hydrogen transfer has a barrier of 43.1~kcal/mol without water. Guided by earlier mechanistic studies, one explicit water molecule was therefore added to L3-2.\autocite{xia2007water,liang2008phosphine} Water protonates the ylide, and the resulting hydroxide removes the adjacent substrate proton, regenerating water and forming the elimination-ready intermediate B5. The two steps have barriers of 11.0 and 6.5~kcal/mol, respectively. From water-free B5, release of PMe$_3$ and cyclopentene U4 has a barrier of 15.2~kcal/mol, below that for the competing pathway to B9 (17.8~kcal/mol).

The hydroxide that enables this proton shuttle also opens a route to catalyst transformation. Instead of removing the substrate proton, it can attack the phosphonium centre in B4 to form a P--OH adduct (B6). Subsequent reactions lead to phosphine oxide, identifying a potential route to catalyst deactivation (Fig.~\ref{fig:cat}b). Additional branches involve hydrogen transfer from a phosphine methyl group in L3-2 (45.2~kcal/mol; Fig.~\ref{fig:cat}d) or formation of phosphorus-containing rings. Similar competing transformations occur in the proline network, including retro-aldol cleavage, oxazolidinone formation, and further C--C addition or cycloaddition of azomethine ylides (Fig.~\ref{fig:cat}e).

\FloatBarrier
\subsection{Focused exploration recovers product precursors and revises intermediate assignments}\label{sec:cation_networks}

\begin{figure}[!htbp]
\centering
\captionsetup{font=footnotesize,labelformat=empty,justification=justified,singlelinecheck=false,skip=4pt}
\includegraphics[width=0.92\textwidth]{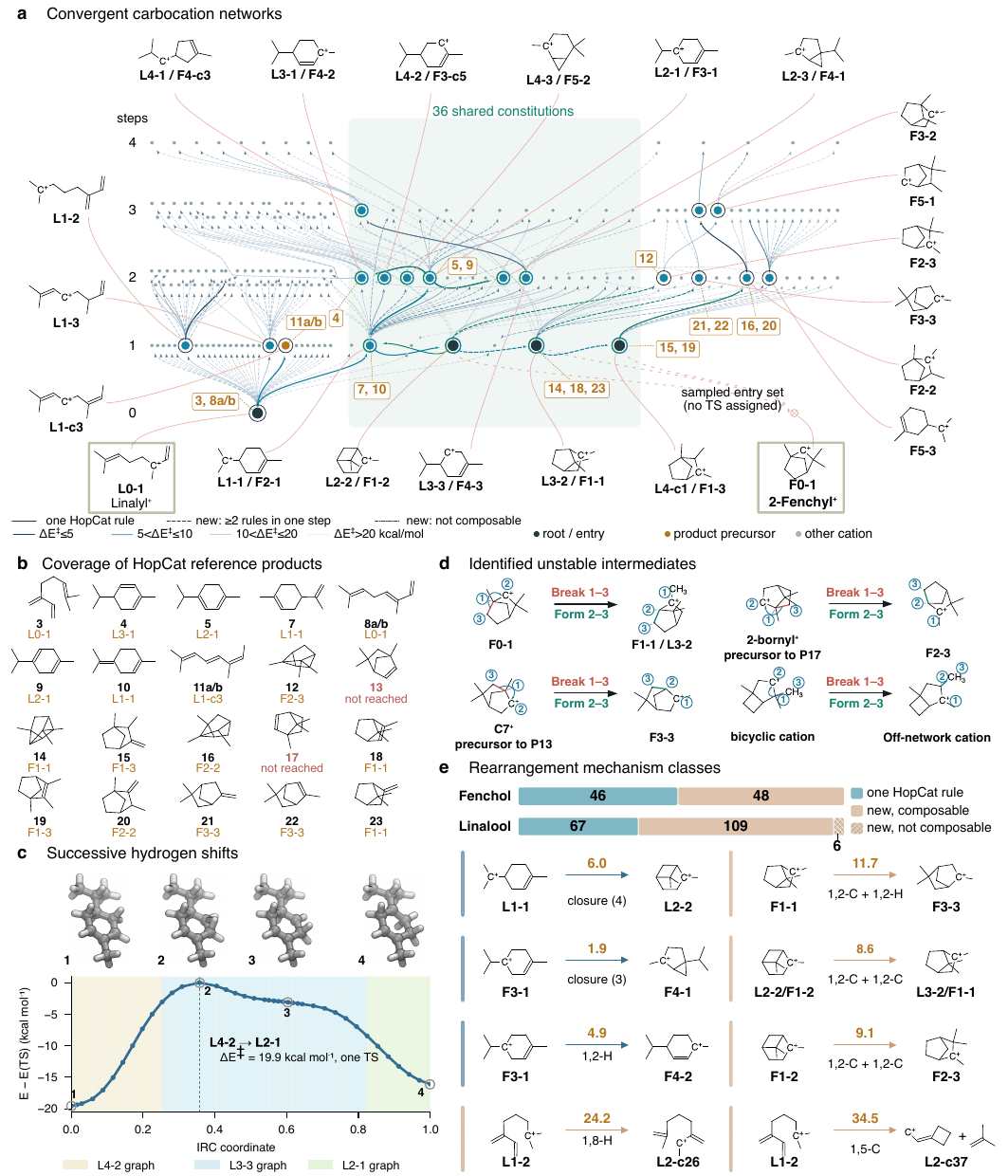}
\caption*{\textbf{Fig. 4 \textbar{} Convergence of carbocation rearrangement networks and identification of unstable intermediates.} (a) Reaction networks from linalyl and fenchol-derived cations: 189 nodes, including 36 shared structures drawn once. F0-1 serves only to initialize Sampler--Simulator entry proposals, whose links have no assigned TSs. Stereoisomers with the same connectivity share one node. L/F labels identify the two explorations, and vertical position denotes displayed network layers. Teal outlines mark expansion sites, while pink curves link molecular structures to their corresponding nodes. Edge width and shade encode local barriers. Solid, dashed, and dash-dotted edges denote single-rule, composable multiple-rule, and non-composable changes under the tested HopCat rule set, respectively. (b) Applying the quenching templates identifies cation precursors for 8/8 non-oxidative linalool and 10/12 fenchol reference products. (c) Carina IRC energy profile and selected structures for L4-2 $\rightarrow$ L2-1. Successive hydrogen shifts cross a single TS and pass through the L3-3 connectivity without an intervening minimum. Shaded regions denote connectivity classes. (d) DFT relaxation of F0-1, C7$^+$, and 2-bornyl$^+$ reproduces their Carina rearrangement outcomes. A sampled bicyclic cation provides an additional control. Red and green bonds mark cleavage and formation; blue arrows indicate inferred electron-pair displacement. (e) Classification of directed source--product pairs, with representative local barriers. \emph{New} includes changes composable from multiple rules and changes not composable under the tested HopCat rule set. Network barriers and IRC energies are Carina-computed gas-phase electronic energies (kcal/mol). Neutral-product quenching is rule-based and does not establish termination barriers or yields.}
\end{figure}

Carbocations can undergo diverse skeletal rearrangements, but identifying which pathways connect them to observed products requires resolving both reaction barriers and intermediate stability. Here, ReactNexus explored linalyl- and fenchol-derived cations through barrier-guided expansion, prioritizing intermediates reached through lower-barrier pathways.\autocite{dewitt2000ratebased,zhao2023deep} The explored networks contain 153 and 72 constitution-level nodes yet recover, under HopCat quenching rules, precursors for 8/8 non-oxidative linalool and 10/12 fenchol reference products, respectively (Fig.~\ref{fig:cation}a,b). This precursor coverage is reached within three expansion rounds. In these systems, a compact explored network therefore accounts for most reference-product precursors. Oxidative p-cymene formation is excluded from the linalool comparison. For context, HopCat reported 3,404 nodes for a taxanine rearrangement and 909,937 nodes for a mixed forward--retro biosynthetic search.\autocite{klucznik2024cationic} Despite their different starting skeletons, the linalyl- and fenchol-derived networks share 36 constitutions, including $\alpha$-terpinyl (L1-1/F2-1) and fenchyl-type (L3-2/F1-1) cations.\par \refstepcounter{figure}\label{fig:cation}

Explicit calculations revise several intermediate assignments in the explored networks. The nominal 2-fenchyl cation F0-1 relaxes to F1-1 in both Carina and DFT optimizations (Fig.~\ref{fig:cation}d). Sampler and Simulator therefore provided alternative entry structures, with F1-1, F1-2, and F1-3 seeding separate expansions. The HopCat-assigned precursors of the two unreached fenchol products, $\delta$-fenchene (13) and bornylene (17), also rearrange: C7$^+$ and 2-bornyl$^+$ relax to F3-3 and camphenyl cation F2-3, respectively, in both methods. Formation of these products through deprotonation coupled to rearrangement remains unresolved because quenching was assigned by rules rather than calculated explicitly. Pathway calculations additionally distinguish transient connectivity from an intervening minimum. Along the Carina-calculated L4-2 $\rightarrow$ L2-1 pathway, two successive hydrogen shifts cross a single TS and pass through the L3-3 connectivity without an intervening minimum (Fig.~\ref{fig:cation}c).

Across both networks, template-free exploration recovers connections consistent with HopCat rules and identifies additional connections beyond the tested rule set (Fig.~\ref{fig:cation}e). Most directed net changes match a single rule or can be composed from multiple rules. The latter include low-barrier connections involving net carbon and hydrogen migration in F1-1 $\rightarrow$ F3-3 (11.7~kcal/mol) and two carbon migrations in F1-2 $\rightarrow$ F2-3 (9.1~kcal/mol). Template-free exploration also finds six connections whose net changes cannot be composed from the tested rules. Their barriers are comparatively high (24.2--51.9~kcal/mol). The absence of these six connections from the identified reference-product routes supports the coverage of the tested HopCat rules for those routes.

\subsection[Network predictions aid experimental interpretation and suggest additional products]{Network predictions aid experimental interpretation and suggest additional products}\label{sec:applications}
We next examined whether ReactNexus could help interpret LC--MS observations from crude reaction mixtures and suggest additional products for experimental investigation.\autocite{abdelgaid2026identification,sakas2024analysis} We focused on the reaction of 2,6-dichloro-4-methylnicotinonitrile (A) with hydrazine (B), which combines two nonequivalent substitution sites, a competing nitrile group, and a nucleophile that remains reactive after substitution. LC--MS monitoring at 70~$^\circ$C revealed M332 signals that could not be accounted for by the expected monosubstitution products alone, leaving their structures and formation pathways unresolved (Fig.~\ref{fig:applications}c).

\refstepcounter{figure}\label{fig:applications}

Starting from A and B, ReactNexus generated a two-layer network of candidate structures and formation pathways (Fig.~\ref{fig:applications}a,b). In the first layer, C2 and C6 substitution give P$_2$ and P$_6$, respectively, both in the M182 family, with Carina electronic barriers of 24.4 and 25.4~kcal/mol. DFT calculations incorporating SMD solvation in DMF\autocite{marenich2009universal} and thermal corrections give corresponding free energies of activation of 15.6 and 16.4~kcal/mol at 298.15~K (Fig.~\SIref{fig:si_multisite_l1_pes}). The relative ordering of the C2 and C6 pathways was sensitive to the computational treatment (Section~S9). Further reaction of these monosubstituted intermediates with A gives the M332 diarylation candidates $P_{332a\text{--}f}$. For each monosubstituted reactant, 1,2-N-diarylation to $P_{332a\text{--}c}$ has lower local electronic barriers than competing 1,1-N-diarylation to $P_{332d\text{--}f}$.

Isolation and $^1$H and $^{13}$C NMR characterization of products formed at 70~$^\circ$C supported four network-predicted structures: P$_6$ and the three diarylation products P$_{332a}$, P$_{332b}$, and P$_{332c}$, with C2/C2, C2/C6, and C6/C6 linkages, respectively (Section~\SIref{si:snar_nmr}). These diarylation structures correspond to the lower-barrier 1,2-N-diarylation pathways in the network. Within the proposed stepwise mechanism, formation of P$_{332a}$ is consistent with further conversion of P$_2$, providing a possible explanation for why P$_2$ was not isolated.

\begin{figure}[!htbp]
\centering
\captionsetup{font=footnotesize,labelformat=empty,justification=justified,singlelinecheck=false,skip=4pt}
\includegraphics[width=0.96\textwidth]{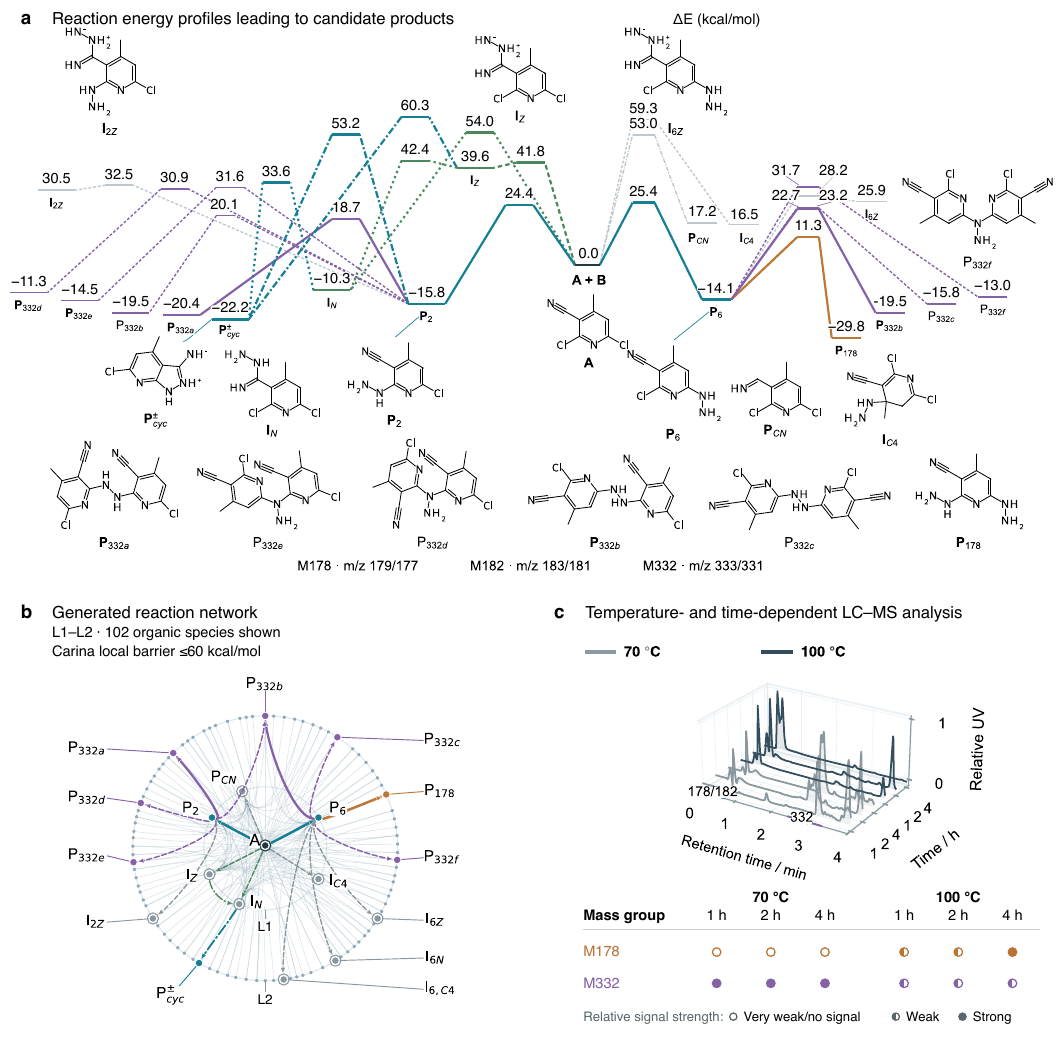}
\caption*{\textbf{Fig. 5 \textbar{} Network predictions and experimental analysis of a multisite reaction.} (a) Carina-computed gas-phase electronic-energy profiles (kcal/mol) and candidate structures from 2,6-dichloro-4-methylnicotinonitrile (A) and hydrazine (B). Teal, purple, and orange pathways indicate the M182 (P$_2$, P$_6$, and $P_{\mathrm{cyc}}^{\pm}$), M332 ($P_{332a\text{--}f}$), and M178 (P$_{178}$) candidates, respectively. Green highlights routes through I$_Z$ and I$_N$. $P_{332a\text{--}c}$ and $P_{332d\text{--}f}$ are 1,2-N- and 1,1-N-diarylation candidates. Second-step reactions use additional A for diarylation and B for further substitution or nitrile addition. Energy profiles are aligned to the displayed parent-product levels; local barriers use each step's own optimized reactant reference. For P$_6$ $\rightarrow$ P$_{178}$, the local electronic barrier is 25.4~kcal/mol relative to the optimized P$_6$ + B complex, corresponding to a plotted TS level of 11.3~kcal/mol. (b) Two-layer network showing 102 organic species. Six small-molecule nodes and their incident edges are omitted from the displayed network. The underlying source network contains 108 species and 122 reaction records with local barriers $\leq$60~kcal/mol. Radial position indicates exploration layer. (c) LC--MS analysis after 1, 2, and 4~h for A (1~equiv.), hydrazine (1.2~equiv.), and K$_2$CO$_3$ (3~equiv.) in DMF at 70 or 100~$^\circ$C. PDA traces at 254~nm are individually normalized. The displayed $m/z$ pairs are 179/177, 183/181, and 333/331 for M178, M182, and M332, respectively. The matrix summarizes M178 and M332 signals: open circles indicate very weak or absent signals, and half-filled and filled circles indicate weak and strong signals, respectively, for qualitative comparison within each mass group. No M178 signal was observed at 70~$^\circ$C.}
\end{figure}

\begin{samepage}
Beyond these characterized products, the network identifies a further-substitution pathway from P$_6$ to P$_{178}$ through reaction with B. No LC--MS signal attributable to M178 was observed at 70~$^\circ$C. Subsequent experiments at 100~$^\circ$C detected signals consistent with this predicted product family; these signals were weak after 1--2~h and more pronounced after 4~h (Fig.~\ref{fig:applications}c).\par
\end{samepage}

The network also identifies a candidate cyclization pathway to $P_{\mathrm{cyc}}^{\pm}$. CuI-promoted cyclization in a related system suggests that modified catalytic conditions could favour the formation of cyclic products, providing a further direction for experimental investigation.\autocite{lavecchia2004pyrazolopyridines}

\section{Conclusion}\label{sec:limitations}
In this work, we establish reaction-network exploration as a generative problem: TS candidates are sampled from reactant graphs alone, and products are inferred from the resulting pathways. This enables exploration without predefined products or reaction templates and can reveal transformations beyond the initial mechanistic hypothesis.

Across the applications, species formed along a reaction pathway participate in subsequent transformations, while the roles of the starting components evolve as reactions proceed. Decomposition products mediate later proton transfers (Fig.~\ref{fig:r1}), a coupling substrate takes part in catalyst activation (Fig.~\ref{fig:cat}a), and the added water that enables catalyst regeneration also opens a potential deactivation route (Fig.~\ref{fig:cat}b). These observations caution against assigning fixed roles to reactants, catalysts, and by-products, and highlight the need to consider competing and subsequent reactions when interpreting product formation. In the multisite substitution, network candidates supported structural interpretation alongside NMR characterization, while subsequent experiments detected LC--MS signals consistent with an additional product family predicted by the network (Fig.~\ref{fig:applications}).

The present networks nevertheless remain sampled descriptions of reaction possibilities rather than quantitative predictions of mixture evolution. Finite sampling and barrier-guided expansion can omit relevant channels, and generation frequencies are not reaction rates. Diradical pathways need broader sampling and validation, and current sampling does not reliably constrain generation to an intended stereochemical configuration. Quantitative kinetics will further require solvent-dependent energetics and validated free-energy corrections, with consistent treatment of low-frequency modes, conformational populations, and concentrations.\autocite{ribeiro2011use,burger2026hip} Extending the chemical scope to metal-containing systems is another priority.

Many of these limitations can be addressed through broader training data and improved model conditioning, alongside advances in energetics and exploration strategies. Building on the present results, these developments could enable more reliable prediction of competing reactions and product formation in complex mixtures. Integrating TS-based network exploration with learning-based reaction prediction\autocite{joung2025electron} and synthesis planning\autocite{maziarz2026chemist} could connect proposed transformations to mechanistic pathways and kinetic competition, supporting more informed reaction design.

\section{Methods}\label{sec:methods}
\rnx{} couples reactant-conditioned TS generation, trajectory prediction, and energetic evaluation for kinetic-guided reaction-network exploration. The following sections describe database construction, model training, and network exploration, with detailed implementations and settings provided in the Supplementary Information.

\subsection[Reaction database construction and partitioning]{Reaction database construction and partitioning}\label{subsec:database}

The core component of RD11E was generated by selecting molecular seeds from PubChem, enumerating candidate products through generic bond edits, and characterizing the resulting reactions using a YARP-derived TS search workflow.\autocite{zhao2021simultaneously,zhao2022algorithmic,zhao2023comprehensive} Specifically, approximately half a million molecular seeds were selected with an emphasis on chemical diversity.\autocite{kim2023pubchem} Eligible seeds contained only H, B, C, N, O, F, Si, P, S, Cl, and Br, with at most 15 heavy atoms, total charges of $-1$, $0$, or $+1$, and spin multiplicities of 1 or 2. Charge states were retained as recorded in PubChem, without generating additional states by protonation or deprotonation. Product enumeration defined the reactant--product pairs used to initiate the TS searches, and a tree-based filter removed unphysical candidates.\autocite{zhao2022conformational} Enumeration was used to construct training examples offline; network generation at inference does not require predefined products.

\begingroup\emergencystretch=3em

TS searches began with reaction conformational sampling to generate three-dimensional reactant--product pairs.\autocite{zhao2022conformational} Double-ended growing string method (GSM) calculations at the GFN2-xTB level then provided initial TS guesses.\autocite{zimmerman2013growing,bannwarth2019gfn2} These guesses were refined by TS optimization, followed by IRC calculations at the same level to identify the reactants and products connected by each optimized TS. This stage yielded approximately four million reactions with valid TSs.

Following screening of the xTB results, approximately one million reactions were selected for DFT refinement. The xTB-optimized TS geometries served as starting points for gas-phase TS optimization at the $\omega$B97X-D4/def2-TZVPD level, followed by IRC calculations at the same level.\autocite{chai2008systematic,caldeweyher2019d4,rappoport2010property} GSM calculations used pyGSM,\autocite{aldaz2018discovery} TS optimization and IRC calculations used pysisyphus,\autocite{steinmetzer2021pysisyphus} and DFT electronic-structure calculations used GPU4PySCF 1.4.1.\autocite{li2025introducing} The functional/dispersion specification was \texttt{wb97x-d4:\allowbreak wb97x-2008}.\par\endgroup

RD11E was supplemented with 52,169 bimolecular reactions from RGD1 and Reaction-QM,\autocite{zhao2023comprehensive,lee2026reactionqm} water-catalysed reactions from previous studies of hydrogen cyanide and glucose reaction networks,\autocite{zhao2023thermally,zhao2023deep} and 53,315 in-house calculations on macrocycle decomposition, click chemistry, and polymer-chain initiation. These additional systems extended coverage to 25 heavy atoms. For these supplemental reactions, existing TS geometries were reoptimized at the same DFT level, followed by IRC calculations.

Before training, records with anomalous endpoints, discontinuous energy profiles, or severely distorted optimization geometries were removed. Retained reactions require positive TS-to-endpoint energy differences no greater than 150~kcal/mol for both IRC endpoints and both optimized endpoints. The database totals and source breakdown are reported separately from the cleaned training populations in Section~\SIref{si:rd11e}.

All three models use the same reaction-level training--test split, with validation data drawn from the training partition. The 1,000 benchmark reactions are excluded from training and validation, although their reactants may occur in these sets. Results stratified by reactant overlap are reported in Tables~S6--S14.

Calculations for the KHP network, IC rearrangement, and proline- and PMe$_3$-catalysed applications were not used for training, and the training data contain no matching reactions from other sources.

\subsection{Sampler construction and training}\label{meth:sampler}
Sampler generates all-atom TS geometries conditioned on the complete reactant graph, total charge, and spin multiplicity, without product information or a three-dimensional reactant geometry. The graph can contain several disconnected molecular components. Sampling does not explicitly enforce stereochemical constraints (Section~\SIref{si:identity_details}). A harmonic prior places springs along reactant bonds and keeps the components in a common spatial region.\autocite{jing2023eigenfold,stark2024harmonic} A flow-matching model transforms independent prior samples into TS candidates, enabling different reaction channels and conformations to be generated from the same input.\autocite{lipman2023flow,liu2023flow}

The model uses the DiTMC transformer, with attention across atom pairs, including pairs from different molecules.\autocite{frank2025sampling} During training, prior samples are aligned to reference TSs, and intermediate geometries are sampled from noise-perturbed interpolations between them. The model learns to recover the reference TS from these intermediate geometries. Either endpoint graph can condition a training TS, exposing both reaction directions. At inference, the learned flow is integrated from an independent prior sample in 50 Euler steps. The full graph features, prior, learning objective, and integration scheme, together with training and sampling algorithms, are specified in Section~\SIref{si:sampler}. Independent prior draws supply sampling diversity; their frequencies are not interpreted as reaction rates or product populations.

\subsection{Simulator construction and inference}\label{meth:simulator}
Simulator predicts a reaction trajectory from a supplied TS geometry, atomic elements, total charge, and spin multiplicity. Its EquiformerV3 backbone predicts displacements from the TS as a function of a normalized progress coordinate, reproducing the supplied structure at the central frame.\autocite{liao2026equiformerv3} This coordinate describes progress along a path, not physical time. Training uses trajectories obtained from DFT IRC calculations, resampled independently to 100 frames on each side of the central TS. The positive branch is oriented towards the lower-energy endpoint during training to avoid inconsistent directional labels.

The objective combines atomic-position and interatomic-distance errors. Perturbing input TS geometries during training provides tolerance to unoptimized candidates. At inference, frames are predicted directly from the supplied geometry without force-based propagation on a specified potential energy surface. Generated, Carina-optimized, and DFT-optimized TS geometries can therefore serve as inputs without requiring a converged saddle point on that surface. The number of predicted frames can be chosen from 3 to 201. Three-frame prediction evaluates the TS and two endpoints for screening; larger frame counts provide finer path resolution. Simulator supplies endpoint assignments and initial geometries for subsequent checks; it does not predict energies. Its architecture, objective, training settings, and endpoint-perception procedure are given in Section~\SIref{si:tracer}.

\subsection{Screening and representative selection}\label{meth:screening}
Generated TS candidates are first clustered by geometrical similarity under the fixed input atom mapping. Up to $k$ geometrically diverse representatives per cluster are selected for Simulator endpoint prediction, where $k$ is configurable. The medoid is selected first, followed by candidates maximizing their minimum distance to the selected set. All members are retained when a cluster contains fewer than $k$ candidates.

Simulator screening retains predictions for which exactly one endpoint matches the input adjacency and the other has different connectivity. Matching preserves the original atom mapping and every explicit hydrogen and fragment (Section~\SIref{si:identity_details}).

Carina single-point energies (SPEs) are used to screen candidates within an energy window relative to the lowest finite candidate SPE in the same input batch. This is a candidate-selection window, not an activation-barrier cutoff. Candidates passing both endpoint and SPE screening are then grouped into pathway families by mapped endpoint states and bond-change signatures. Different bond-change patterns remain separate even when they connect the same reactants and products. Within each family, the lowest-SPE candidate seeds farthest-point sampling, which adds geometrically diverse representatives. This second selection preserves alternative TS arrangements within a pathway family and is distinct from the initial clustering.

Selected representatives can undergo TS refinement and endpoint verification as described in Section~\ref{meth:verification}. Refined connections are assigned from their calculated endpoints, which can differ from the initial predictions. Distance definitions, selection limits, and exploration-specific settings are provided in Section~\SIref{si:workflow_screening}.

\subsection{Carina and stationary-point verification}\label{meth:verification}
Nexus-Verifier uses Carina, an EquiformerV3 reactive MLIP with charge and multiplicity embeddings, for energetic evaluation and pathway refinement.\autocite{liao2026equiformerv3} It was trained on cleaned structures from reactions in the RD11E training partition using energy and force targets, without Hessian labels. Elemental reference energies are added to the predicted atomic energy to recover the total energy, from which forces and Hessians are obtained by differentiation. Full architecture, training settings, and evaluation populations are provided in Sections~\SIref{si:carina} and~\SIref{si:model_evaluation}.

For network exploration, TS optimization uses redundant internal coordinates and restricted-step image rational-function optimization in pysisyphus, with native Gaussian \texttt{gau} convergence criteria, up to 50 steps, and Hessian recalculation every five steps.\autocite{steinmetzer2021pysisyphus} TS candidates are assessed using both optimizer convergence and the presence of a single significant imaginary frequency below $-10$~cm$^{-1}$. Candidates that reach the optimization-step limit may proceed to IRC calculations but remain labelled unconverged.

IRC calculations use the Carina potential in both directions, with endpoint optimization and common-reactant searches configured for each exploration. Raw terminal frames are distinguished from optimized endpoints and are not assumed to be minima without frequency calculations. Any connectivity change during endpoint optimization is attributed to that relaxation. Numerical settings and the treatment of unconverged candidates are detailed in Section~\SIref{si:stationary_verification}, with IC conformational sampling described in Section~\SIref{si:ic_conformers}.

\subsection{Global kinetic-guided network growth}\label{subsec:policy}
Exploration beyond the reactions of the initial reactants requires deciding which product sets to expand next. Each accepted product set, including all co-products and spectators, becomes a candidate reactant set. Each reactant set is expanded on its own, and products of different reactant sets are not recombined.  Unexpanded candidates are kept in a persistent global pool, so that candidates from earlier rounds are considered together with newly generated ones.

Network expansion is ranked using local electronic TS heights relative to a common optimized reactant complex for each input, including all explicit molecules (Section~\SIref{si:energy_details}). These Carina-computed energy differences represent electronic barriers for converged TSs and provisional heights for nonstationary candidates. Let $p$ be a route from the initial reactant set $r$ to a node $v$, $\ell_p$ its number of reaction steps, and $b_{(1)}(p)\geq b_{(2)}(p)\geq\cdots\geq b_{(\ell_p)}(p)$ the local electronic heights of these steps in descending order. The heuristic expansion score of $v$ is
\begin{equation}
S(v)=\min_{p\in\mathcal{P}_{L_{\max}}(r,v)}\sum_{i=1}^{\ell_p}10^{-(i-1)}b_{(i)}(p),
\label{eq:navigation_score}
\end{equation}
where $\mathcal{P}_{L_{\max}}(r,v)$ contains the simple routes from $r$ to $v$ with at most $L_{\max}$ reaction steps. This path-length limit is distinct from the number of expansion rounds. A lower score indicates a higher priority. The largest height receives the greatest weight, with decreasing weights for the remaining heights. This score determines exploration priority, and newly discovered routes can lower the score of an existing candidate.

After $r$ is expanded, each round selects the $n$ lowest-scoring candidates from the global pool. Unselected candidates remain available for later rounds. Once the prescribed calculations for all selected inputs in a round have finished, accepted connections are added to the network, eligible new reactant sets enter the candidate pool, and the scores are updated. Previously completed calculations with matching inputs and compatible settings can be reused only when the corresponding input is selected for expansion.

Exploration-specific settings and stopping criteria are described in Section~\SIref{si:navigation_details}.

\FloatBarrier

\section*{Data and code availability}
Static images of the reaction networks are provided in the Supplementary Information. Interactive HTML views are supplied as separate supplementary files. The reaction database and the trained models will be released upon acceptance of the manuscript. \rnx{} will be available shortly for online use and network visualization through the MIRA platform at \href{\reportcodeurl}{\reportcodeurl}.

\begin{samepage}
\section*{Acknowledgement}
We thank Heather J. Kulik for useful and insightful discussions. We acknowledge the NVIDIA DevTech and ALCHEMI teams for assistance with operator optimization to accelerate EquiformerV3, and Wen Yan and colleagues at ByteDance for supporting the operation and maintenance of high-throughput GPU4PySCF calculations. We also thank the entire Deep Principle team for discussion and support.\par
\end{samepage}

\begingroup\emergencystretch=2em\printbibliography\par\endgroup

@article{unsleber2020exploration,
  title = {{The Exploration of Chemical Reaction Networks}},
  author = {Unsleber, Jan P. and Reiher, Markus},
  year = {2020},
  journal = {Annu. Rev. Phys. Chem.},
  volume = {71},
  number = {1},
  pages = {121--142},
  doi = {10.1146/annurev-physchem-071119-040123},
  url = {https://doi.org/10.1146/annurev-physchem-071119-040123}
}

@article{dewyer2018methods,
  title = {{Methods for exploring reaction space in molecular systems}},
  author = {Dewyer, Amanda L. and Argüelles, Alonso J. and Zimmerman, Paul M.},
  year = {2018},
  journal = {WIREs Comput. Mol. Sci.},
  volume = {8},
  number = {2},
  pages = {e1354},
  doi = {10.1002/wcms.1354},
  url = {https://doi.org/10.1002/wcms.1354}
}

@article{steiner2022autonomous,
  title = {{Autonomous Reaction Network Exploration in Homogeneous and Heterogeneous Catalysis}},
  author = {Steiner, Miguel and Reiher, Markus},
  year = {2022},
  journal = {Top. Catal.},
  volume = {65},
  number = {1-4},
  pages = {6--39},
  doi = {10.1007/s11244-021-01543-9},
  url = {https://doi.org/10.1007/s11244-021-01543-9}
}

@article{wen2023chemical,
  title = {{Chemical reaction networks and opportunities for machine learning}},
  author = {Wen, Mingjian and Spotte-Smith, Evan Walter Clark and Blau, Samuel M. and McDermott, Matthew J. and Krishnapriyan, Aditi S. and Persson, Kristin A.},
  year = {2023},
  journal = {Nat. Comput. Sci.},
  volume = {3},
  number = {1},
  pages = {12--24},
  doi = {10.1038/s43588-022-00369-z},
  url = {https://doi.org/10.1038/s43588-022-00369-z}
}

@article{gao2016reaction,
  title = {{Reaction Mechanism Generator: Automatic construction of chemical kinetic mechanisms}},
  author = {Gao, Connie W. and Allen, Joshua W. and Green, William H. and West, Richard H.},
  year = {2016},
  journal = {Comput. Phys. Commun.},
  volume = {203},
  pages = {212--225},
  doi = {10.1016/j.cpc.2016.02.013},
  url = {https://doi.org/10.1016/j.cpc.2016.02.013}
}

@article{liu2021rmg,
  title = {{Reaction Mechanism Generator v3.0: Advances in Automatic Mechanism Generation}},
  author = {Liu, Mengjie and Grinberg Dana, Alon and Johnson, Matthew S. and Goldman, Mark J. and Jocher, Agnes and Payne, A. Mark and Grambow, Colin A. and Han, Kehang and Yee, Nathan W. and Mazeau, Emily J. and Blondal, Katrin and West, Richard H. and Goldsmith, C. Franklin and Green, William H.},
  year = {2021},
  journal = {J. Chem. Inf. Model.},
  volume = {61},
  number = {6},
  pages = {2686--2696},
  doi = {10.1021/acs.jcim.0c01480},
  url = {https://doi.org/10.1021/acs.jcim.0c01480}
}

@article{broadbelt1994computer,
  title = {{Computer Generated Pyrolysis Modeling: On-the-Fly Generation of Species, Reactions, and Rates}},
  author = {Broadbelt, Linda J. and Stark, Scott M. and Klein, Michael T.},
  year = {1994},
  journal = {Ind. Eng. Chem. Res.},
  volume = {33},
  number = {4},
  pages = {790--799},
  doi = {10.1021/ie00028a003},
  url = {https://doi.org/10.1021/ie00028a003}
}

@article{klucznik2024cationic,
  title = {{Computational prediction of complex cationic rearrangement outcomes}},
  author = {Klucznik, Tomasz and Syntrivanis, Leonidas-Dimitrios and Baś, Sebastian and Mikulak-Klucznik, Barbara and Moskal, Martyna and Szymkuć, Sara and Mlynarski, Jacek and Gadina, Louis and Beker, Wiktor and Burke, Martin D. and Tiefenbacher, Konrad and Grzybowski, Bartosz A.},
  year = {2024},
  journal = {Nature},
  volume = {625},
  number = {7995},
  pages = {508--515},
  doi = {10.1038/s41586-023-06854-3},
  url = {https://doi.org/10.1038/s41586-023-06854-3}
}

@article{wang2014discovering,
  title = {{Discovering chemistry with an ab initio nanoreactor}},
  author = {Wang, Lee-Ping and Titov, Alexey and McGibbon, Robert and Liu, Fang and Pande, Vijay S. and Martínez, Todd J.},
  year = {2014},
  journal = {Nat. Chem.},
  volume = {6},
  number = {12},
  pages = {1044--1048},
  doi = {10.1038/nchem.2099},
  url = {https://doi.org/10.1038/nchem.2099}
}

@article{wang2016automated,
  title = {{Automated Discovery and Refinement of Reactive Molecular Dynamics Pathways}},
  author = {Wang, Lee-Ping and McGibbon, Robert T. and Pande, Vijay S. and Martinez, Todd J.},
  year = {2016},
  journal = {J. Chem. Theory Comput.},
  volume = {12},
  number = {2},
  pages = {638--649},
  doi = {10.1021/acs.jctc.5b00830},
  url = {https://doi.org/10.1021/acs.jctc.5b00830}
}

@article{zhang2024exploring,
  title = {{Exploring the frontiers of condensed-phase chemistry with a general reactive machine learning potential}},
  author = {Zhang, Shuhao and Makoś, Małgorzata Z. and Jadrich, Ryan B. and Kraka, Elfi and Barros, Kipton and Nebgen, Benjamin T. and Tretiak, Sergei and Isayev, Olexandr and Lubbers, Nicholas and Messerly, Richard A. and Smith, Justin S.},
  year = {2024},
  journal = {Nat. Chem.},
  volume = {16},
  number = {5},
  pages = {727--734},
  doi = {10.1038/s41557-023-01427-3},
  url = {https://doi.org/10.1038/s41557-023-01427-3}
}

@article{grimme2019exploration,
  title = {{Exploration of Chemical Compound, Conformer, and Reaction Space with Meta-Dynamics Simulations Based on Tight-Binding Quantum Chemical Calculations}},
  author = {Grimme, Stefan},
  year = {2019},
  journal = {J. Chem. Theory Comput.},
  volume = {15},
  number = {5},
  pages = {2847--2862},
  doi = {10.1021/acs.jctc.9b00143},
  url = {https://doi.org/10.1021/acs.jctc.9b00143}
}

@article{pracht2020automated,
  title = {{Automated exploration of the low-energy chemical space with fast quantum chemical methods}},
  author = {Pracht, Philipp and Bohle, Fabian and Grimme, Stefan},
  year = {2020},
  journal = {Phys. Chem. Chem. Phys.},
  volume = {22},
  number = {14},
  pages = {7169--7192},
  doi = {10.1039/c9cp06869d},
  url = {https://doi.org/10.1039/c9cp06869d}
}

@misc{ichino2024systematic,
  title = {{Systematic and unbiased pathway exploration by artificial force application to a generic neural network potential}},
  author = {Ichino, Tomoya and Nabata, Hitoshi and Matsumoto, Kota and Hasegawa, Taisuke and Maeda, Satoshi},
  year = {2024},
  howpublished = {ChemRxiv preprint},
  publisher = {American Chemical Society (ACS)},
  doi = {10.26434/chemrxiv-2024-9h20v}
}

@article{staub2026accurate,
  title = {{An Accurate and Efficient Reaction Path Search with Iteratively Trained Neural Network Potential: Answering the Passerini Mechanism Controversy}},
  author = {Staub, Ruben and Harabuchi, Yu and Seraphim, Carine and Varnek, Alexandre and Maeda, Satoshi},
  year = {2026},
  journal = {J. Chem. Theory Comput.},
  volume = {22},
  number = {1},
  pages = {422--440},
  doi = {10.1021/acs.jctc.5c01293},
  url = {https://doi.org/10.1021/acs.jctc.5c01293}
}

@article{shang2013stochastic,
  title = {{Stochastic Surface Walking Method for Structure Prediction and Pathway Searching}},
  author = {Shang, Cheng and Liu, Zhi-Pan},
  year = {2013},
  journal = {J. Chem. Theory Comput.},
  volume = {9},
  number = {3},
  pages = {1838--1845},
  doi = {10.1021/ct301010b},
  url = {https://doi.org/10.1021/ct301010b}
}

@article{zimmerman2013growing,
  title = {{Growing string method with interpolation and optimization in internal coordinates: Method and examples}},
  author = {Zimmerman, Paul M.},
  year = {2013},
  journal = {J. Chem. Phys.},
  volume = {138},
  number = {18},
  pages = {184102},
  doi = {10.1063/1.4804162},
  url = {https://doi.org/10.1063/1.4804162}
}

@article{zimmerman2013automated,
  title = {{Automated discovery of chemically reasonable elementary reaction steps}},
  author = {Zimmerman, Paul M.},
  year = {2013},
  journal = {J. Comput. Chem.},
  volume = {34},
  number = {16},
  pages = {1385--1392},
  doi = {10.1002/jcc.23271},
  url = {https://doi.org/10.1002/jcc.23271}
}

@article{grambow2018unimolecular,
  title = {{Unimolecular Reaction Pathways of a $\gamma$-Ketohydroperoxide from Combined Application of Automated Reaction Discovery Methods}},
  author = {Grambow, Colin A. and Jamal, Adeel and Li, Yi-Pei and Green, William H. and Zádor, Judit and Suleimanov, Yury V.},
  year = {2018},
  journal = {J. Am. Chem. Soc.},
  volume = {140},
  number = {3},
  pages = {1035--1048},
  doi = {10.1021/jacs.7b11009},
  url = {https://doi.org/10.1021/jacs.7b11009}
}

@article{unsleber2022chemoton,
  title = {{Chemoton 2.0: Autonomous Exploration of Chemical Reaction Networks}},
  author = {Unsleber, Jan P. and Grimmel, Stephanie A. and Reiher, Markus},
  year = {2022},
  journal = {J. Chem. Theory Comput.},
  volume = {18},
  number = {9},
  pages = {5393--5409},
  doi = {10.1021/acs.jctc.2c00193},
  url = {https://doi.org/10.1021/acs.jctc.2c00193}
}

@article{zhao2021simultaneously,
  title = {{Simultaneously improving reaction coverage and computational cost in automated reaction prediction tasks}},
  author = {Zhao, Qiyuan and Savoie, Brett M.},
  year = {2021},
  journal = {Nat. Comput. Sci.},
  volume = {1},
  number = {7},
  pages = {479--490},
  doi = {10.1038/s43588-021-00101-3},
  url = {https://doi.org/10.1038/s43588-021-00101-3}
}

@article{zhao2022algorithmic,
  title = {{Algorithmic Explorations of Unimolecular and Bimolecular Reaction Spaces}},
  author = {Zhao, Qiyuan and Savoie, Brett M.},
  year = {2022},
  journal = {Angew. Chem. Int. Ed.},
  volume = {61},
  number = {46},
  pages = {e202210693},
  doi = {10.1002/anie.202210693},
  url = {https://doi.org/10.1002/anie.202210693}
}

@article{zhao2022conformational,
  title = {{Conformational Sampling for Transition State Searches on a Computational Budget}},
  author = {Zhao, Qiyuan and Hsu, Hsuan-Hao and Savoie, Brett M.},
  year = {2022},
  journal = {J. Chem. Theory Comput.},
  volume = {18},
  number = {5},
  pages = {3006--3016},
  doi = {10.1021/acs.jctc.2c00081},
  url = {https://doi.org/10.1021/acs.jctc.2c00081}
}

@article{zhao2023comprehensive,
  title = {{Comprehensive exploration of graphically defined reaction spaces}},
  author = {Zhao, Qiyuan and Vaddadi, Sai Mahit and Woulfe, Michael and Ogunfowora, Lawal A. and Garimella, Sanjay S. and Isayev, Olexandr and Savoie, Brett M.},
  year = {2023},
  journal = {Sci. Data},
  volume = {10},
  number = {1},
  pages = {145},
  doi = {10.1038/s41597-023-02043-z},
  url = {https://doi.org/10.1038/s41597-023-02043-z}
}

@article{zhao2023deep,
  title = {{Deep reaction network exploration of glucose pyrolysis}},
  author = {Zhao, Qiyuan and Savoie, Brett M.},
  year = {2023},
  journal = {Proc. Natl Acad. Sci. USA},
  volume = {120},
  number = {34},
  pages = {e2305884120},
  doi = {10.1073/pnas.2305884120},
  url = {https://doi.org/10.1073/pnas.2305884120}
}

@article{zhao2023thermally,
  title = {{Thermally Accessible Prebiotic Pathways for Forming Ribonucleic Acid and Protein Precursors from Aqueous Hydrogen Cyanide}},
  author = {Zhao, Qiyuan and Garimella, Sanjay S. and Savoie, Brett M.},
  year = {2023},
  journal = {J. Am. Chem. Soc.},
  volume = {145},
  number = {11},
  pages = {6135--6143},
  doi = {10.1021/jacs.2c11857},
  url = {https://doi.org/10.1021/jacs.2c11857}
}

@article{zhao2025graphically,
  title = {{Graphically Defined Model Reactions Are Extensible, Accurate, and Systematically Improvable}},
  author = {Zhao, Qiyuan and Singla, Veerupaksh and Hsu, Hsuan-Hao and Savoie, Brett M.},
  year = {2025},
  journal = {J. Chem. Theory Comput.},
  volume = {21},
  number = {13},
  pages = {6572--6583},
  doi = {10.1021/acs.jctc.5c00279},
  url = {https://doi.org/10.1021/acs.jctc.5c00279}
}

@article{unke2021machine,
  title = {{Machine Learning Force Fields}},
  author = {Unke, Oliver T. and Chmiela, Stefan and Sauceda, Huziel E. and Gastegger, Michael and Poltavsky, Igor and Schütt, Kristof T. and Tkatchenko, Alexandre and Müller, Klaus-Robert},
  year = {2021},
  journal = {Chem. Rev.},
  volume = {121},
  number = {16},
  pages = {10142--10186},
  doi = {10.1021/acs.chemrev.0c01111},
  url = {https://doi.org/10.1021/acs.chemrev.0c01111}
}

@inproceedings{batatia2022mace,
  title = {{MACE: Higher Order Equivariant Message Passing Neural Networks for Fast and Accurate Force Fields}},
  author = {Batatia, Ilyes and Kovacs, David P and Simm, Gregor and Ortner, Christoph and Csanyi, Gabor},
  year = {2022},
  booktitle = {Advances in Neural Information Processing Systems}, volume = {35},
  pages = {11423--11436},
  doi = {10.52202/068431-0830}
}

@misc{rhodes2025orbv3,
  title = {{Orb-v3: atomistic simulation at scale}},
  author = {Rhodes, Benjamin and Vandenhaute, Sander and Šimkus, Vaidotas and Gin, James and Godwin, Jonathan and Duignan, Tim and Neumann, Mark},
  year = {2025},
  eprint = {2504.06231},
  archivePrefix = {arXiv},
  publisher = {arXiv},
  doi = {10.48550/arxiv.2504.06231},
  url = {https://arxiv.org/abs/2504.06231}
}

@article{yuan2024analytical,
  title = {{Analytical ab initio hessian from a deep learning potential for transition state optimization}},
  author = {Yuan, Eric C.-Y. and Kumar, Anup and Guan, Xingyi and Hermes, Eric D. and Rosen, Andrew S. and Zádor, Judit and Head-Gordon, Teresa and Blau, Samuel M.},
  year = {2024},
  journal = {Nat. Commun.},
  volume = {15},
  number = {1},
  pages = {8865},
  doi = {10.1038/s41467-024-52481-5},
  url = {https://doi.org/10.1038/s41467-024-52481-5}
}

@article{cui2025horm,
  title = {{A Large Scale Molecular Hessian Database for Optimizing Reactive Machine Learning Interatomic Potentials}},
  author = {Cui, Taoyong and Han, Yunhong and Jia, Haojun and Duan, Chenru and Zhao, Qiyuan},
  year = {2026},
  journal = {Sci. Data},
  volume = {13},
  number = {1},
  pages = {37},
  doi = {10.1038/s41597-025-06350-5},
  url = {https://doi.org/10.1038/s41597-025-06350-5}
}

@article{zhao2025harnessing,
  title = {{Harnessing Machine Learning to Enhance Transition State Search with Interatomic Potentials and Generative Models}},
  author = {Zhao, Qiyuan and Han, Yunhong and Zhang, Duo and Wang, Jiaxu and Zhong, Peichen and Cui, Taoyong and Yin, Bangchen and Cao, Yirui and Jia, Haojun and Duan, Chenru},
  year = {2025},
  journal = {Adv. Sci.},
  volume = {12},
  number = {34},
  pages = {e06240},
  doi = {10.1002/advs.202506240},
  url = {https://doi.org/10.1002/advs.202506240}
}

@article{duan2023accurate,
  title = {{Accurate transition state generation with an object-aware equivariant elementary reaction diffusion model}},
  author = {Duan, Chenru and Du, Yuanqi and Jia, Haojun and Kulik, Heather J.},
  year = {2023},
  journal = {Nat. Comput. Sci.},
  volume = {3},
  number = {12},
  pages = {1045--1055},
  doi = {10.1038/s43588-023-00563-7},
  url = {https://doi.org/10.1038/s43588-023-00563-7}
}

@article{kim2024diffusion,
  title = {{Diffusion-based generative AI for exploring transition states from 2D molecular graphs}},
  author = {Kim, Seonghwan and Woo, Jeheon and Kim, Woo Youn},
  year = {2024},
  journal = {Nat. Commun.},
  volume = {15},
  number = {1},
  pages = {341},
  doi = {10.1038/s41467-023-44629-6},
  url = {https://doi.org/10.1038/s41467-023-44629-6}
}

@article{duan2025optimal,
  title = {{Optimal transport for generating transition states in chemical reactions}},
  author = {Duan, Chenru and Liu, Guan-Horng and Du, Yuanqi and Chen, Tianrong and Zhao, Qiyuan and Jia, Haojun and Gomes, Carla P. and Theodorou, Evangelos A. and Kulik, Heather J.},
  year = {2025},
  journal = {Nat. Mach. Intell.},
  volume = {7},
  number = {4},
  pages = {615--626},
  doi = {10.1038/s42256-025-01010-0},
  url = {https://doi.org/10.1038/s42256-025-01010-0}
}

@article{galustian2025goflow,
  title = {{GoFlow: efficient transition state geometry prediction with flow matching and E(3)-equivariant neural networks}},
  author = {Galustian, Leonard and Mark, Konstantin and Karwounopoulos, Johannes and Kovar, Maximilian P.-P. and Heid, Esther},
  year = {2025},
  journal = {Digit. Discov.},
  volume = {4},
  number = {12},
  pages = {3492--3501},
  doi = {10.1039/d5dd00283d},
  url = {https://doi.org/10.1039/d5dd00283d}
}

@misc{sun2026unified,
  title = {{A Unified Generative Framework for Scalable Chemical Reaction Network Exploration}},
  author = {Sun, Zechang and Hu, Chenxi and Lin, Kailai and Li, Jin and Cao, Changsu and Lv, Dingshun and Chen, Ji and Ren, Weiluo and Pham, Hung Q.},
  year = {2026},
  eprint = {2606.21002},
  archivePrefix = {arXiv},
  publisher = {arXiv},
  doi = {10.48550/arxiv.2606.21002},
  url = {https://arxiv.org/abs/2606.21002}
}

@article{jalan2013new,
  title = {{New Pathways for Formation of Acids and Carbonyl Products in Low-Temperature Oxidation: The Korcek Decomposition of $\gamma$-Ketohydroperoxides}},
  author = {Jalan, Amrit and Alecu, Ionut M. and Meana-Pañeda, Rubén and Aguilera-Iparraguirre, Jorge and Yang, Ke R. and Merchant, Shamel S. and Truhlar, Donald G. and Green, William H.},
  year = {2013},
  journal = {J. Am. Chem. Soc.},
  volume = {135},
  number = {30},
  pages = {11100--11114},
  doi = {10.1021/ja4034439},
  url = {https://doi.org/10.1021/ja4034439}
}

@article{oliveira2017inverse,
  title = {{Inverse electron demand Diels--Alder reactions in chemical biology}},
  author = {Oliveira, B. L. and Guo, Z. and Bernardes, G. J. L.},
  year = {2017},
  journal = {Chem. Soc. Rev.},
  volume = {46},
  number = {16},
  pages = {4895--4950},
  doi = {10.1039/c7cs00184c},
  url = {https://doi.org/10.1039/c7cs00184c}
}

@article{list2000proline,
  title = {{Proline-Catalyzed Direct Asymmetric Aldol Reactions}},
  author = {List, Benjamin and Lerner, Richard A. and Barbas, Carlos F.},
  year = {2000},
  journal = {J. Am. Chem. Soc.},
  volume = {122},
  number = {10},
  pages = {2395--2396},
  doi = {10.1021/ja994280y},
  url = {https://doi.org/10.1021/ja994280y}
}

@article{bahmanyar2001transition,
  title = {{Transition States of Amine-Catalyzed Aldol Reactions Involving Enamine Intermediates: Theoretical Studies of Mechanism, Reactivity, and Stereoselectivity}},
  author = {Bahmanyar, S. and Houk, K. N.},
  year = {2001},
  journal = {J. Am. Chem. Soc.},
  volume = {123},
  number = {45},
  pages = {11273--11283},
  doi = {10.1021/ja011403h},
  url = {https://doi.org/10.1021/ja011403h}
}

@article{bahmanyar2003quantum,
  title = {{Quantum Mechanical Predictions of the Stereoselectivities of Proline-Catalyzed Asymmetric Intermolecular Aldol Reactions}},
  author = {Bahmanyar, S. and Houk, K. N. and Martin, Harry J. and List, Benjamin},
  year = {2003},
  journal = {J. Am. Chem. Soc.},
  volume = {125},
  number = {9},
  pages = {2475--2479},
  doi = {10.1021/ja028812d},
  url = {https://doi.org/10.1021/ja028812d}
}

@article{abdelgaid2026identification, title={Identification of Solid-Electrolyte Interphase Species by Joint Characterization of Li-Ion Battery Chemistry by Mass Spectrometry and Electrochemical Reaction Networks}, volume={148}, ISSN={1520-5126}, url={https://doi.org/10.1021/jacs.6c04755}, DOI={10.1021/jacs.6c04755}, number={16}, journal={J. Am. Chem. Soc.}, publisher={American Chemical Society (ACS)}, author={Abdelgaid, Mona and Hvidsten, Oliver and Sombret, Théo and Kherchiche, Egon and Maillard, Julien and Gajan, Antonin and Bernard, Patrick and Kaźmierczak, Kamila and Araya-Polo, Mauricio and Salvato Vallverdu, Germain and Afonso, Carlos and Giusti, Pierre and Persson, Kristin A.}, year={2026}, pages={17493--17507} }

@article{seumer2026practical, title={Practical Computational Exploration of Chemical Reaction Space: Assessing the Chemical Stability of Active Pharmaceutical Ingredients Under Acidic Conditions}, ISSN={1521-3773}, url={https://doi.org/10.1002/anie.4407622}, DOI={10.1002/anie.4407622}, journal={Angew. Chem. Int. Ed.}, publisher={Wiley}, author={Seumer, Julius and Rasmussen, Maria Harris and Brea, Oriana and Muuronen, Mikko and Göller, Andreas H. and Jensen, Jan H.}, year={2026},
  pages = {e4407622},
  volume = {65},
  number = {44}
}

@article{caldeweyher2019d4, title={A generally applicable atomic-charge dependent London dispersion correction}, volume={150}, ISSN={1089-7690}, url={https://doi.org/10.1063/1.5090222}, DOI={10.1063/1.5090222}, number={15}, journal={J. Chem. Phys.}, publisher={AIP Publishing}, author={Caldeweyher, Eike and Ehlert, Sebastian and Hansen, Andreas and Neugebauer, Hagen and Spicher, Sebastian and Bannwarth, Christoph and Grimme, Stefan}, year={2019},
  pages = {154122}
}

@article{rappoport2010property, title={Property-optimized Gaussian basis sets for molecular response calculations}, volume={133}, ISSN={1089-7690}, url={https://doi.org/10.1063/1.3484283}, DOI={10.1063/1.3484283}, number={13}, journal={J. Chem. Phys.}, publisher={AIP Publishing}, author={Rappoport, Dmitrij and Furche, Filipp}, year={2010},
  pages = {134105}
}

@article{steinmetzer2021pysisyphus, title={pysisyphus: Exploring potential energy surfaces in ground and excited states}, volume={121}, ISSN={1097-461X}, url={https://doi.org/10.1002/qua.26390}, DOI={10.1002/qua.26390}, number={3}, journal={Int. J. Quantum Chem.}, publisher={Wiley}, author={Steinmetzer, Johannes and Kupfer, Stephan and Gräfe, Stefanie}, year={2021},
  pages = {e26390}
}

@inproceedings{lipman2023flow,
  title = {Flow Matching for Generative Modeling},
  author = {Lipman, Yaron and Chen, Ricky T. Q. and Ben-Hamu, Heli and Nickel, Maximilian and Le, Matt},
  booktitle = {The Eleventh International Conference on Learning Representations},
  year = {2023},
  url = {https://openreview.net/forum?id=PqvMRDCJT9t}
}

@inproceedings{liu2023flow,
  title = {Flow Straight and Fast: Learning to Generate and Transfer Data with Rectified Flow},
  author = {Liu, Xingchao and Gong, Chengyue and Liu, Qiang},
  booktitle = {The Eleventh International Conference on Learning Representations},
  year = {2023},
  url = {https://openreview.net/forum?id=gWxpdtQpiYV}
}

@article{li2025introducing, title={Introducing GPU Acceleration into the Python-Based Simulations of Chemistry Framework}, volume={129}, ISSN={1520-5215}, url={https://doi.org/10.1021/acs.jpca.4c05876}, DOI={10.1021/acs.jpca.4c05876}, number={5}, journal={J. Phys. Chem. A}, publisher={American Chemical Society (ACS)}, author={Li, Rui and Sun, Qiming and Zhang, Xing and Chan, Garnet Kin-Lic}, year={2025}, pages={1459--1468} }

@article{bannwarth2019gfn2,
  title={GFN2-xTB---An Accurate and Broadly Parametrized Self-Consistent Tight-Binding Quantum Chemical Method with Multipole Electrostatics and Density-Dependent Dispersion Contributions},
  author={Bannwarth, Christoph and Ehlert, Sebastian and Grimme, Stefan},
  journal={J. Chem. Theory Comput.},
  volume={15},
  number={3},
  pages={1652--1671},
  year={2019},
  publisher={ACS Publications},
  doi={10.1021/acs.jctc.8b01176},
  url={https://doi.org/10.1021/acs.jctc.8b01176}
}

@article{kim2023pubchem,
  title={PubChem 2023 update},
  author={Kim, Sunghwan and Chen, Jie and Cheng, Tiejun and Gindulyte, Asta and He, Jia and He, Siqian and Li, Qingliang and Shoemaker, Benjamin A. and Thiessen, Paul A. and Yu, Bo and Zaslavsky, Leonid and Zhang, Jian and Bolton, Evan E.},
  journal={Nucleic Acids Res.},
  volume={51},
  number={D1},
  pages={D1373--D1380},
  year={2023},
  doi={10.1093/nar/gkac956},
  publisher={Oxford University Press},
  url={https://doi.org/10.1093/nar/gkac956}
}

@misc{liao2026equiformerv3,
  title = {{EquiformerV3}: Scaling Efficient, Expressive, and General {SE(3)}-Equivariant Graph Attention Transformers},
  author = {Liao, Yi-Lun and Hoffman, Alexander J. and Shen, Sabrina C. and Duval, Alexandre and Norwood, Sam Walton and Smidt, Tess},
  year = {2026}, eprint = {2604.09130}, archivePrefix = {arXiv},
  doi = {10.48550/arXiv.2604.09130}, url = {https://arxiv.org/abs/2604.09130}
}

@article{ribeiro2011use,
  title = {Use of Solution-Phase Vibrational Frequencies in Continuum Models for the Free Energy of Solvation},
  author = {Ribeiro, Raphael F. and Marenich, Aleksandr V. and Cramer, Christopher J. and Truhlar, Donald G.},
  journal = {J. Phys. Chem. B},
  volume = {115}, number = {49}, pages = {14556--14562}, year = {2011},
  doi = {10.1021/jp205508z}, url = {https://doi.org/10.1021/jp205508z}
}

@article{maeda2018implementation,
  title = {Implementation and performance of the artificial force induced reaction method in the {GRRM17} program},
  author = {Maeda, Satoshi and Harabuchi, Yu and Takagi, Makito and Saita, Kenichiro and Suzuki, Kimichi and Ichino, Tomoya and Sumiya, Yosuke and Sugiyama, Kanami and Ono, Yuriko},
  journal = {J. Comput. Chem.},
  volume = {39}, number = {4}, pages = {233--251}, year = {2018},
  doi = {10.1002/jcc.25106}, url = {https://doi.org/10.1002/jcc.25106}
}

@article{sumiya2020rate,
  title = {Rate Constant Matrix Contraction Method for Systematic Analysis of Reaction Path Networks},
  author = {Sumiya, Yosuke and Maeda, Satoshi},
  journal = {Chem. Lett.},
  volume = {49}, number = {5}, pages = {553--564}, year = {2020},
  doi = {10.1246/cl.200092}, url = {https://doi.org/10.1246/cl.200092}
}

@article{hayashi2022silico,
  title = {In silico reaction screening with difluorocarbene for {N}-difluoroalkylative dearomatization of pyridines},
  author = {Hayashi, Hiroki and Katsuyama, Hitomi and Takano, Hideaki and Harabuchi, Yu and Maeda, Satoshi and Mita, Tsuyoshi},
  journal = {Nat. Synth.},
  volume = {1}, number = {10}, pages = {804--814}, year = {2022},
  doi = {10.1038/s44160-022-00128-y}, url = {https://doi.org/10.1038/s44160-022-00128-y}
}

@inproceedings{frank2025sampling,
  title     = {Sampling {3D} Molecular Conformers with Diffusion Transformers},
  author    = {Frank, J. Thorben and Ripken, Winfried and Lied, Gregor and M{\"u}ller, Klaus-Robert and Unke, Oliver T. and Chmiela, Stefan},
  booktitle = {Advances in Neural Information Processing Systems},
  volume    = {38},
  year      = {2025},
  pages     = {187081--187131},
  doi       = {10.52202/085713-5626}
}

@article{kabsch1976solution,
  title   = {A solution for the best rotation to relate two sets of vectors},
  author  = {Kabsch, Wolfgang},
  journal = {Acta Crystallogr. A},
  volume  = {32},
  number  = {5},
  pages   = {922--923},
  year    = {1976}
}

@article{jing2023eigenfold,
  title   = {{EigenFold}: Generative Protein Structure Prediction with Diffusion Models},
  author  = {Jing, Bowen and Erives, Ezra and Pao-Huang, Peter and Corso, Gabriele and Berger, Bonnie and Jaakkola, Tommi},
  journal = {arXiv preprint arXiv:2304.02198},
  year    = {2023}
}

@inproceedings{stark2024harmonic,
  title     = {Harmonic Self-Conditioned Flow Matching for Joint Multi-Ligand Docking and Binding Site Design},
  author    = {St{\"a}rk, Hannes and Jing, Bowen and Barzilay, Regina and Jaakkola, Tommi},
  booktitle = {Proceedings of the 41st International Conference on Machine Learning},
  series    = {Proceedings of Machine Learning Research},
  volume    = {235},
  year      = {2024},
  pages = {46468--46494}
}

@inproceedings{klein2023equivariant,
  title     = {Equivariant Flow Matching},
  author    = {Klein, Leon and Kr{\"a}mer, Andreas and No{\'e}, Frank},
  booktitle = {Advances in Neural Information Processing Systems},
  volume    = {36},
  year      = {2023},
  pages = {59886--59910},
  doi = {10.52202/075280-2617}
}

@article{tong2024improving,
  title   = {Improving and generalizing flow-based generative models with minibatch optimal transport},
  author  = {Tong, Alexander and Fatras, Kilian and Malkin, Nikolay and Huguet, Guillaume and Zhang, Yanlei and Rector-Brooks, Jarrid and Wolf, Guy and Bengio, Yoshua},
  journal = {Transactions on Machine Learning Research},
  year    = {2024}
}

@inproceedings{loshchilov2019decoupled,
  title     = {Decoupled Weight Decay Regularization},
  author    = {Loshchilov, Ilya and Hutter, Frank},
  booktitle = {International Conference on Learning Representations},
  year      = {2019}
}

@inproceedings{loshchilov2017sgdr,
  title     = {{SGDR}: Stochastic Gradient Descent with Warm Restarts},
  author    = {Loshchilov, Ilya and Hutter, Frank},
  booktitle = {International Conference on Learning Representations},
  year      = {2017}
}

@inproceedings{hoogeboom2022equivariant,
  title     = {Equivariant Diffusion for Molecule Generation in {3D}},
  author    = {Hoogeboom, Emiel and Satorras, V{\'i}ctor Garcia and Vignac, Cl{\'e}ment and Welling, Max},
  booktitle = {Proceedings of the 39th International Conference on Machine Learning},
  series    = {Proceedings of Machine Learning Research},
  volume    = {162},
  pages     = {8867--8887},
  year      = {2022}
}

@inproceedings{tancik2020fourier,
  title     = {Fourier Features Let Networks Learn High Frequency Functions in Low Dimensional Domains},
  author    = {Tancik, Matthew and Srinivasan, Pratul P. and Mildenhall, Ben and Fridovich-Keil, Sara and Raghavan, Nithin and Singhal, Utkarsh and Ramamoorthi, Ravi and Barron, Jonathan T. and Ng, Ren},
  booktitle = {Advances in Neural Information Processing Systems},
  volume    = {33},
  year      = {2020},
  pages = {7537--7547}
}

@article{bensberg2023concentration,
  title = {Concentration-Flux-Steered Mechanism Exploration with an Organo\-catalysis Application},
  author = {Bensberg, Moritz and Reiher, Markus},
  journal = {Isr. J. Chem.},
  year = {2023},
  volume = {63},
  number = {7-8},
  pages = {e202200123},
  doi = {10.1002/ijch.202200123},
  url = {https://doi.org/10.1002/ijch.202200123}
}

@article{liu2023optimal,
  title = {An optimal {Fe--C} coordination ensemble for hydrocarbon chain growth: a full {Fischer--Tropsch} synthesis mechanism from machine learning},
  author = {Liu, Qian-Yu and Chen, Dongxiao and Shang, Cheng and Liu, Zhi-Pan},
  journal = {Chem. Sci.},
  year = {2023},
  volume = {14},
  number = {35},
  pages = {9461--9475},
  doi = {10.1039/D3SC02054A},
  url = {https://doi.org/10.1039/D3SC02054A}
}

@article{burger2026hip,
  title = {{HIP}: {Hessian Interatomic Potentials} without derivatives},
  author = {Burger, Andreas and Thiede, Luca and Rønne, Nikolaj and Vijaykumar, Nandita and Bernales, Varinia and Vegge, Tejs and Bhowmik, Arghya and Aspuru-Guzik, Alán},
  journal = {Nat. Comput. Sci.},
  year = {2026},
  doi = {10.1038/s43588-026-01066-x},
  url = {https://doi.org/10.1038/s43588-026-01066-x}
}

@article{anstine2026aimnet2rxn,
  title = {{AIMNet2-rxn}: A machine-learned potential for generalized reaction modeling on a millions-of-pathways scale},
  author = {Anstine, Dylan M. and Zhao, Qiyuan and Zubatyuk, Roman and Zhang, Shuhao and Singla, Veerupaksh and Nikitin, Filipp and Savoie, Brett M. and Isayev, Olexandr},
  journal = {Sci. Adv.},
  year = {2026},
  volume = {12},
  number = {41},
  pages = {eaea1557},
  doi = {10.1126/sciadv.aea1557},
  url = {https://doi.org/10.1126/sciadv.aea1557}
}

@article{tuo2026flow,
  title = {{Flow matching for reaction pathway generation}},
  author = {Tuo, Ping and Chen, Jiale and Li, Ju},
  year = {2026},
  journal = {Nat. Commun.},
  volume = {17},
  pages = {8769},
  doi = {10.1038/s41467-026-75654-w},
  url = {https://doi.org/10.1038/s41467-026-75654-w}
}

@article{xia2007water,
  author = {Xia, Yuanzhi and Liang, Yong and Chen, Yuanyuan and Wang, Ming and Jiao, Lei and Huang, Feng and Liu, Song and Li, Yahong and Yu, Zhi-Xiang},
  title = {An Unexpected Role of a Trace Amount of Water in Catalyzing Proton Transfer in Phosphine-Catalyzed (3 + 2) Cycloaddition of Allenoates and Alkenes},
  journal = {J. Am. Chem. Soc.},
  year = {2007},
  volume = {129},
  number = {12},
  pages = {3470--3471},
  doi = {10.1021/ja068215h}
}

@article{liang2008phosphine,
  author = {Liang, Yong and Liu, Song and Xia, Yuanzhi and Li, Yahong and Yu, Zhi-Xiang},
  title = {Mechanism, Regioselectivity, and the Kinetics of Phosphine-Catalyzed [3+2] Cycloaddition Reactions of Allenoates and Electron-Deficient Alkenes},
  journal = {Chem. Eur. J.},
  year = {2008},
  volume = {14},
  number = {14},
  pages = {4361--4373},
  doi = {10.1002/chem.200701725}
}

@article{lee2026reactionqm,
  title = {A Comprehensive Dataset of Chemical Reactions Covering Second and Third Row Elements with Million-Scale Quantum Chemical Calculations},
  author = {Lee, Kyunghoon and Kang, Sungwoo and Kang, Beomgyu and Son, Won-Joon and Lee, Seungmin and Kim, Dae Sin and Kim, Woo Youn},
  journal = {Sci. Data},
  year = {2026},
  volume = {13},
  pages = {983},
  doi = {10.1038/s41597-026-07325-w}
}

@article{lee2019unexpected,
  author = {Lee, Chung Whan and Taylor, Buck L. H. and Petrova, Galina P. and Patel, Ashay and Morokuma, Keiji and Houk, K. N. and Stoltz, Brian M.},
  title = {An Unexpected Ireland--Claisen Rearrangement Cascade During the Synthesis of the Tricyclic Core of Curcusone C: Mechanistic Elucidation by Trial-and-Error and Automatic Artificial Force-Induced Reaction (AFIR) Computations},
  journal = {J. Am. Chem. Soc.},
  year = {2019},
  volume = {141},
  number = {17},
  pages = {6995--7004},
  doi = {10.1021/jacs.9b01146}
}

@article{young2021autode,
  author = {Young, Tom A. and Silcock, Joseph J. and Sterling, Alistair J. and Duarte, Fernanda},
  title = {{autodE}: Automated Calculation of Reaction Energy Profiles---Application to Organic and Organometallic Reactions},
  journal = {Angew. Chem. Int. Ed.},
  year = {2021},
  volume = {60},
  number = {8},
  pages = {4266--4274},
  doi = {10.1002/anie.202011941}
}

@article{allemann2010computational,
  title = {Computational investigations of the stereoselectivities of proline-related catalysts for aldol reactions},
  author = {Allemann, Christophe and Um, Joann M. and Houk, K. N.},
  year = {2010},
  journal = {J. Mol. Catal. A Chem.},
  volume = {324},
  number = {1-2},
  pages = {31--38},
  doi = {10.1016/j.molcata.2010.03.020},
  url = {https://doi.org/10.1016/j.molcata.2010.03.020}
}

@article{dewitt2000ratebased,
  title = {Computer Generation of Reaction Mechanisms Using Quantitative Rate Information: Application to Long-Chain Hydrocarbon Pyrolysis},
  author = {De Witt, Matthew J. and Dooling, David J. and Broadbelt, Linda J.},
  year = {2000},
  journal = {Ind. Eng. Chem. Res.},
  volume = {39},
  number = {7},
  pages = {2228--2237},
  doi = {10.1021/ie990608k},
  url = {https://doi.org/10.1021/ie990608k}
}

@article{lavecchia2004pyrazolopyridines,
  author = {Lavecchia, G. and Berteina-Raboin, S. and Guillaumet, G.},
  title = {Synthesis and functionalisation of 1H-pyrazolo[3,4-b]pyridines involving copper and palladium-promoted coupling reactions},
  journal = {Tetrahedron Lett.},
  year = {2004},
  volume = {45},
  number = {11},
  pages = {2389--2392},
  doi = {10.1016/j.tetlet.2004.01.067},
  url = {https://doi.org/10.1016/j.tetlet.2004.01.067}
}

@article{sakas2024analysis,
  title = {MS and NMR Analysis of Isotopically Labeled Chloramination Disinfection Byproducts: Hyperlinks and Chemical Reactions},
  author = {Sakas, Justinas and Kitson, Ezra and Bell, Nicholle G. A. and Uhr{\'i}n, Du{\v s}an},
  journal = {Anal. Chem.},
  year = {2024},
  volume = {96},
  number = {21},
  pages = {8263--8272},
  doi = {10.1021/acs.analchem.3c03888},
  url = {https://doi.org/10.1021/acs.analchem.3c03888}
}

@article{aldaz2018discovery,
  author = {Aldaz, Cody and Kammeraad, Joshua A. and Zimmerman, Paul M.},
  title = {Discovery of conical intersection mediated photochemistry with growing string methods},
  journal = {Phys. Chem. Chem. Phys.},
  year = {2018},
  volume = {20},
  pages = {27394--27405},
  doi = {10.1039/C8CP04703K}
}

@article{xu2026automated,
  author = {Xu, Li-Cheng and An, Junyi and Liu, Weiqi and Shi, Yun-Fei and Ji, Chong-Lei and Cao, Fenglei and Qi, Yuan},
  title = {Automated transition state generation for mechanistic exploration in organic synthesis},
  journal = {Nat. Commun.},
  year = {2026},
  volume = {17},
  pages = {10254},
  doi = {10.1038/s41467-026-77230-8},
  url = {https://doi.org/10.1038/s41467-026-77230-8}
}

@article{chai2008systematic,
  author = {Chai, Jeng-Da and Head-Gordon, Martin},
  title = {Systematic optimization of long-range corrected hybrid density functionals},
  journal = {J. Chem. Phys.},
  volume = {128},
  pages = {084106},
  year = {2008},
  doi = {10.1063/1.2834918}
}

@article{marenich2009universal,
  author = {Marenich, Aleksandr V. and Cramer, Christopher J. and Truhlar, Donald G.},
  title = {Universal Solvation Model Based on Solute Electron Density and on a Continuum Model of the Solvent Defined by the Bulk Dielectric Constant and Atomic Surface Tensions},
  journal = {J. Phys. Chem. B},
  volume = {113},
  pages = {6378--6396},
  year = {2009},
  doi = {10.1021/jp810292n}
}

@article{joung2025electron,
  author = {Joung, Joonyoung F. and Fong, Mun Hong and Casetti, Nicholas and Liles, Jordan P. and Dassanayake, Ne S. and Coley, Connor W.},
  title = {Electron flow matching for generative reaction mechanism prediction},
  journal = {Nature},
  year = {2025},
  volume = {645},
  pages = {115--123},
  doi = {10.1038/s41586-025-09426-9}
}

@article{maziarz2026chemist,
  author = {Maziarz, Krzysztof and Liu, Guoqing and Pultar, Felix and Gardner, John and Gensch, Tobias and Helie, Jean and Misztela, Hubert and Tripp, Austin and Li, Junren and Kornev, Aleksei and Gaiński, Piotr and Hoefling, Holger and Fortunato, Mike and Gupta, Rishi and Baxter, Andrew and Poole, Darren L. and Elward, Jennifer M. and Krzyzanowski, Adrian and Pogány, Peter and Pickett, Stephen D. and Wall, Ian D. and Bishop, Christopher M. and Humphreys, Philip G. and Lumley, James A. and Wiesenfeldt, Mario P. and Segler, Marwin H. S.},
  title = {Chemist-aligned retrosynthesis by ensembling diverse inductive bias models},
  journal = {Nature},
  year = {2026},
  doi = {10.1038/s41586-026-11160-9}
}

@misc{mcinnes2018umap,
  author = {McInnes, Leland and Healy, John and Melville, James},
  title = {{UMAP}: Uniform Manifold Approximation and Projection for Dimension Reduction},
  year = {2018},
  eprint = {1802.03426},
  archivePrefix = {arXiv},
  primaryClass = {stat.ML},
  doi = {10.48550/arXiv.1802.03426}
}

\end{document}


\begin{center}
    {\Large\bfseries Supporting Information \par}
    \vspace{1em}
    {\large\bfseries \reporttitle\par}
    \vspace{1em}
    \reportauthors\par\vspace{0.8em}
    {\small\reportaffiliations\par}
\end{center}

\setcounter{tocdepth}{1}
\begingroup\makeatletter\renewcommand{\l@section}{\@dottedtocline{1}{0em}{3em}}\renewcommand{\l@subsection}{\@dottedtocline{2}{1.5em}{4em}}\makeatother\tableofcontents\endgroup

\clearpage

\section{Reaction database}\label{si:database_construction}\label{si:database_summary}\label{si:rd11e}

\subsection{Construction and chemical scope}
We assembled a reaction database designed for breadth in elements, charge and spin states, system size, and number of components. Database construction is summarized in Fig.~\Mainref{fig:overview}b. From PubChem we selected approximately half a million molecular seeds containing only H, B, C, N, O, F, Si, P, S, Cl, and Br, covering open-shell and closed-shell systems in neutral, cationic, and anionic states with up to 15 heavy atoms.\autocite{kim2023pubchem} For each compound, candidate reactions were generated by random bond changes involving up to three broken and three formed bonds, extending the b2f2 and b3f3 elementary reaction steps used in YARP,\autocite{zhao2021simultaneously,zhao2022algorithmic} and unphysical candidates were removed with a tree-based filter.\autocite{zhao2022conformational} Multi-molecular reactions were enumerated between pairs of reactants and with explicit catalysts or solvent molecules. Bimolecular examples were selected from RGD1 and Reaction-QM.\autocite{zhao2023comprehensive,lee2026reactionqm} Water-catalysed examples came from previous hydrogen cyanide and glucose studies,\autocite{zhao2023thermally,zhao2023deep} while in-house calculations on macrocycle decomposition, click chemistry, and polymer-chain initiation extended coverage to reaction systems with up to 25 heavy atoms. 

Core reactions were characterized with the YARP-derived reaction characterization workflow. In the first phase, reaction conformational sampling,\autocite{zhao2022conformational} GFN2-xTB,\autocite{bannwarth2019gfn2} growing string calculations,\autocite{zimmerman2013growing} TS optimization, and IRC were performed on CPUs, yielding approximately 4 million reactions with valid TSs. Growing-string calculations used pyGSM in double-ended mode (DE-GSM),\autocite{zimmerman2013growing,aldaz2018discovery} and TS optimization and IRC calculations were performed with pysisyphus.\autocite{steinmetzer2021pysisyphus} In the second phase, 1 million of these reactions were selected for refinement by DFT TS optimization and IRC at the $\omega$B97X-D4/def2-TZVPD level using GPU4PySCF 1.4.1 in the gas phase.\autocite{chai2008systematic,caldeweyher2019d4,rappoport2010property,li2025introducing} The functional/dispersion specification was \texttt{wb97x-d4:wb97x-2008}, explicitly identifying the D4 parameter set. Supplemental reactions from external datasets and earlier calculations entered this DFT stage directly: their existing TS geometries were reoptimized, followed by IRC calculations at the same level. Database totals below precede training-data cleaning and selection. 

\subsection{Dataset composition}
RD11E contains \textbf{1,113,366 reaction records} and \textbf{130,325,537 structures}. The reaction systems contain up to 25 heavy atoms, with total charges of $-1$, $0$, or $+1$ and spin multiplicities of 1 or 2. 

\begingroup\emergencystretch=1em
All structures include atomic coordinates and an electronic energy. Forces are available for 124,296,911 structures and Hessians for 4,972,426. Table~\ref*{tab:rd11e_census} summarizes the structures and available labels from TS optimization, IRC calculations, and endpoint optimization, including the stored TS and endpoint structures.\par\endgroup 

\begin{table}[!htbp]
\centering
\caption{Summary of structures and available properties in RD11E.}
\label{tab:rd11e_census}
\small
\setlength{\tabcolsep}{8pt}
\begin{tabular}{lrrr}
\toprule
Calculation or structure & Structures & With forces & With Hessians \\
\midrule
TS optimization (TSOPT) & 10,974,879 & 8,431,578 & 3,859,060 \\
Structures from IRC calculations & 74,275,122 & 74,162,708 & 0 \\
Forward endpoint optimization (FOPT) & 19,765,682 & 18,252,478 & 0 \\
Backward endpoint optimization (BOPT) & 19,748,210 & 18,239,568 & 0 \\
TS structure & 1,113,366 & 935,939 & 1,113,366 \\
IRC forward endpoint & 1,113,366 & 1,111,470 & 0 \\
IRC backward endpoint & 1,113,366 & 1,111,470 & 0 \\
Optimized reactant endpoint & 1,110,787 & 1,025,599 & 0 \\
Optimized product endpoint & 1,110,759 & 1,026,101 & 0 \\
\midrule
\textbf{Total} & \textbf{130,325,537} & \textbf{124,296,911} & \textbf{4,972,426} \\
\bottomrule
\end{tabular}
\end{table}

\subsection{\textcolor{black}{Data cleaning and partitioning}}\label{si:overlap_scope}
Before model training, RD11E data are cleaned to remove records with anomalous endpoint geometries, discontinuous IRC energy profiles, or severely distorted structures during TS optimization. Reactions are retained only if the energy differences between the TS and all four endpoints (IRC reactant, IRC product, optimized reactant, and optimized product) are positive and no greater than 150~kcal/mol.

All three models use the same reaction-level training--test split, with validation data drawn from the training partition. \textcolor{black}{Reaction records are mutually exclusive across the training, validation, and test sets.}

Calculations for the KHP network, IC rearrangement, and proline- and PMe$_3$-catalysed applications were not used for training, and the training data contain no matching reactions from other sources. The glucose and four-step hydrogen cyanide networks contribute training examples; the glucose network is not used for validation.

\subsection{Reaction and energy distributions}
\begin{figure}[H]
\centering
\includegraphics[width=\textwidth]{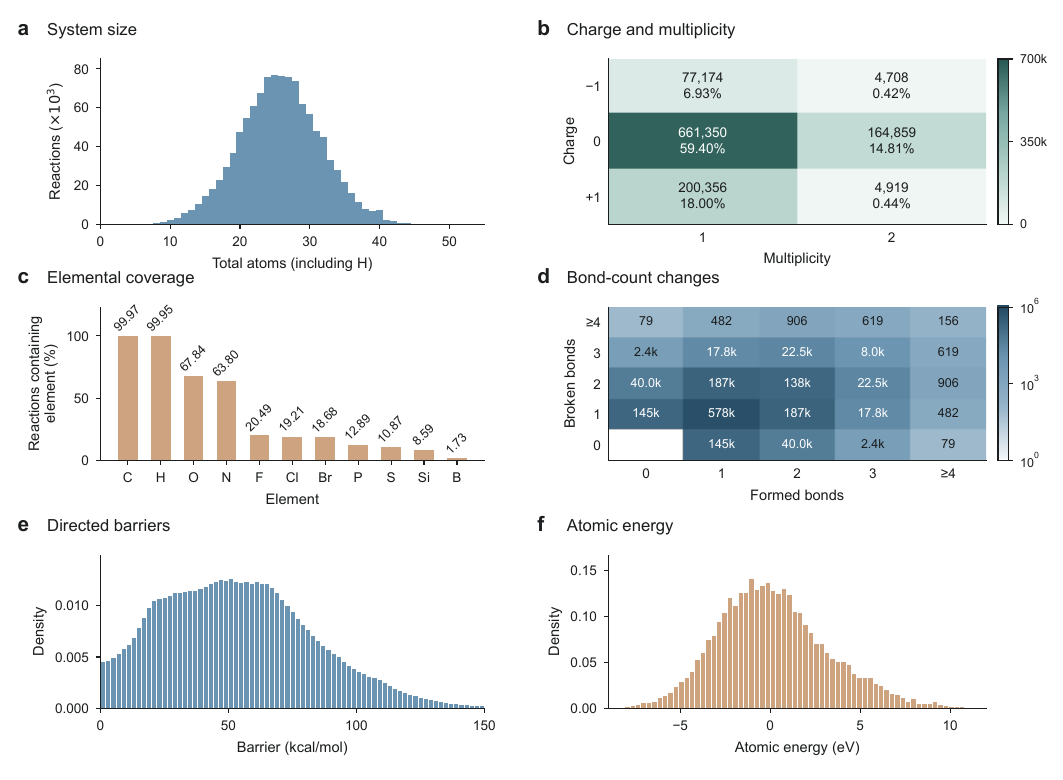}
\captionsetup{font=small}
\caption{\textbf{Fig.~\thefigure{} \textbar{} Composition and reaction statistics of RD11E.}
\textbf{a.} Distribution of total atom count, including hydrogen.
\textbf{b.} Charge and spin multiplicity across the 1,113,366 reaction records; cells show counts and percentages.
\textbf{c.} Fraction of reaction records containing each element; categories overlap.
\textbf{d.} Bond-count changes for 1,559,119 unique directed endpoint-graph pairs with consistent bond annotations. Both directions are included; counts use a logarithmic color scale.
\textbf{e.} Forward and reverse electronic barriers for 1,084,080 reactions (2,168,160 values) satisfying the energy criteria described above.
\textbf{f.} Atomic energies for 236,881 structures from 1,997 sampled reactions satisfying the same energy criteria. The displayed range contains 99.87\% of the structures, and the density is normalized over the full sample.}
\label{fig:rd11e_overview}
\end{figure}

Fig.~\ref*{fig:rd11e_overview} summarizes the composition and reaction characteristics of RD11E. Panels a--c show the distributions of system size, charge, spin multiplicity, and elemental composition, while panel d shows the numbers of bonds broken and formed during reactions.

The energy distributions are shown in Fig.~\ref*{fig:rd11e_overview}e-f. Forward and reverse barriers are calculated relative to the optimized reactant and product energies, respectively. These are electronic energies without zero-point or thermal corrections. Panel f shows the atomic energy distribution, using the definition in Section~\ref*{si:carina_outputs} and equal weights for the sampled structures.

\subsection{Molecular chemical space}
We use UMAP\autocite{mcinnes2018umap} to visualize approximately 250,000 structures with distinct molecular graphs from a random sample of RD11E.

\begin{figure}[!htbp]
\centering
\includegraphics[width=\textwidth]{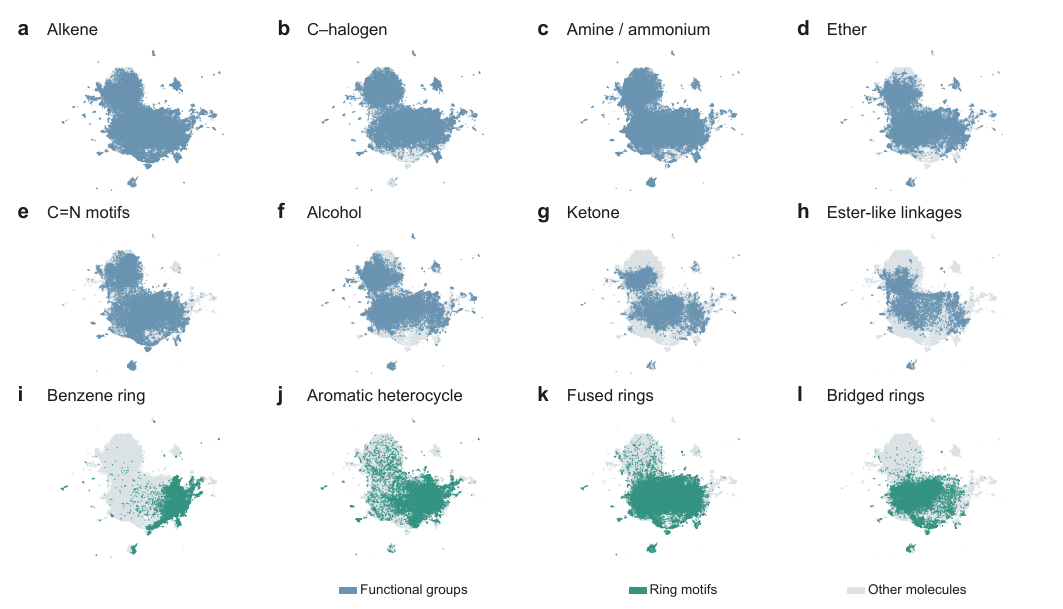}
\captionsetup{font=small}
\caption{\textbf{Fig.~\thefigure{} \textbar{} Functional groups and ring systems in RD11E molecular chemical space.}
All panels show the same UMAP embedding. Colored points indicate molecules matching the named category, and gray points indicate other molecules. Blue denotes functional groups and teal denotes ring systems. The C=N category includes imine-, oxime-, and hydrazone-type motifs, while ester-like linkages include carbonate and carbamate motifs.}
\label{fig:rd11e_chemical_space}
\end{figure}

Molecules can belong to more than one category in Fig.~\ref*{fig:rd11e_chemical_space}.

\FloatBarrier

\section{Sampler architecture and training}\label{si:sampler}

\begin{figure}[!htbp]
\centering
\includegraphics[width=0.65\textwidth]{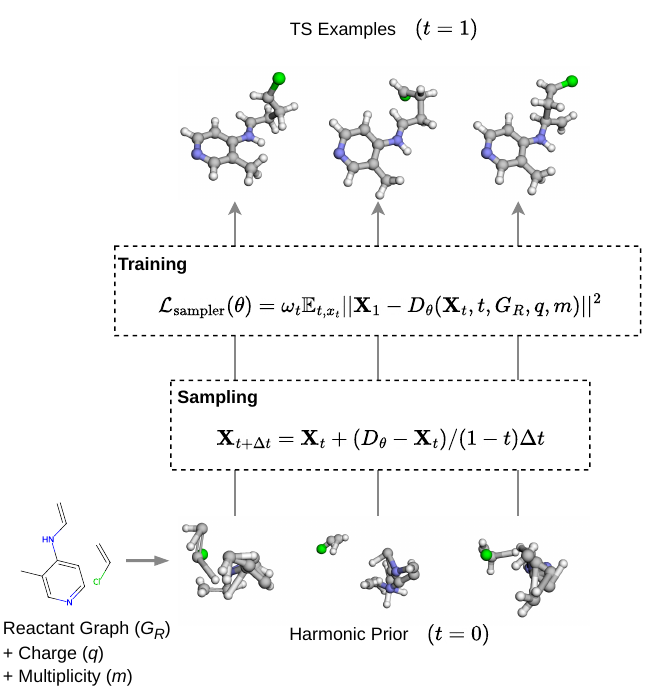}
\caption[Overview of Nexus-Sampler]{\textbf{Fig.~\thefigure{} \textbar{} Overview of \nsampler{}.} The reactant graph $G_R$, with total charge $q$ and spin multiplicity $m$, is the only input. During training, a sample $\mathbf{X}_0$ of the harmonic prior on $G_R$ is aligned to the reference TS $\mathbf{X}_1$, an intermediate structure $\mathbf{X}_t$ is drawn from the noise-smoothed interpolation between them, and the DiTMC network $D_\theta$, conditioned on $G_R$, $q$, and $m$, learns to recover $\mathbf{X}_1$. At inference, a fresh prior sample is integrated in 50 Euler steps along $v_\theta=(D_\theta-\mathbf{X}_t)/(1-t)$; independent prior samples yield different TS conformers and reaction channels.}
\label{fig:si_sampler}
\end{figure}

\nsampler{} generates three-dimensional TS candidates for all atoms of a reacting system from its molecular graph, total charge, and spin multiplicity. It combines flow matching from a harmonic prior on the reactant graph (Sections~\ref*{si:sampler_prior} and~\ref*{si:sampler_fm}) with the DiTMC transformer developed for molecular conformer generation\autocite{frank2025sampling}. Generating TSs differs from generating conformers in four respects. The target is a saddle point rather than a minimum; the conditioning graph may contain several molecules and must specify the charge, spin multiplicity and radical centers; the graph of either endpoint is a valid condition for the same TS; and the generated distribution should cover several reaction channels. Figure~\ref*{fig:si_sampler} summarizes the model, Table~\ref*{tab:si_sampler_hparams} lists all hyperparameters, and Algorithms~\ref*{alg:si_sampler_train} and~\ref*{alg:si_sampler_sample} give the training step and the sampler.

\subsection{Notation}\label{si:sampler_notation}
A reacting system has $n$ atoms with atomic numbers $Z_1,\dots,Z_n$. Its reactant graph $G_R=(V,E_R)$ has one node per atom, hydrogens included, and one edge per covalent bond. A system containing several molecules is represented as a disconnected graph, with each molecule forming a connected component. Node $i$ carries a feature vector $\mathbf{a}_i$, bond $(i,j)$ carries $\mathbf{b}_{ij}$, and $q$ and $m$ are the total charge and spin multiplicity. The $i$th row $\mathbf{x}_i$ of a geometry $\mathbf{X}\in\mathbb{R}^{n\times3}$ is the position of atom $i$ in \AA{}, with atoms ordered as in $G_R$. Because the TS shares this order, no atom matching is required. Translations are removed by working in the zero-centroid subspace $\mathcal{X}=\{\mathbf{X}:\sum_i\mathbf{x}_i=\mathbf{0}\}$, with orthogonal projection $\Pi\mathbf{X}=\mathbf{X}-\tfrac{1}{n}\mathbf{1}\mathbf{1}^{\top}\mathbf{X}$. A rotation $\mathbf{R}\in\mathrm{SO}(3)$ acts as $\mathbf{X}\mapsto\mathbf{X}\mathbf{R}^{\top}$, and $\|\cdot\|$ is the Frobenius norm.

\subsection{Training records.} \nsampler{} is trained on the DFT-refined reactions of the database (Section~S1). Each record provides the $\omega$B97X-D4/def2-TZVPD TS $\mathbf{X}^\ddagger$, the optimized IRC endpoints $\mathbf{X}_R$ and $\mathbf{X}_P$, and $q$ and $m$. The records are split into 541688, 67741, and 67788 reactions for training, validation, and testing.

\subsection{Graph featurization}\label{si:sampler_data}

\paragraph{Endpoint symmetrization.} A TS is the saddle point between two minima and is therefore a valid target for the graph of either. Whenever a record is drawn during training, the conditioning graph is perceived from $\mathbf{X}_R$ or $\mathbf{X}_P$ with equal probability, while the target TS is unchanged. This doubles the number of distinct graph--TS pairs and trains the model on reverse as well as forward reactions.

\paragraph{Node and edge features.} Each atom is described by the 64-dimensional vector $\mathbf{a}_i$ of Table~\ref*{tab:si_sampler_features}. The element is encoded separately, as the identity of the atom token. The total charge and spin multiplicity are appended to every node vector, $\tilde{\mathbf{a}}_i=[\mathbf{a}_i,q,m]$, so that the electronic state is specified explicitly even where the formal charges or radical sites perceived from a geometry are ambiguous. Each bond enters as two directed edges with a six-dimensional one-hot vector $\mathbf{b}_{ij}$ over single, double, triple, aromatic, and other bonds, plus one class for non-bonded pairs.

\begin{table}[!htbp]
\centering
\caption{\textbf{Atom features of the reactant graph.} One-hot encodings include an additional ``other'' category for values outside the listed range, which is counted in the dimension.}
\label{tab:si_sampler_features}
\begin{threeparttable}
\begin{tabular}{@{}llr@{}}
\toprule
Feature & Values & Dimension \\
\midrule
Chirality tag & unspecified, tetrahedral CW, tetrahedral CCW, other & 5 \\
Total degree & 0--10 & 12 \\
Formal charge & $-5$ to $+5$ & 12 \\
Implicit valence & 0--6 & 8 \\
Attached hydrogens & 0--8 & 10 \\
Hybridization & sp, sp$^2$, sp$^3$, sp$^3$d, sp$^3$d$^2$ & 6 \\
Radical electrons & 0--4 & 6 \\
Aromatic & no, yes & 2 \\
In ring & no, yes & 2 \\
Ring count & integer & 1 \\
\midrule
Total & & 64 \\
\bottomrule
\end{tabular}
\end{threeparttable}
\end{table}

\subsection{Harmonic prior}\label{si:sampler_prior}
The prior is the Gaussian distribution of a spring network on the reactant graph.\autocite{jing2023eigenfold,stark2024harmonic} Let $\mathbf{L}_{\mathrm{bond}}=\sum_{(i,j)\in E_R}(\mathbf{e}_i-\mathbf{e}_j)(\mathbf{e}_i-\mathbf{e}_j)^{\top}$ be the Laplacian of the bonds and $\mathbf{L}_{K}=n\mathbf{I}-\mathbf{1}\mathbf{1}^{\top}$ that of the complete graph, where $\mathbf{e}_i$ is the $i$th unit vector. Then
\begin{equation}
p_0(\mathbf{X}_0\mid G_R)\propto\exp\left[-\tfrac{1}{2}\operatorname{tr}\left(\mathbf{X}_0^{\top}\mathbf{L}\,\mathbf{X}_0\right)\right],\quad \mathbf{X}_0\in\mathcal{X},\qquad \mathbf{L}=\kappa_{\mathrm{b}}\mathbf{L}_{\mathrm{bond}}+\kappa_{\mathrm{p}}\mathbf{L}_{K},
\label{eq:si_prior}
\end{equation}
so each Cartesian column of $\mathbf{X}_0$ is independently distributed as $\mathcal{N}(\mathbf{0},\mathbf{L}^{+})$. The spring constants are set through the effective resistance $\mathcal{R}_{ij}=(\mathbf{e}_i-\mathbf{e}_j)^{\top}\mathbf{L}^{+}(\mathbf{e}_i-\mathbf{e}_j)$, which gives $\mathbb{E}\|\mathbf{x}_i-\mathbf{x}_j\|^2=3\mathcal{R}_{ij}$. An isolated bond has resistance $1/\kappa_{\mathrm{b}}$, and the complete graph alone gives every pair the resistance $2/(n\kappa_{\mathrm{p}})$. We therefore choose
\begin{equation}
\kappa_{\mathrm{b}}=\frac{3}{r_{\mathrm{b}}^{2}},\qquad \kappa_{\mathrm{p}}=\frac{6}{n\,r_{\mathrm{p}}^{2}},\qquad r_{\mathrm{b}}=\SI{1.5}{\angstrom},\quad r_{\mathrm{p}}=\SI{20}{\angstrom}.
\label{eq:si_prior_constants}
\end{equation}
With these constants, $r_{\mathrm{b}}$ is an upper bound on the root-mean-square bond length, which rings and the all-pair springs shorten further, and the all-pair term alone would separate atoms by $r_{\mathrm{p}}$ on average, independently of system size. The stiff bond springs set the local structure. The weak all-pair springs fix the overall extent and confine the molecules of a multi-component input to a common region, whose relative positions bond springs alone would leave undetermined. Springs between atoms two bonds apart are deliberately omitted. A harmonic pair potential is minimal at coincidence, so such springs would compress the structure rather than open its valence angles.

The harmonic prior starts the flow closer to the data than an isotropic Gaussian does. Centered TSs in the training set have a root-mean-square Cartesian coordinate of \SI{1.70}{\angstrom}, so a unit Gaussian is 1.7 times too compact. Because it places atoms independently, its closest atom pair is on average \SI{0.31}{\angstrom} apart, against \SI{1.03}{\angstrom} in TSs. Harmonic-prior samples reproduce a mean bonded distance of \SI{1.30}{\angstrom} and lie at a Kabsch-aligned RMSD of \SI{2.54}{\angstrom} from their target TS, compared with \SI{2.95}{\angstrom} for the unit Gaussian, which shortens the transport that the flow must learn.

\subsection{Flow-matching objective}\label{si:sampler_fm}

\paragraph{Probability path.} Let $\mathbf{X}_1=\Pi\mathbf{X}^\ddagger$ be a reference TS and $\mathbf{X}_0\sim p_0(\cdot\mid G_R)$ a prior sample. $\mathbf{X}_0$ is first superimposed onto $\mathbf{X}_1$ by the proper rotation and translation that minimize $\|\mathbf{X}_0\mathbf{R}^{\top}+\mathbf{1}\bm{\tau}^{\top}-\mathbf{X}_1\|$ (Kabsch algorithm\autocite{kabsch1976solution}) and is then re-centered. Because atoms are labeled by $G_R$, no permutation needs to be matched. The alignment removes the rotational component of the transport, so that the network learns changes in shape rather than reorientations, and it shortens the straight-line paths between paired structures.\autocite{klein2023equivariant} The aligned pair is then rotated jointly by a uniformly random rotation. With $t\sim\mathcal{U}[0,1]$ drawn for each reaction and $\bm{\epsilon}=\Pi\tilde{\bm{\epsilon}}$, where $\tilde{\bm{\epsilon}}\in\mathbb{R}^{n\times3}$ has independent standard normal entries, the intermediate structure is
\begin{equation}
\mathbf{X}_t=\Pi\left[(1-t)\,\mathbf{X}_0+t\,\mathbf{X}_1+\sigma\bm{\epsilon}\right],\qquad \sigma=\SI{0.5}{\angstrom}.
\label{eq:si_interpolant}
\end{equation}
The noise of constant width smooths every intermediate distribution, including both endpoints.\autocite{tong2024improving} The network is thus trained on a band of structures around the straight-line paths rather than on the paths alone, which makes it robust to the deviations that accumulate during numerical integration.

\paragraph{Parameterization and loss.} Following DiTMC,\autocite{frank2025sampling} the network $D_\theta(\mathbf{X}_t,t\mid G_R,q,m)\in\mathcal{X}$ predicts the TS rather than a velocity. The velocity used for sampling is
\begin{equation}
v_\theta(\mathbf{X}_t,t)=\frac{D_\theta(\mathbf{X}_t,t\mid G_R,q,m)-\mathbf{X}_t}{1-t},
\label{eq:si_velocity}
\end{equation}
and the loss for one reaction, averaged over a minibatch, is
\begin{equation}
\ell=\frac{w(t)}{3n}\left\|D_\theta(\mathbf{X}_t,t\mid G_R,q,m)-\mathbf{X}_1\right\|^{2},\qquad w(t)=\frac{1}{\left[1-\min(t,t_{\mathrm{c}})\right]^{2}},\quad t_{\mathrm{c}}=0.9.
\label{eq:si_loss}
\end{equation}

\subsection{Training}\label{si:sampler_training}
\nsampler{} is implemented in PyTorch with PyTorch Geometric and PyTorch Lightning and trained with distributed data parallelism on 8 GPUs (NVIDIA H20), with 64 reactions per GPU in each minibatch. Systems of different size are padded to the largest in the minibatch for the transformer, whereas the graph conditioner operates on the sparse bond graph. Parameters are optimized with AdamW\autocite{loshchilov2019decoupled} ($\beta_1=0.95$, $\beta_2=0.999$, weight decay 0.01). The learning rate rises linearly from \num{1e-5} to \num{1e-4} over the first \num{5e4} steps and then decays along a single cosine\autocite{loshchilov2017sgdr} to \num{1e-5} at \num{5e6} steps. Gradients are clipped adaptively,\autocite{hoogeboom2022equivariant} limiting the global norm to $\min(1.5\mu+2s,\,100)$, where $\mu$ and $s$ are the mean and standard deviation of the 50 most recent gradient norms. An exponential moving average (EMA) of the parameters, with decay $\min[0.999,(1+k)/(10+k)]$ after $k$ updates, is maintained from step \num{1000} and used for validation, checkpointing, and sampling. Training proceeds in epochs of \num{5000} steps for up to \num{1000} epochs. The checkpoint with the lowest validation loss is used throughout, and matrix multiplications use TF32 precision. Table~\ref*{tab:si_sampler_hparams} lists all settings, and Algorithm~\ref*{alg:si_sampler_train} summarizes one training step.

\begin{table}[!htbp]
\centering
\caption{\textbf{Hyperparameters of \nsampler{}.}}
\label{tab:si_sampler_hparams}
\small
\setlength{\tabcolsep}{4pt}
\begin{tabular}{@{}lll@{}}
\toprule
Component & Setting & Value \\
\midrule
\multirow{3}{*}{Inputs}
 & Atom features $\tilde{\mathbf{a}}_i$ & 64 + $q$ + $m$ \\
 & Bond features $\mathbf{b}_{ij}$ & 6 (one-hot) \\
 & Endpoint graph probability & 0.5 each \\
\midrule
\multirow{2}{*}{Harmonic prior}
 & Bond length scale $r_{\mathrm{b}}$ & \SI{1.5}{\angstrom} \\
 & All-pair extent $r_{\mathrm{p}}$ & \SI{20}{\angstrom} \\
\midrule
\multirow{5}{*}{Flow matching}
 & Smoothing noise $\sigma$ & \SI{0.5}{\angstrom} \\
 & Time distribution & $\mathcal{U}[0,1]$ \\
 & Loss weight & $[1-\min(t,0.9)]^{-2}$ \\
 & Alignment of $\mathbf{X}_0$ to $\mathbf{X}_1$ & Kabsch, proper rotations \\
 & Rotation augmentation & uniform on $\mathrm{SO}(3)$ \\
\midrule
\multirow{6}{*}{Network (DiTMC+rPE-L)}
 & Transformer blocks & 8 \\
 & Heads $\times$ head width & $12\times32$ \\
 & Hidden width $H$ / FFN width & 384 / 1536 \\
 & Graph-conditioner layers & 3 (width 384) \\
 & Graph-distance embeddings & 512 ($s_{\infty}=510$) \\
 & Parameters & 28.3M \\
\midrule
\multirow{9}{*}{Training}
 & Optimizer & AdamW, $\beta_1=0.95$, $\beta_2=0.999$ \\
 & Weight decay & 0.01 \\
 & Learning rate & \num{1e-5} $\rightarrow$ \num{1e-4} $\rightarrow$ \num{1e-5} \\
 & Warm-up / cosine horizon & \num{5e4} / \num{5e6} steps \\
 & Minibatch & 64 reactions per GPU $\times$ 8 GPUs \\
 & Steps per epoch / maximum epochs & \num{5000} / \num{1000} \\
 & Gradient clipping & adaptive, $\min(1.5\mu+2s,100)$ over 50 steps \\
 & EMA decay / start & 0.999 / step \num{1000} \\
 & Model selection & lowest validation loss (EMA weights) \\
\midrule
Sampling & Integrator & Euler, $K=50$ uniform steps \\
\bottomrule
\end{tabular}
\end{table}

\begin{algorithm}[!htbp]
\caption{Training of \nsampler{}. The loop body is written for one reaction and runs in parallel over the minibatch. All structures are centered, and $\textsc{Kabsch}(\mathbf{X}_0,\mathbf{X}_1)$ returns $\mathbf{X}_0$ after the proper rotation that best superimposes it onto $\mathbf{X}_1$.}
\label{alg:si_sampler_train}
\begin{algorithmic}[1]
\Require training reactions, each with TS $\mathbf{X}^\ddagger$, optimized endpoints $\mathbf{X}_R$ and $\mathbf{X}_P$, charge $q$, and multiplicity $m$
\Repeat
  \State Draw a minibatch $\mathcal{B}$ of reactions
  \ForAll{reactions in $\mathcal{B}$}
    \State $G\gets$ graph of $\mathbf{X}_R$ or of $\mathbf{X}_P$, each with probability $1/2$ \Comment{endpoint symmetrization}
    \State $\mathbf{X}_1\gets\Pi\mathbf{X}^\ddagger$ \Comment{target TS}
    \State $\mathbf{X}_0\sim p_0(\cdot\mid G)$ \Comment{harmonic prior, Eq.~\ref*{eq:si_prior}}
    \State $\mathbf{X}_0\gets\textsc{Kabsch}(\mathbf{X}_0,\mathbf{X}_1)$ \Comment{superimpose prior sample on TS}
    \State Rotate $\mathbf{X}_0$ and $\mathbf{X}_1$ by the same random rotation \Comment{augmentation}
    \State $t\sim\mathcal{U}[0,1]$;\quad $\bm{\epsilon}\sim\mathcal{N}(\mathbf{0},\mathbf{I})$, centered
    \State $\mathbf{X}_t\gets(1-t)\,\mathbf{X}_0+t\,\mathbf{X}_1+\sigma\bm{\epsilon}$ \Comment{Eq.~\ref*{eq:si_interpolant}}
    \State $\hat{\mathbf{X}}_1\gets D_\theta(\mathbf{X}_t,t\mid G,q,m)$ \Comment{predicted TS}
    \State $\ell\gets w(t)\,\|\hat{\mathbf{X}}_1-\mathbf{X}_1\|^{2}/(3n)$ \Comment{Eq.~\ref*{eq:si_loss}}
  \EndFor
  \State Update $\theta$ by AdamW on the mean of $\ell$ over $\mathcal{B}$; update the EMA of $\theta$
\Until{the step budget is reached}
\State \Return the EMA parameters with the lowest validation loss
\end{algorithmic}
\end{algorithm}

\subsection{Sampling}\label{si:sampler_sampling}
A TS is generated by integrating $\mathrm{d}\mathbf{X}_t/\mathrm{d}t=v_\theta(\mathbf{X}_t,t)$ (Eq.~\ref*{eq:si_velocity}) with $K=50$ explicit Euler steps on the uniform grid $t_k=k/K$. Integration starts from a prior sample $\mathbf{X}^{(0)}\sim p_0(\cdot\mid G_R)$, without alignment or added smoothing noise, and each step reads
\begin{equation}
\mathbf{X}^{(k+1)}=\Pi\left[\mathbf{X}^{(k)}+\frac{1}{K-k}\left(D_\theta(\mathbf{X}^{(k)},t_k\mid G_R,q,m)-\mathbf{X}^{(k)}\right)\right],\qquad k=0,\dots,K-1.
\label{eq:si_euler}
\end{equation}
Each step thus moves the structure a fraction $1/(K-k)$ of the way towards the current prediction. The last step has unit size and returns the prediction $D_\theta(\mathbf{X}^{(K-1)},t_{K-1})$ itself, so the sample carries no residual smoothing noise. A sample costs $K$ network evaluations. Because the integration is deterministic, independent prior samples are the only source of diversity among TS conformers and reaction channels. Inputs are batched over systems and over repeated draws for the same system, and coordinates are returned in \AA{}, centered, in the atom order of $G_R$. Algorithm~\ref*{alg:si_sampler_sample} summarizes the sampler; the clustering and selection of samples within \rnx{} are described in Section~\Mainref{sec:framework} of the main text.

\begin{algorithm}[!htbp]
\caption{Sampling $N$ TS candidates with \nsampler{}. The $N$ samples are independent and run in parallel.}
\label{alg:si_sampler_sample}
\begin{algorithmic}[1]
\Require reactant graph $G_R$; charge $q$; multiplicity $m$; number of samples $N$; number of steps $K=50$
\For{$s=1,\dots,N$}
  \State $\mathbf{X}\sim p_0(\cdot\mid G_R)$ \Comment{start from the harmonic prior}
  \For{$k=0,\dots,K-1$}
    \State $t\gets k/K$
    \State $\hat{\mathbf{X}}_1\gets D_\theta(\mathbf{X},t\mid G_R,q,m)$ \Comment{predicted TS}
    \State $\mathbf{v}\gets(\hat{\mathbf{X}}_1-\mathbf{X})/(1-t)$ \Comment{velocity, Eq.~\ref*{eq:si_velocity}}
    \State $\mathbf{X}\gets\mathbf{X}+\mathbf{v}/K$ \Comment{Euler step}
  \EndFor
  \State $\mathbf{X}^\ddagger_s\gets\mathbf{X}$ \Comment{the last step lands on $\hat{\mathbf{X}}_1$}
\EndFor
\State \Return $\{\mathbf{X}^\ddagger_s\}_{s=1}^{N}$
\end{algorithmic}
\end{algorithm}

\FloatBarrier

\section{Simulator architecture and training}\label{si:tracer}

\begin{figure}[htbp]
\centering
\includegraphics[width=\textwidth]{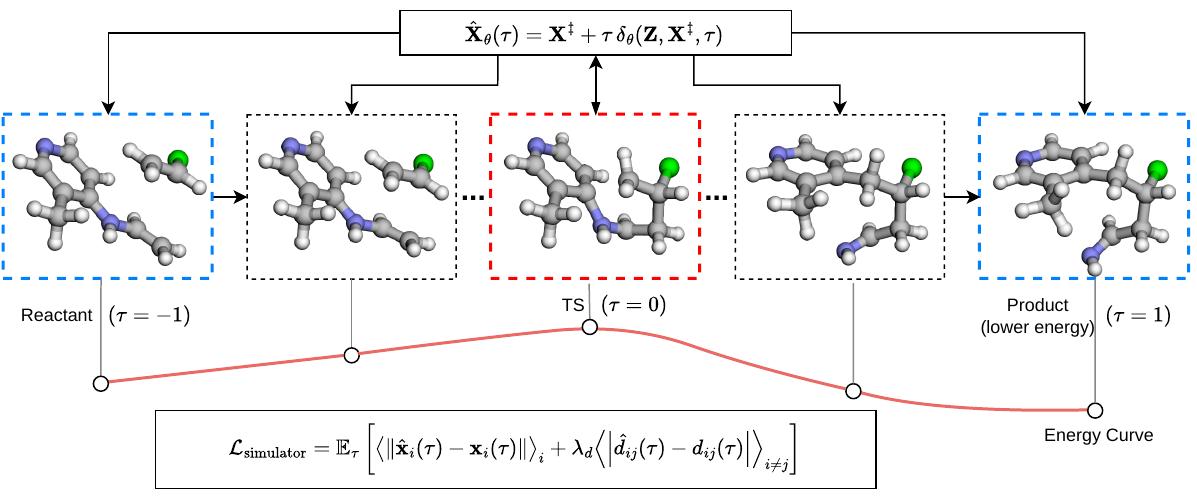}
\caption[Overview of Nexus-Simulator]{\textbf{Fig.~\thefigure{} \textbar{} Overview of \ntracer{}.} The input TS (red box, $\tau=0$) is mapped to every frame of the reaction path by $\hat{\mathbf{X}}_\theta(\tau)=\mathbf{X}^\ddagger+\tau\,\bm{\delta}_\theta(\mathbf{Z},\mathbf{X}^\ddagger,\tau)$, with the two IRC endpoints (blue boxes) at $\tau=\pm1$. All frames are predicted from the same TS and differ only in $\tau$. The model is trained by minimizing the per-atom position error and the interatomic-distance error $D$ between predicted and reference frames, weighted by $\lambda_d$. The energy profile is shown for orientation only; \ntracer{} predicts geometries, not energies.}
\label{fig:si_tracer}
\end{figure}

\nsampler{} proposes TSs without products, so the species that each TS connects must be identified before it can enter a network. \ntracer{} predicts this information directly: from a TS geometry, the elements, the total charge, and the spin multiplicity, it generates the reaction path through the TS into the two adjacent basins. Unlike \nsampler{}, it is a deterministic model that returns one trajectory per TS. Figure~\ref*{fig:si_tracer} summarizes the model, Table~\ref*{tab:si_tracer_hparams} lists all hyperparameters, and Algorithms~\ref*{alg:si_tracer_train} and~\ref*{alg:si_tracer_sample} give the training procedure and the generation of a trajectory.

\subsection{Training data and path coordinate}\label{si:tracer_data}

\paragraph{Trajectories.} \ntracer{} is trained on database trajectories obtained from DFT IRC calculations at the $\omega$B97X-D4/def2-TZVPD level. Each trajectory consists of the TS frame and the two IRC branches, with an energy for every frame. Each branch is independently interpolated in Cartesian coordinates to $T=100$ frames, with the original TS structure inserted between the branches. Every reaction is thus represented by $2T+1=201$ frames $\{\mathbf{X}_t\}_{t=-T}^{T}$ with the TS at $t=0$. Coordinates are expressed in the Cartesian frame of the TS and centered on the TS centroid, and no frame is aligned separately. The endpoints are the terminal IRC frames rather than separately optimized minima, so they lie near but not at the bottom of their basins.

\paragraph{Progress coordinate.} Frame $t$ is assigned the coordinate $\tau=t/T\in[-1,1]$, normalized separately on each side of the TS. A coordinate defined over the raw frame index would place the TS at a reaction-dependent value, because native IRC branches differ in length. Identical inputs $(\mathbf{X}^\ddagger,\tau)$ would then have targets at different stages of the reaction, and the model would average them.

\paragraph{Orientation.} From the TS geometry alone, which side of a saddle point is called reactant or product is a convention, and the stored frame order is not a consistent function of geometry. Training on the stored order would present nearly identical inputs with targets on opposite branches, which drives the prediction towards their average, the TS itself. We therefore orient $\tau$ by the IRC endpoint energies: if $E_{-T}<E_{T}$, the frame order is reversed, so that the positive branch always ends in the lower-energy endpoint. Only the two endpoint energies are used, and only during training. At inference, the positive branch is the model's prediction of the lower-energy basin. Which branch leads back to the input reactant is determined afterwards by comparing the endpoint graphs with $G_R$ (Section~\ref*{si:tracer_inference}). For nearly thermoneutral reactions, the orientation label is less informative, and the assignment of branches relies on this graph comparison alone.

\subsection{Model}\label{si:tracer_model}

\paragraph{Residual parameterization.} The predicted frame at coordinate $\tau$ is
\begin{equation}
\hat{\mathbf{X}}_\theta(\tau)=\tilde{\mathbf{X}}^\ddagger+\tau\,\bm{\delta}_\theta\left(\mathbf{Z},\tilde{\mathbf{X}}^\ddagger,\tau\mid q,m\right),
\label{eq:si_tracer_head}
\end{equation}
where $\tilde{\mathbf{X}}^\ddagger$ is the centered input TS and $\bm{\delta}_\theta\in\mathbb{R}^{n\times3}$ is the network output. The reaction path is a small deformation of the TS: on average it spans \SI{0.30}{\angstrom} per atom, whereas atoms lie \SI{2.78}{\angstrom} from the molecular centroid. Predicting absolute coordinates would require the network to reproduce the identity map to within a few percent before any reaction displacement could be resolved. The residual form removes this burden, and the factor $\tau$ makes $\hat{\mathbf{X}}_\theta(0)=\tilde{\mathbf{X}}^\ddagger$ exact. For small $|\tau|$, $\hat{\mathbf{X}}_\theta(\tau)\approx\tilde{\mathbf{X}}^\ddagger+\tau\,\bm{\delta}_\theta(\mathbf{Z},\tilde{\mathbf{X}}^\ddagger,0)$, so the two branches leave the TS in opposite directions along a common vector. This vector plays the role of the transition vector from which an IRC is started, but it is inferred from the geometry without a Hessian. Every frame is predicted independently from the same input and differs only in $\tau$; the path is continuous because $\bm{\delta}_\theta$ is a smooth function of $\tau$.

\begingroup\emergencystretch=2em\paragraph{Backbone.} $\bm{\delta}_\theta$ is an \textcolor{black}{EquiformerV3} network\autocite{liao2026equiformerv3} with 12 transformer blocks. Node features are spherical-harmonic expansions up to degree $l_{\max}=4$ (order $m_{\max}=2$) with 32 channels, initialized from an embedding of the element. Edges connect atom pairs within each reacting system, including pairs belonging to different molecular components. Distances are expanded in 600 Gaussian radial basis functions on $[0,\SI{5}{\angstrom}]$ and multiplied by a smooth polynomial envelope that vanishes at \SI{5}{\angstrom}, so each block mixes information within \SI{5}{\angstrom} and the receptive field grows with depth. The coordinate $\tau$ is encoded by sine and cosine features at eight frequencies spaced logarithmically from 1 to 64 cycles per unit $\tau$, which counteracts the spectral bias of a plain MLP and lets the network resolve the rapid geometric changes near the TS.\autocite{tancik2020fourier} These features are mapped by an MLP to 8 channels, and $(q,m)$ by a second MLP to another 8 channels. Both are concatenated as additional channels to the node features before the first block. An equivariant graph-attention head with a single output channel produces the network output, whose degree-one components give the displacement $\bm{\delta}_{\theta,i}$ of each atom. Stochastic depth (rate 0.05) and attention-weight dropout (rate 0.1) are applied during training. The network has 27.2M parameters.\par\endgroup

\subsection{Training objective}\label{si:tracer_loss}

\paragraph{Frame sampling and input noise.} In each step, one frame is drawn per reaction: a branch is chosen with probability $1/2$, and a frame uniformly within that branch, so that both branches are equally represented and $\tau$ is approximately uniform on the frame grid. The input TS is perturbed by centroid-free Gaussian noise, $\tilde{\mathbf{X}}^\ddagger=\mathbf{X}^\ddagger+\Pi\bm{\eta}$ with $\bm{\eta}\sim\mathcal{N}(\mathbf{0},\sigma_c^2\mathbf{I})$ and $\sigma_c=\SI{0.1}{\angstrom}$. In \rnx{}, \ntracer{} receives TSs generated by \nsampler{}, which deviate from optimized TSs; the noise exposes the model to such deviations during training. No noise is added at inference.

\paragraph{Loss.} With $\hat{\mathbf{x}}_i$ and $\mathbf{x}_i$ the predicted and reference positions of atom $i$ in the sampled frame, and $\hat{d}_{ij}$ and $d_{ij}$ the corresponding interatomic distances, the loss for one reaction is
\begin{equation}
\ell=\frac{1}{n}\sum_{i=1}^{n}\left\|\hat{\mathbf{x}}_i-\mathbf{x}_i\right\|+\lambda_d\,\frac{1}{n(n-1)}\sum_{i\neq j}\left|\hat{d}_{ij}-d_{ij}\right|,\qquad \lambda_d=10,
\label{eq:si_tracer_loss}
\end{equation}
averaged over the reactions in a minibatch. A constant $\epsilon=10^{-8}$ is added under every square root to keep gradients finite at zero error. Both terms use unsquared errors, which limits the influence of the occasional poorly described frame. The position term fixes the path in Cartesian space. Endpoint connectivity, however, is judged from interatomic distances, and a Cartesian error of about \SI{0.5}{\angstrom} can create or break a bond in the endpoint graph while contributing little to the position loss. The distance term is invariant to rigid motions and penalizes such bond-length errors directly. At $\lambda_d=10$ it accounts for most of the total loss, so checkpoints are selected by the validation position loss. 

\subsection{Training}\label{si:tracer_training}
\ntracer{} is implemented in PyTorch with PyTorch Geometric and PyTorch Lightning and trained with distributed data parallelism on 8 GPUs, with 16 reactions per GPU in each minibatch. Parameters are optimized with AdamW\autocite{loshchilov2019decoupled} ($\beta_1=0.95$, $\beta_2=0.999$, weight decay \num{1e-8}). The learning rate follows cosine annealing with warm restarts:\autocite{loshchilov2017sgdr} every cycle of \num{5e5} steps starts at its peak value and decays to \num{1e-8}. The peak is \num{5e-4} in the first cycle and is multiplied by 0.05 at each restart, and there is no warm-up. Gradient clipping and the exponential moving average of the parameters are as for \nsampler{} (Section~\ref*{si:sampler_training}). Training proceeds in epochs of \num{5000} steps for up to \num{1000} epochs. The checkpoint with the lowest validation position loss is used throughout. Table~\ref*{tab:si_tracer_hparams} lists all settings, and Algorithm~\ref*{alg:si_tracer_train} summarizes training.

\begin{table}[htbp]
\centering
\caption{\textbf{Hyperparameters of \ntracer{}.}}
\label{tab:si_tracer_hparams}
\small
\setlength{\tabcolsep}{4pt}
\begin{tabular}{@{}lll@{}}
\toprule
Component & Setting & Value \\
\midrule
\multirow{4}{*}{Data}
 & Frames per branch $T$ / total & 100 / 201 (linear resampling) \\
 & Progress coordinate & $\tau=t/T$, normalized per branch \\
 & Orientation & $\tau>0$ towards the lower-energy endpoint \\
 & Input-TS noise $\sigma_c$ & \SI{0.1}{\angstrom} (training only) \\
\midrule
\multirow{8}{*}{\shortstack[l]{Network\\(\textcolor{black}{EquiformerV3})}}
 & Transformer blocks & 12 \\
 & $l_{\max}$ / $m_{\max}$ & 4 / 2 \\
 & Node channels & 32 (+8 for $\tau$, +8 for $q,m$) \\
 & Attention heads / hidden channels & 8 / 64 \\
 & FFN hidden channels & 128 \\
 & Radial basis / envelope cutoff & 600 Gaussians / \SI{5}{\angstrom} \\
 & $\tau$ Fourier frequencies & 8, log-spaced from 1 to 64 \\
 & Stochastic depth / attention dropout & 0.05 / 0.1 \\
\midrule
\multirow{2}{*}{Loss}
 & Position term & mean per-atom error norm \\
 & Distance weight $\lambda_d$ & 10 \\
\midrule
\multirow{9}{*}{Training}
 & Optimizer & AdamW, $\beta_1=0.95$, $\beta_2=0.999$ \\
 & Weight decay & \num{1e-8} \\
 & Learning rate & cosine with warm restarts, \num{5e-4} $\rightarrow$ \num{1e-8} \\
 & Cycle length / peak decay per restart & \num{5e5} steps / 0.05 \\
 & Minibatch & 16 reactions per GPU x 8 GPUs \\
 & Frames per reaction per step & 1 \\
 & Steps per epoch / maximum epochs & \num{5000} / \num{1000} \\
 & EMA decay / start & 0.999 / step \num{1000} \\
 & Model selection & lowest validation position loss (EMA weights) \\
\midrule
Inference & Frames & 3--201; TS and both endpoints included \\
\bottomrule
\end{tabular}
\end{table}

\begin{algorithm}[!htbp]
\caption{Training of \ntracer{}. The loop body is written for one reaction and runs in parallel over the minibatch. $\ell_{\mathrm{pos}}$ and $\ell_{\mathrm{dist}}$ are the position and distance terms of Eq.~\ref*{eq:si_tracer_loss}.}
\label{alg:si_tracer_train}
\begin{algorithmic}[1]
\Require training reactions, each with interpolated frames from IRC calculations $\{\mathbf{X}_t\}_{t=-T}^{T}$ (TS $\mathbf{X}^\ddagger=\mathbf{X}_0$), endpoint energies $E_{-T}$ and $E_{T}$, elements $\mathbf{Z}$, charge $q$, and multiplicity $m$
\Repeat
  \State Draw a minibatch $\mathcal{B}$ of reactions
  \ForAll{reactions in $\mathcal{B}$}
    \State Subtract the centroid of $\mathbf{X}^\ddagger$ from every frame
    \If{$E_{-T}<E_{T}$} \Comment{$+\tau$ leads to the lower-energy end}
      \State Reverse the frame order, $\mathbf{X}_t\gets\mathbf{X}_{-t}$
    \EndIf
    \State $s\sim\mathcal{U}\{-1,+1\}$;\quad $t\sim\mathcal{U}\{0,\dots,T\}$;\quad $\tau\gets s\,t/T$ \Comment{random branch and frame}
    \State $\tilde{\mathbf{X}}^\ddagger\gets\mathbf{X}^\ddagger+\sigma_c\bm{\eta}$,\quad $\bm{\eta}\sim\mathcal{N}(\mathbf{0},\mathbf{I})$, centered \Comment{perturbed input TS}
    \State $\hat{\mathbf{X}}\gets\tilde{\mathbf{X}}^\ddagger+\tau\,\bm{\delta}_\theta(\mathbf{Z},\tilde{\mathbf{X}}^\ddagger,\tau\mid q,m)$ \Comment{Eq.~\ref*{eq:si_tracer_head}}
    \State $\ell\gets\ell_{\mathrm{pos}}(\hat{\mathbf{X}},\mathbf{X}_{st})+\lambda_d\,\ell_{\mathrm{dist}}(\hat{\mathbf{X}},\mathbf{X}_{st})$ \Comment{target frame $\mathbf{X}_{st}$}
  \EndFor
  \State Update $\theta$ by AdamW on the mean of $\ell$ over $\mathcal{B}$; update the EMA of $\theta$
\Until{the step budget is reached}
\State \Return the EMA parameters with the lowest validation position loss
\end{algorithmic}
\end{algorithm}

\subsection{Inference and endpoint assignment}\label{si:tracer_inference}
A trajectory is generated by evaluating Eq.~\ref*{eq:si_tracer_head} at selected progress coordinates in $[-1,1]$, starting from the centered input TS without noise. The number of predicted frames can be chosen from 3 to 201, including the TS and both endpoints. A dense prediction uses $\tau_t=t/T$ for $t=-T,\dots,T$ with $T=100$; fewer coordinates provide a coarser path, with $\tau=-1,0,+1$ sufficient for endpoint screening. The frames are independent and can be evaluated in parallel. Each requested frame requires one network evaluation and no energy or force calculation, and the output is deterministic. Endpoint graphs are built from interatomic distances as in YARP:\autocite{zhao2021simultaneously,zhao2022algorithmic} atoms $i$ and $j$ are bonded when $d_{ij}<1.2\,(r_i+r_j)$, with $r_i$ the covalent radius ($1.5\,(r_i+r_j)$ for H--H contacts), and the longest bonds of over-coordinated atoms are removed. The two endpoint graphs define the channel of the TS. Because the positive branch is oriented towards the lower-energy endpoint rather than away from the input reactant, the endpoint whose graph matches the input reactant graph $G_R$ identifies the reactant side. The predicted endpoints approximate terminal IRC frames and serve as starting structures for subsequent endpoint optimizations. Algorithm~\ref*{alg:si_tracer_sample} summarizes the procedure.

\begin{algorithm}[!htbp]
\caption{Generating a dense reaction trajectory with \ntracer{}. The $2T+1$ frames are independent and evaluated in parallel batches. Fewer progress coordinates can be requested for a coarser path or endpoint screening. \textsc{Graph} perceives bonds from interatomic distances as described in the text.}
\label{alg:si_tracer_sample}
\begin{algorithmic}[1]
\Require TS $\mathbf{X}^\ddagger$, for example from \nsampler{}; elements $\mathbf{Z}$; charge $q$; multiplicity $m$; reactant graph $G_R$; $T=100$
\State $\mathbf{X}^\ddagger\gets\Pi\mathbf{X}^\ddagger$ \Comment{center; no noise at inference}
\For{$t=-T,\dots,T$}
  \State $\tau\gets t/T$
  \State $\hat{\mathbf{X}}_t\gets\mathbf{X}^\ddagger+\tau\,\bm{\delta}_\theta(\mathbf{Z},\mathbf{X}^\ddagger,\tau\mid q,m)$ \Comment{Eq.~\ref*{eq:si_tracer_head}}
\EndFor
\State $G_{-T}\gets\textsc{Graph}(\hat{\mathbf{X}}_{-T})$;\quad $G_{T}\gets\textsc{Graph}(\hat{\mathbf{X}}_{T})$ \Comment{endpoint graphs}
\If{$G_{T}=G_R$} \Comment{reactant predicted at $\tau=+1$}
  \State Reverse the frame order and swap $G_{-T}$ and $G_{T}$
\EndIf
\State \Return trajectory $\{\hat{\mathbf{X}}_t\}_{t=-T}^{T}$ and channel $G_{-T}\to G_{T}$
\end{algorithmic}
\end{algorithm}

\paragraph{Use within network exploration.}
For phosphine exploration, Simulator evaluates the TS and both endpoints ($\tau=-1,0,+1$) before and after TS optimization. IRC calculations and endpoint optimization determine refined connectivity.

\FloatBarrier

\section{Carina architecture and training}\label{si:carina}
\label{si:carina_architecture}
Carina uses the EquiformerV3 architecture\autocite{liao2026equiformerv3} with charge and multiplicity embeddings. The embeddings are combined and added to the scalar ($l=0$) atomic features before the first transformer block.

\label{si:carina_outputs}
The model predicts the atomic energy, defined as $E_{\mathrm{atomic}}=E_{\mathrm{DFT}}-\sum_i w_{Z_i}$, where $w_{Z_i}$ is the fixed reference energy for the element of atom $i$. Adding the elemental reference energies recovers the total energy. Hessians were not used as training targets at this stage.

\label{si:carina_training_objective}
\label{si:carina_training}
Carina was trained on structures belonging to reactions in the RD11E training partition, after data cleaning and removal of geometrically near-duplicate structures. Structures from validation and test reactions were excluded from training. Architecture and training settings are summarized in Table~\ref*{tab:carina_hyperparameters}.

\begin{table}[!htbp]
\centering
\caption{\textbf{Hyperparameters of Carina.}}
\label{tab:carina_hyperparameters}
\small
\begin{tabular}{@{}lll@{}}
\toprule
Component & Setting & Value \\
\midrule
\multirow{10}{*}{\shortstack[l]{Network\\(\textcolor{black}{EquiformerV3})}}
 & Transformer blocks / parameters & 8 / 55.85M \\
 & $l_{\max}$ / $m_{\max}$ / node channels & 6 / 2 / 128 \\
 & Attention heads / hidden channels & 8 / 32 \\
 & Coefficient / value channels & 64 / 16 \\
 & FFN / edge hidden channels & 512 / 128 \\
 & Radial basis / envelope cutoff & 64 Gaussians / \SI{6}{\angstrom} \\
 & Cutoff envelope exponent & 5 \\
 & Charge / multiplicity embedding & 128 / 128 \\
 & Conditioning projection & $256\rightarrow128$, SiLU; added to $l=0$ \\
 & Stochastic depth / attention dropout & 0.05 / 0.1 \\
\midrule
\multirow{3}{*}{Loss}
 & Energy term & Mean per-atom absolute error \\
 & Force term & Mean per-atom error norm \\
 & Energy:force weights & $20:50\rightarrow80:20$ \\
\midrule
\multirow{8}{*}{Training}
 & Optimizer & AdamW, $\beta_1=0.9$, $\beta_2=0.95$ \\
 & Weight decay & \num{0.001} \\
 & Learning rate & Warm-up, cosine decay; \num{1e-4} $\rightarrow$ \num{1e-6} \\
 & Microbatch & 12 structures per GPU $\times$ 8 GPUs \\
 & Gradient accumulation / batch size & 4 microbatches / 384 structures \\
 & Precision & FP32 \\
 & Maximum gradient norm & 100 \\
 & EMA decay & 0.95 \\
\bottomrule
\end{tabular}
\end{table}

\FloatBarrier

\section{Model evaluation and benchmark definitions}\label{si:model_evaluation}\label{si:benchmark_details}

\subsection{Single-step generation and TS refinement}
A common set of 1,000 reference reactions was selected from the test set for both reactant-only TS generation and the reactive-MLIP benchmark. \textcolor{black}{Although the reference reaction records are excluded from training and validation, reactant matching to Sampler training data yields two subsets of 563 and 437 records, with and without overlap, respectively.} For TS generation, only the corresponding reactant graphs were supplied to \nsampler{}, \textcolor{black}{with ten samples per reference reaction, giving 10,000 TS candidates.} Reference products were retained for evaluation and were not supplied as generation inputs. All 10,000 initial candidates are used as the denominator for optimization, frequency, IRC, and input-connectivity outcomes (Fig.~\Mainref{fig:overview}c). \textcolor{black}{For Sampler and Simulator evaluation, endpoint matching preserves total charge and allows resonance-equivalent representations. Input-reactant matching includes returns to the same complete reactant set; productive-connection recovery additionally requires a distinct product under this criterion.} The 93.6\% valid-TS population contains 9,360 structures. RMSDs in Fig.~\Mainref{fig:overview}d compare each generated structure with its own DFT-optimized counterpart after alignment, separately for heavy atoms and all atoms. The evaluated systems are dominated by 10--15 heavy atoms.

\label{si:sampler_eval}
Fig.~\ref*{fig:si_model_evaluation}a-b shows the refinement distributions for the valid DFT TS population described in the main text. Heavy-atom and all-atom RMSDs are calculated using separate Kabsch alignments with fixed atom correspondence and without reflections. The change of electronic energy ($E_{\mathrm{gen}}-E_{\mathrm{DFT\,TS}}$) is evaluated at the DFT level for both geometries. 

\subsection{Simulator connectivity and geometry}
For Simulator connectivity evaluation, the same 8,622 candidates are used in all three input scenarios. The common comparison set requires completed DFT IRC calculations and a converged Carina optimization, together with the corresponding generated and DFT-optimized TS geometries. Each of these three geometries is supplied separately to \ntracer{}. \emph{Raw} agreement uses all 8,622 candidates as the denominator and requires both predicted endpoint graphs to match the DFT IRC endpoint graphs. For each input scenario, the \emph{Screened} population contains only predictions with assignable endpoint graphs, at least one matching the input reactant and a distinct product at the other endpoint. Screened agreement uses the number retained in that scenario as the denominator and the number of retained predictions matching both DFT IRC endpoints as the numerator. 
Optimization convergence does not establish a first-order saddle point: 8,013 of the Carina-optimized geometries are strict TSs, while 609 (7.06\% of 8,622) fail the strict criterion (converged optimization, exactly one imaginary frequency, below $-50$~cm$^{-1}$). All optimization-converged Carina geometries in the common comparison set are eligible as Simulator inputs, including geometries that fail this frequency criterion. Agreement with DFT IRC therefore reflects input geometry quality and differences between the two potential energy surfaces as well as Simulator prediction. 
For a comparison on the same potential energy surface, Simulator and IRC endpoints are evaluated from the same Carina TS. Among 7,967 candidates with a strict Carina TS and accepted Carina and DFT IRCs, both endpoint graphs agree between Simulator and Carina IRC in 7,227 cases (90.71\%). Of these, 798 candidates have mutually matching Simulator and Carina IRC endpoints that both differ from the DFT IRC endpoints.

\label{si:tracer_eval}
The reference-TS benchmark in Fig.~\ref*{fig:si_model_evaluation}c compares predicted endpoint graphs with database DFT IRC endpoints. The predicted endpoints are matched to the reference endpoints in either trajectory direction.

\textcolor{black}{For the geometry comparison, 7,912 candidates from the common set of 8,622 have a unique DFT IRC endpoint matching the input reactant and a distinct product under the same resonance-equivalent criterion.} Predictions from all three TS inputs are compared with these same DFT IRC terminal geometries, without subsequent endpoint optimization. Graph-mismatched predictions remain in the analysis. Heavy-atom RMSDs are computed after separately aligning each predicted endpoint to its assigned reference as in Section~\ref*{si:sampler_eval}. R and P observations are pooled with equal weight (Fig.~\ref*{fig:si_model_evaluation}d).

\subsection{Reactive-MLIP and energy--force benchmarks}
The reactive-potential benchmark uses the same 1,000 reference reactions selected for the Sampler benchmark and follows the reported GSM, TS-optimization, and IRC procedure.\autocite{zhao2025harnessing} The specified reactant and product input geometries were obtained by optimizing the corresponding IRC endpoints. Thus each selected reference reaction provides a reactant graph for Sampler generation and an optimized reactant/product geometry pair for MLIP pathway searches. The reference-reaction recovery criterion requires both specified reactant and product to be recovered, corresponding to the \emph{Intended} label in Fig.~\Mainref{fig:overview}f.

\begin{figure}[!htbp]
\centering
\includegraphics[width=\textwidth]{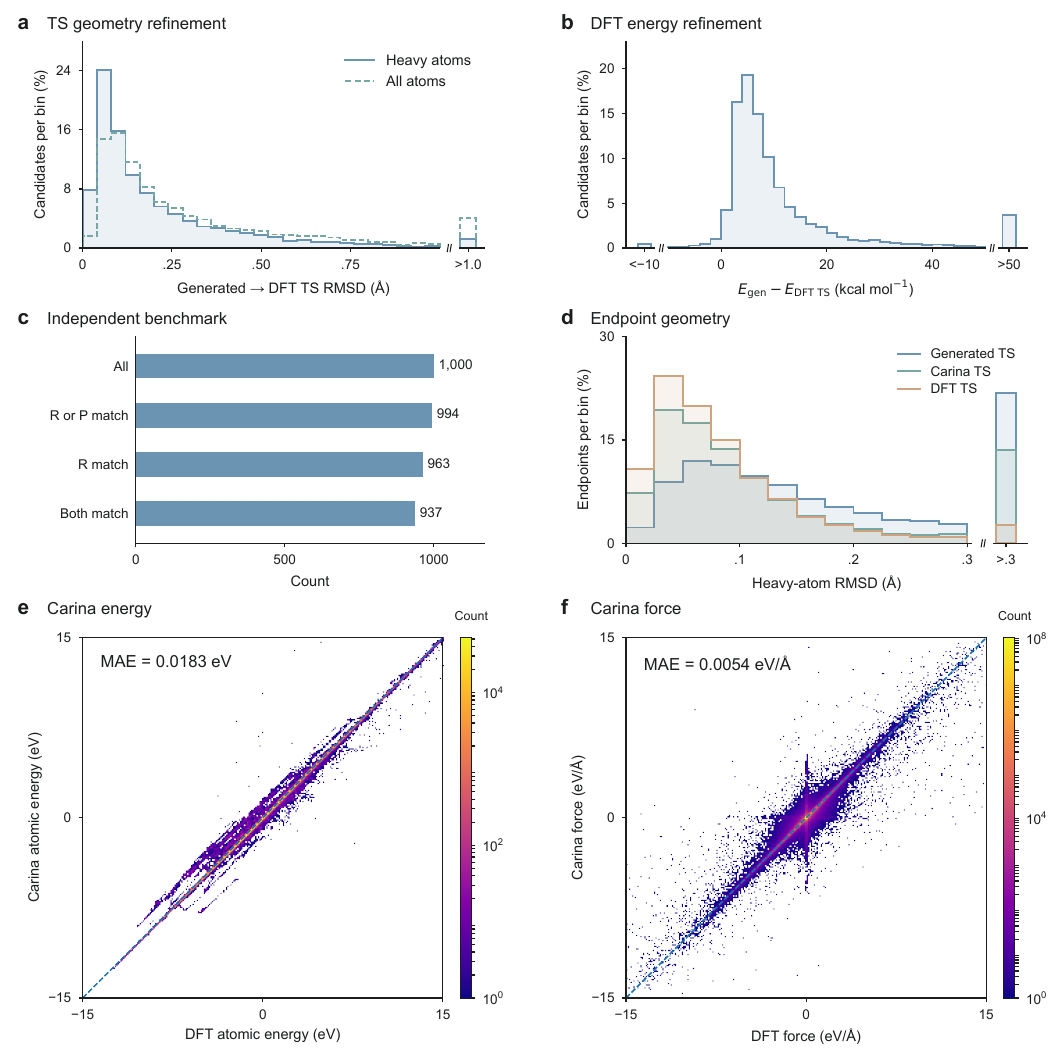}
\captionsetup{font=small}
\caption[Supplementary evaluation of the three models]{\textbf{Fig.~\thefigure{} \textbar{} Supplementary evaluation of Nexus-Sampler, Nexus-Simulator, and Carina.}
\textbf{a.} Heavy-atom and all-atom RMSD between generated and DFT-optimized geometries for 9,360 strict TSs.
\textbf{b.} DFT electronic-energy change upon TS refinement for the same candidates. Separated edge bins in a and b pool the indicated tails.
\textbf{c.} Endpoint-graph recovery from 1,000 database DFT TS inputs. 
\textbf{d.} Pooled R/P endpoint heavy-atom RMSD for generated, Carina-optimized, and DFT-optimized TS inputs, using the same 7,912 candidates and 15,824 endpoint observations per input type. Histograms are normalized over all endpoint observations. The final bin includes all values above 0.3~\AA{}.
\textbf{e-f.} Carina EMA atomic energies and forces against DFT, using 3,525,035 atomic energies and 259,825,277 forces. Colors show logarithmic bin counts on separate scales; dashed lines indicate equality.}
\label{fig:si_model_evaluation}
\end{figure}

\label{si:carina_eval}
The test set contains 31,209 RD11E test reactions and 3,526,174 structures, including structures from unconverged trajectories. Fig.~\ref*{fig:si_model_evaluation}e-f compares Carina EMA predictions with DFT atomic energies and Cartesian forces. Atomic energy is defined in Section~\ref*{si:carina_outputs}.

The comparisons include atomic energies within $[-15,15]$~eV and forces within $[-15,15]$~eV\,\AA$^{-1}$, with the two ranges applied independently. MAEs are calculated for the data shown.

\FloatBarrier
\begin{table}[H]
\centering
\caption{Sampler outcomes by reactant overlap (Fig.~\Mainref{fig:overview}c).}
\label{tab:overlap_sampler}
\small
\setlength{\tabcolsep}{5pt}
\renewcommand{\arraystretch}{1.12}
\begin{tabular*}{\textwidth}{@{}p{\dimexpr\textwidth-195pt\relax}rrr@{}}
\toprule
Metric & \makebox[55pt][r]{Overlap} & \makebox[55pt][r]{Disjoint} & \makebox[55pt][r]{Overall} \\
\midrule
Generated candidates ($n$) & 5,630 & 4,370 & 10,000 \\
TSOPT pass (\%) & 95.77 & 96.16 & 95.94 \\
TS valid (\%) & 93.55 & 93.66 & 93.60 \\
IRC pass (\%) & 90.28 & 91.65 & 90.88 \\
R match (\%) & \textcolor{black}{86.32} & \textcolor{black}{86.68} & \textcolor{black}{86.48} \\
R $\ne$ P (\%) & \textcolor{black}{83.07} & \textcolor{black}{83.62} & \textcolor{black}{83.31} \\
\bottomrule
\end{tabular*}
\par\vspace{5pt}
\begin{minipage}{\textwidth}\footnotesize\raggedright
\textcolor{black}{Overlap and Disjoint denote the presence or absence of reactant matches in Sampler training data.} Percentages use all generated candidates within each group.
\end{minipage}
\end{table}

\begin{table}[H]
\centering
\caption{TS refinement by reactant overlap (Fig.~\Mainref{fig:overview}d; Fig.~\ref*{fig:si_model_evaluation}a--b).}
\label{tab:overlap_sampler_refinement}
\small
\setlength{\tabcolsep}{5pt}
\renewcommand{\arraystretch}{1.12}
\begin{tabular*}{\textwidth}{@{}p{\dimexpr\textwidth-265pt\relax}p{60pt}rrr@{}}
\toprule
Metric & Statistic & \makebox[55pt][r]{Overlap} & \makebox[55pt][r]{Disjoint} & \makebox[55pt][r]{Overall} \\
\midrule
\multicolumn{2}{@{}l}{Evaluated TSs ($n$)} & 5,267 & 4,093 & 9,360 \\
\addlinespace[3pt]
Heavy-atom RMSD (\AA{}) & Mean & 0.202 & 0.226 & 0.212 \\
 & Median & 0.120 & 0.141 & 0.128 \\
 & P90 & 0.489 & 0.531 & 0.503 \\
\addlinespace[3pt]
All-atom RMSD (\AA{}) & Mean & 0.288 & 0.324 & 0.304 \\
 & Median & 0.177 & 0.210 & 0.189 \\
 & P90 & 0.674 & 0.752 & 0.706 \\
\addlinespace[3pt]
$\Delta E$ (kcal mol$^{-1}$) & Mean & 14.74 & 18.32 & 16.30 \\
 & Median & 6.58 & 7.58 & 6.94 \\
 & P90 & 21.56 & 28.06 & 24.07 \\
\bottomrule
\end{tabular*}
\par\vspace{5pt}
\begin{minipage}{\textwidth}\footnotesize\raggedright
RMSDs compare generated and DFT-optimized TSs. P90 is the 90th percentile; $\Delta E=E_{\mathrm{gen}}-E_{\mathrm{DFT\,TS}}$.
\end{minipage}
\end{table}

\begin{table}[H]
\centering
\caption{Simulator agreement with DFT IRC by reactant overlap (Fig.~\Mainref{fig:overview}e).}
\label{tab:overlap_connectivity}
\small
\setlength{\tabcolsep}{5pt}
\renewcommand{\arraystretch}{1.12}
\begin{tabular*}{\textwidth}{@{}p{\dimexpr\textwidth-265pt\relax}p{60pt}rrr@{}}
\toprule
TS input & Metric & \makebox[55pt][r]{Overlap} & \makebox[55pt][r]{Disjoint} & \makebox[55pt][r]{Overall} \\
\midrule
\multicolumn{2}{@{}l}{Evaluated candidates ($n$)} & 4,802 & 3,820 & 8,622 \\
\addlinespace[3pt]
Generated & Raw (\%) & 85.55 & 81.15 & 83.60 \\
 & Retained ($n$) & \textcolor{black}{4,629} & 3,600 & \textcolor{black}{8,229} \\
 & Screened (\%) & \textcolor{black}{87.28} & 84.28 & \textcolor{black}{85.96} \\
\addlinespace[3pt]
Carina-optimized & Raw (\%) & 81.80 & 78.06 & 80.14 \\
 & Retained ($n$) & 4,087 & 3,224 & 7,311 \\
 & Screened (\%) & 92.12 & 89.27 & 90.86 \\
\addlinespace[3pt]
DFT-optimized & Raw (\%) & 92.11 & 89.90 & 91.13 \\
 & Retained ($n$) & 4,484 & 3,512 & 7,996 \\
 & Screened (\%) & 93.84 & 92.97 & 93.46 \\
\bottomrule
\end{tabular*}
\par\vspace{5pt}
\begin{minipage}{\textwidth}\footnotesize\raggedright
Agreement requires both endpoints to match DFT IRC endpoints. Raw and Screened percentages use all evaluated and retained candidates, respectively, within each group and input type.
\end{minipage}
\end{table}

\begin{table}[H]
\centering
\caption{Simulator endpoint recovery from reference DFT TSs by reactant overlap (Fig.~\ref*{fig:si_model_evaluation}c).}
\label{tab:overlap_reference_recovery}
\small
\setlength{\tabcolsep}{5pt}
\renewcommand{\arraystretch}{1.12}
\begin{tabular*}{\textwidth}{@{}p{\dimexpr\textwidth-195pt\relax}rrr@{}}
\toprule
Metric & \makebox[55pt][r]{Overlap} & \makebox[55pt][r]{Disjoint} & \makebox[55pt][r]{Overall} \\
\midrule
Evaluated reactions ($n$) & 563 & 437 & 1,000 \\
R or P match (\%) & 99.47 & 99.31 & 99.40 \\
R match (\%) & 97.34 & 94.97 & 96.30 \\
Both match (\%) & 95.20 & 91.76 & 93.70 \\
\bottomrule
\end{tabular*}
\end{table}

\begin{table}[H]
\centering
\caption{Simulator endpoint heavy-atom RMSD (\AA{}) by reactant overlap (Fig.~\ref*{fig:si_model_evaluation}d).}
\label{tab:overlap_geometry}
\small
\setlength{\tabcolsep}{5pt}
\renewcommand{\arraystretch}{1.12}
\begin{tabular*}{\textwidth}{@{}p{\dimexpr\textwidth-265pt\relax}p{60pt}rrr@{}}
\toprule
TS input & Statistic & \makebox[55pt][r]{Overlap} & \makebox[55pt][r]{Disjoint} & \makebox[55pt][r]{Overall} \\
\midrule
\multicolumn{2}{@{}l}{Evaluated candidates ($n$)} & 4,428 & 3,484 & 7,912 \\
\multicolumn{2}{@{}l}{Evaluated endpoints ($n$)} & 8,856 & 6,968 & 15,824 \\
\addlinespace[3pt]
Generated & Mean & 0.204 & 0.225 & 0.213 \\
 & Median & 0.136 & 0.150 & 0.142 \\
 & P90 & 0.467 & 0.506 & 0.484 \\
\addlinespace[3pt]
Carina-optimized & Mean & 0.157 & 0.173 & 0.164 \\
 & Median & 0.082 & 0.090 & 0.085 \\
 & P90 & 0.411 & 0.435 & 0.420 \\
\addlinespace[3pt]
DFT-optimized & Mean & 0.086 & 0.095 & 0.090 \\
 & Median & 0.066 & 0.070 & 0.068 \\
 & P90 & 0.164 & 0.190 & 0.177 \\
\bottomrule
\end{tabular*}
\par\vspace{5pt}
\begin{minipage}{\textwidth}\footnotesize\raggedright
RMSDs pool R/P endpoints relative to DFT IRC geometries. P90 is the 90th percentile.
\end{minipage}
\end{table}

\begin{table}[H]
\centering
\caption{Carina reactive-MLIP benchmark by reactant overlap (Fig.~\Mainref{fig:overview}f).}
\label{tab:overlap_carina}
\small
\setlength{\tabcolsep}{5pt}
\renewcommand{\arraystretch}{1.12}
\begin{tabular*}{\textwidth}{@{}p{\dimexpr\textwidth-195pt\relax}rrr@{}}
\toprule
Metric & \makebox[55pt][r]{Overlap} & \makebox[55pt][r]{Disjoint} & \makebox[55pt][r]{Overall} \\
\midrule
Evaluated reactions ($n$) & 563 & 437 & 1,000 \\
GSM pass (\%) & 99.47 & 99.54 & 99.50 \\
TS valid (\%) & 83.66 & 82.61 & 83.20 \\
Intended (\%) & \textcolor{black}{73.71} & 74.83 & \textcolor{black}{74.20} \\
\bottomrule
\end{tabular*}
\end{table}

\begin{table}[H]
\centering
\caption{Carina TS validity and IRC acceptance by reactant overlap.}
\label{tab:overlap_carina_ts}
\small
\setlength{\tabcolsep}{5pt}
\renewcommand{\arraystretch}{1.12}
\begin{tabular*}{\textwidth}{@{}p{\dimexpr\textwidth-195pt\relax}rrr@{}}
\toprule
Metric & \makebox[55pt][r]{Overlap} & \makebox[55pt][r]{Disjoint} & \makebox[55pt][r]{Overall} \\
\midrule
Evaluated candidates ($n$) & 4,802 & 3,820 & 8,622 \\
Strict TS (\%) & 92.71 & 93.22 & 92.94 \\
Non-strict geometries (\%) & 7.29 & 6.78 & 7.06 \\
\addlinespace[4pt]
Evaluated strict TSs ($n$) & 4,452 & 3,561 & 8,013 \\
Carina IRC not accepted (\%) & 0.56 & 0.59 & 0.57 \\
\bottomrule
\end{tabular*}
\par\vspace{5pt}
\begin{minipage}{\textwidth}\footnotesize\raggedright
TS percentages use all evaluated candidates; IRC non-acceptance percentages use strict TSs within each group.
\end{minipage}
\end{table}

\begin{table}[H]
\centering
\caption{Simulator agreement with DFT IRC for Carina-optimized inputs.}
\label{tab:overlap_dft_irc}
\small
\setlength{\tabcolsep}{5pt}
\renewcommand{\arraystretch}{1.12}
\begin{tabular*}{\textwidth}{@{}p{\dimexpr\textwidth-195pt\relax}rrr@{}}
\toprule
Metric & \makebox[55pt][r]{Overlap} & \makebox[55pt][r]{Disjoint} & \makebox[55pt][r]{Overall} \\
\midrule
Evaluated strict TSs ($n$) & 4,452 & 3,561 & 8,013 \\
Both match (\%) & 83.83 & 80.40 & 82.30 \\
\addlinespace[4pt]
Evaluated non-strict geometries ($n$) & 350 & 259 & 609 \\
Both match (\%) & 56.00 & 45.95 & 51.72 \\
\bottomrule
\end{tabular*}
\end{table}

\begin{table}[H]
\centering
\caption{Simulator agreement with Carina IRC for Carina-optimized inputs.}
\label{tab:overlap_carina_irc}
\small
\setlength{\tabcolsep}{5pt}
\renewcommand{\arraystretch}{1.12}
\begin{tabular*}{\textwidth}{@{}p{\dimexpr\textwidth-195pt\relax}rrr@{}}
\toprule
Metric & \makebox[55pt][r]{Overlap} & \makebox[55pt][r]{Disjoint} & \makebox[55pt][r]{Overall} \\
\midrule
Candidates with accepted IRCs ($n$) & 4,427 & 3,540 & 7,967 \\
Carina IRC agreement (\%) & 91.96 & 89.15 & 90.71 \\
Carina IRC agreement, DFT IRC disagreement (\%) & 9.67 & 10.45 & 10.02 \\
\bottomrule
\end{tabular*}
\par\vspace{5pt}
\begin{minipage}{\textwidth}\footnotesize\raggedright
Agreement requires both endpoints to match. Both percentages use all evaluated candidates within each group; the second is a subset of the first.
\end{minipage}
\end{table}

\subsection{Related generative approaches and RMSD conventions}\label{si:literature_baselines}
For comparison, we express structural deviations using the per-atom convention,
\begin{equation}
\mathrm{RMSD}_{\mathrm{atom}}=\sqrt{\frac{1}{N}\sum_{i=1}^{N}\|\Delta\mathbf r_i\|^2}
=\sqrt{3}\,\sqrt{\frac{1}{3N}\sum_{i=1}^{N}\sum_{\alpha=x,y,z}(\Delta r_{i\alpha})^2},
\end{equation}
where $N$ is the number of included atoms after alignment and atom matching. The second square root is the coordinate-wise RMSE returned by the historical pymatgen matchers used in React-OT and specified in GoFlow's Supplementary Eq.~(2). Pymatgen subsequently corrected these matchers to use the per-atom convention.\footnote{See the RMSD correction in \href{https://github.com/materialsproject/pymatgen-core/releases/tag/v2026.5.18}{pymatgen-core v2026.5.18}.}

React-OT takes known reactant and product geometries as input and reports a mean reference-TS RMSD of 0.103~\AA{} on Transition1x; the converted value is 0.178~\AA{}.\autocite{duan2025optimal} GoFlow-25 takes reactant and product graphs and reports 0.18~\AA{} on RDB7, corresponding to 0.312~\AA{} under the same normalization conversion. Its prediction is selected from 25 generated samples without reference-TS access.\autocite{galustian2025goflow} The literature values measure deviations from reference TSs, whereas our values measure structural changes during DFT optimization of generated candidates. TSDiff's original D--MAE measures interatomic-distance errors and is not subject to this $\sqrt{3}$ conversion.\autocite{kim2024diffusion}

MolGEN's TS generator is conditioned on known reactant and product geometries. Its Table~1 reports valid-TS fractions of 87.46\% for Transition1x and 89.20\% for the combined-data FM/fine-tuned model, with the evaluated generated TS guesses as denominator; validity requires one imaginary frequency, not recovery of the specified endpoints. In its reactant-started KHP workflow, 1,000 generated products are reduced to 59 distinct, optimized candidates before TS generation. Of these, 40/59 (67.8\%) yield valid TSs and 14/59 (23.7\%) recover the specified R/P pair by IRC. These fractions use the 59 selected candidates, not all 1,000 product proposals.\autocite{tuo2026flow}

For completeness, TSDiff is a known-R/P-graph baseline rather than a reactant-only generator. On its original test set, 9,289/9,576 generated guesses (eight per reaction) optimize to single-imaginary-frequency TSs, or 97.0\%. IRC is evaluated on one randomly selected geometry per reaction: 998/1,197 (83.4\%) recover both specified endpoints.\autocite{kim2024diffusion} These results use different evaluation populations and are not directly comparable with our benchmark.

\FloatBarrier

\section{Network exploration workflow}\label{si:network_workflow}

\subsection{Candidate screening and representative selection}\label{si:workflow_screening}
Rapid exploration uses Simulator endpoint predictions and Carina single-point energies. The refinement workflow additionally applies Carina TS optimization and connectivity calculations to selected candidates.

\paragraph{Geometrical reduction before endpoint prediction.}
Sampler generates a batch of TS candidates for each input. Complete-linkage clustering uses heavy-atom RMSDs after fixed-mapping Kabsch alignment with proper rotations and no reflection. Within each cluster, the medoid is selected first; subsequent candidates maximize their minimum distance to the selected set, retaining up to $k$ representatives or all members of smaller clusters. The phosphine top-three exploration used the settings below.

\begin{table}[!htbp]
\centering\small
\caption*{\textbf{Screening settings for the phosphine top-three exploration.}}
\begin{tabular}{@{}p{0.48\textwidth}p{0.47\textwidth}@{}}
\toprule
Setting & Value \\
\midrule
Generated candidates per input & 1,000 \\
Initial clustering cutoff / representatives & 0.15~\AA{} / $k=3$ per cluster \\
SPE window above the batch minimum & 50~kcal/mol \\
Representatives per pathway family & Up to five, including the minimum-SPE candidate \\
\bottomrule
\end{tabular}
\end{table}

\paragraph{Simulator screening and pathway assignment.}
Simulator predicts endpoints for each geometrical representative. Candidates are retained when exactly one endpoint matches the input reactants and the other has different connectivity (Section~\ref*{si:identity_details}); unresolved endpoints are excluded. Mapped endpoint states and net bond-change signatures define pathway families. In the phosphine workflow, these states retain adjacency, bond-electron and formal-charge information, total charge, spin, isotopes, and sampled stereochemical variants. Different bond-change patterns form separate families, with geometrical diversity retained within each family.

\paragraph{SPE-based screening and representative selection.}
Carina SPEs are evaluated for the geometrical representatives. The energy window is measured from the lowest finite SPE in the input batch. Candidates passing both endpoint and energy screening are grouped into pathway families. Within each family, farthest-point sampling uses fixed-mapping Kabsch RMSDs over all heavy atoms and hydrogens participating in the mapped bond changes. Selection starts from the lowest-SPE candidate and adds candidates maximizing their minimum distance to the selected set, up to the representative limit. This stage has no additional RMSD cutoff.

\paragraph{Refinement, node consolidation, and continued growth.}
In the refinement workflow, Carina TS optimization is followed by a second Simulator evaluation, IRC calculations, and the endpoint/common-reactant operations configured for that exploration. Connections are assigned from the calculated endpoints. Equivalent complete product states are consolidated for navigation while separate TS records are retained. Eligible unexpanded states enter the global pool described in Section~\ref*{si:navigation_details}.

\subsection{Stationary-point and endpoint verification}\label{si:stationary_verification}
The following settings apply to network exploration; benchmark criteria are given in Section~\ref*{si:benchmark_details}. Forces and Hessians are obtained by differentiating the Carina energy.\autocite{zhao2025harnessing} TS optimization uses redundant internal coordinates and the restricted-step image rational-function optimizer in pysisyphus,\autocite{steinmetzer2021pysisyphus} with native Gaussian \texttt{gau} convergence criteria, \texttt{trust\_radius=0.2}, at most 50 steps, and analytical Hessian recalculation every five steps. The final geometry is assessed for one significant imaginary frequency below $-10$~cm$^{-1}$. Optimization and frequency status are retained separately for cycle-limited candidates continued to IRC.

IRC integration uses Carina with the Euler predictor--corrector, \texttt{step\_length=0.2} and at most 200 cycles per direction. Endpoint optimization and common-reactant searches follow the settings of each exploration. A minimum assignment requires endpoint optimization and frequency analysis; connectivity changes during relaxation are assigned to that stage.

\subsection{Energy references and levels of verification}\label{si:energy_details}
Unless otherwise specified, network arrows give local Carina electronic-energy differences,
\begin{equation}
\Delta E_e^\ddagger=E_{\rm Carina}(\mathbf{X}_e^\ddagger)-E_{\rm Carina}(\mathbf{X}_{R,j}^{\rm ref}),
\label{eq:local_barrier}
\end{equation}
where $\mathbf{X}_{R,j}^{\rm ref}$ is the common optimized reactant reference for input $j$, including all explicit molecules. The difference is an electronic barrier for a converged TS and a provisional height for a nonstationary candidate. Negative and non-finite heights are excluded from navigation scoring. Independent inputs, including water-containing and water-free systems, have separate references; their local values are not a common-zero energy profile.

Carina describes gas-phase clusters with explicit molecules. Bulk-solvent effects and free-energy corrections are specified separately where evaluated. DFT single points, stationary-point optimization, and IRC checks are distinguished by the operations performed; DFT-verified connections require both TS and connectivity checks. The reference level is $\omega$B97X-D4/def2-TZVPD with GPU4PySCF,\autocite{li2025introducing} with application-specific levels and solvent treatments stated in the corresponding sections.

For the proline TS comparison in Fig.~\Mainref{fig:cat}c, $\omega$B97X-D4/def2-TZVPD/SMD(DMSO) single-point energies and DFT Hessians on Carina-optimized geometries were combined with qRRHO corrections at 298.15~K, a 1~M standard state, and a 100~cm$^{-1}$ low-frequency cutoff.\autocite{marenich2009universal,ribeiro2011use} The geometries were not reoptimized at the DFT level.

Water-containing and water-free B5 inputs in the phosphine system were sampled separately, with 2,000 candidates each. Dissociated water was retained in the overall chemical balance. Barriers use separate common optimized reactant complexes for the two inputs; water association/dissociation free energies were not evaluated.

\subsection{Mapped identity and stereochemistry}\label{si:identity_details}
Endpoint admission compares adjacency matrices from stored geometries under the original atom mapping, including every explicit hydrogen and fragment. A productive connection has one endpoint matching the input reactants and another with different connectivity. Navigation groups additionally account for constitution and electronic state, with fixed-mapping equality recorded separately from element/isotope-labelled graph isomorphism.

Although graph features include stereochemical information, sampling does not explicitly enforce stereochemical constraints. Stereocentres and alkene configurations are assigned afterwards from stored three-dimensional geometries and fixed mapped ligands, distinguishing local parity from changes in CIP naming. The general endpoint gate has no stereochemical term. Product diastereomers remain separate for selectivity analysis. In achiral systems, whole-system mirror-related paths may share ranking evidence after global reflection and atom-mapping verification; other stereochemical paths remain separate.

\subsection{Global candidate pool and network expansion}\label{si:navigation_details}
KHP uses $n=5$, IC uses $n=1$, and the top-three catalytic explorations use $n=3$. IC exploration stopped upon recovery of INT5. The phosphine exploration allowed up to eight reaction steps and targeted release of intact neutral PMe$_3$ from a bound intermediate.

\FloatBarrier

\section{Ireland--Claisen rearrangement}\label{si:ic_application}

\subsection{Transition-state conformational sampling}\label{si:ic_conformers}
Supplementary refinement used Sampler candidates with the target mapped bond-change signatures. CREST sampling\autocite{pracht2020automated,zhao2025graphically} started from the initially verified TS for each of the five reactions, using two restraint strengths on the reactive core. Up to 20 low-energy conformers per ensemble were selected. Selected conformers were refined directly or after reactive-core-fixed preoptimization. All restraints were released before Carina TS optimization and IRC calculations.

\begin{table}[!htbp]
\centering\small
\caption*{\textbf{IC conformational sampling and comparison populations.}}
\begin{tabular}{@{}p{0.68\textwidth}r@{}}
\toprule
Population & Count \\
\midrule
Initial Sampler candidates & 5,000 \\
Sampler candidates selected for supplementary refinement & 540 \\
Conformers generated by constrained CREST sampling & 364 \\
CREST conformers selected for refinement & 194 \\
CREST verification inputs across the two refinement branches & 388 \\
Retained optimization records: Sampler / CREST & 80 / 146 \\
Combined retained records & 226 \\
Geometrical clusters: shared / Sampler only / CREST only & 13 / 27 / 26 \\
\bottomrule
\end{tabular}
\end{table}

The 388 verification inputs include refinement of the same conformers through both branches. The plotted 226 records match the target connection, mapped bond-change signature, and endpoint stereochemistry. Complete-linkage clustering at 0.30~\AA{} heavy-atom RMSD, using fixed atom mapping and proper rotations without reflection, gives 66 geometrical clusters.

For each reaction, $\Delta E=E_{\mathrm{TS,candidate}}-E_{\mathrm{TS,initial}}$ compares candidates with the lowest TS retained in the initial exploration. The central values in Fig.~\Mainref{fig:r1}e report the reduction with the opposite sign: 0.09, 0.36, 1.53, 0.54, and 1.35~kcal/mol for IC1--IC5. The original per-step optimized reactant reference is fixed, so these Carina TS-energy reductions equal the electronic-barrier reductions. The lowest retained TSs for IC1, IC2, and IC5 come from CREST sampling, and those for IC3 and IC4 from supplementary Sampler refinement.

\subsection{Single-point energy comparison}\label{si:ic_dft_comparison}
Single-point energies for the five adopted IC TSs and their unchanged per-step common-reactant geometries were evaluated using B3LYP (Gaussian/VWN3), zero-damped D3, and the 6-311++G(2d,2p) basis, with IEF-PCM n-hexane solvation, using GPU4PySCF 1.8.1. The input structures were taken directly from the Carina calculations, without DFT geometry optimization. Density fitting used def2-universal-jkfit, with grid level 4 and an SCF tolerance of $10^{-9}$~Eh. The PCM dielectric constant was 1.8819, with a modified Bondi cavity and Lebedev order 29; D3 used zero damping without the three-body ATM term. Each local electronic barrier is the TS energy minus its corresponding complete common-reactant energy.

Lee and colleagues used B3LYP-D3BJ/6-311++G(2d,2p) single-point calculations with IEF-PCM hexane solvation on geometries optimized at the B3LYP/6-31G(d) level in the gas phase.\autocite{lee2019unexpected} Their reported values are transcribed from the comparison in Zhao and colleagues.\autocite{zhao2022conformational} Both calculations use the same B3LYP functional, 6-311++G(2d,2p) basis, and IEF-PCM hexane solvent model. The archived GPU4PySCF calculations use zero-damped D3 on Carina geometries, whereas the Gaussian reference uses D3BJ on DFT-optimized geometries.

\begin{table}[htbp]
\centering
\small
\caption{Local electronic barriers for the five-step IC sequence (kcal/mol).}
\label{tab:ic_dft_single_points}
\begin{tabular}{lrrr}
\toprule
Step & Carina & DFT single point & Lee et al. \\
\midrule
IC1 & 31.89 & 24.03 & 23.9 \\
IC2 & 15.49 & 16.48 & 15.6 \\
IC3 & 27.65 & 14.57 & 10.6 \\
IC4 & 39.64 & 21.45 & 18.1 \\
IC5 & 8.38 & 9.58 & 12.9 \\
\bottomrule
\end{tabular}
\end{table}

The additional branches use the following optimized reactant references. Reference and TS energies are on the plotted initial-R scale; all values are in kcal/mol.

\begin{table}[!htbp]
\centering\small
\caption*{\textbf{Energy references for the additional IC branches.}}
\begin{tabular}{@{}lrrrr@{}}
\toprule
Connection & Displayed source & Optimized reference & TS & Local barrier \\
\midrule
P2 $\rightarrow$ INT2 & 11.17 & 9.81875 & 40.08572 & 30.26697 \\
P3 $\rightarrow$ P6 & $-3.69$ & $-3.06273$ & 15.67397 & 18.73670 \\
\bottomrule
\end{tabular}
\end{table}

The displayed P6 structure was optimized to 12.61~kcal/mol on the initial-R scale and verified as a positive-frequency minimum. From a separately optimized P6 structure, the unplotted ring-opening pathway to P3 has a local barrier of 3.87~kcal/mol, using the same reactant reference as the 48.64~kcal/mol pathway in Fig.~\ref*{fig:si_ic_network}.

\subsection{Supplementary energy and network figures}\label{si:ic_figures}

\begin{figure}[H]
\centering
\includegraphics[width=\textwidth]{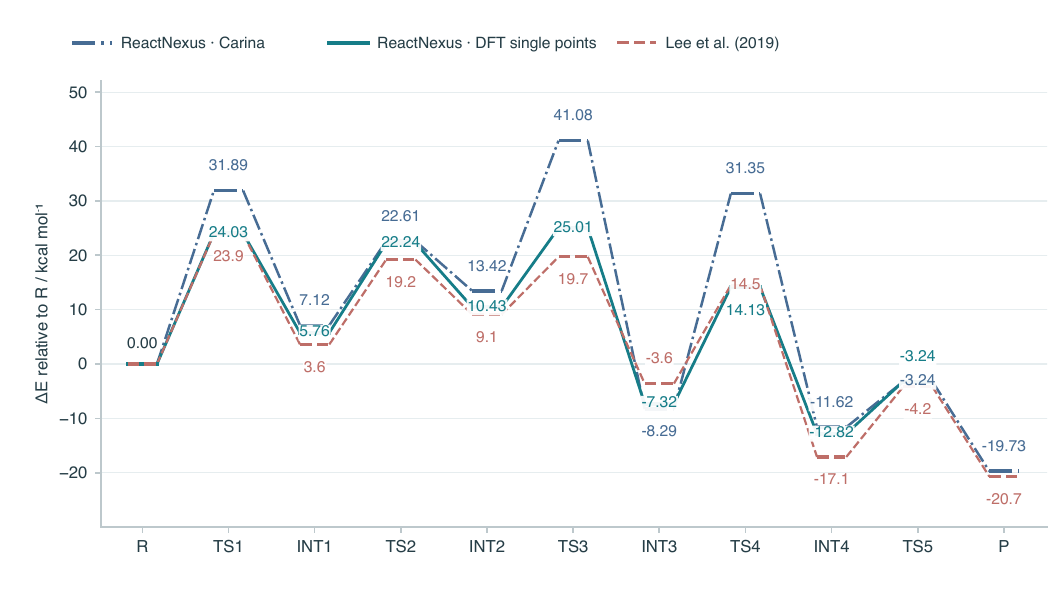}
\caption{\textbf{Fig. \thefigure\ \textbar{} IC energetic comparison with earlier calculations.} Carina, fixed-geometry DFT single-point, and Lee et al.\ profiles use their respective initial R as zero. The Carina and DFT curves use the same five TS and step-specific common-reactant geometries. The calculations differ in D3 damping and input geometries, as described in Section~\ref*{si:ic_dft_comparison}. Lines show the reference sequence at the level of molecular connectivity. Energies are electronic differences in kcal/mol.}
\label{fig:si_ic_energies}
\end{figure}

\begin{figure}[H]
\centering
\includegraphics[width=\textwidth]{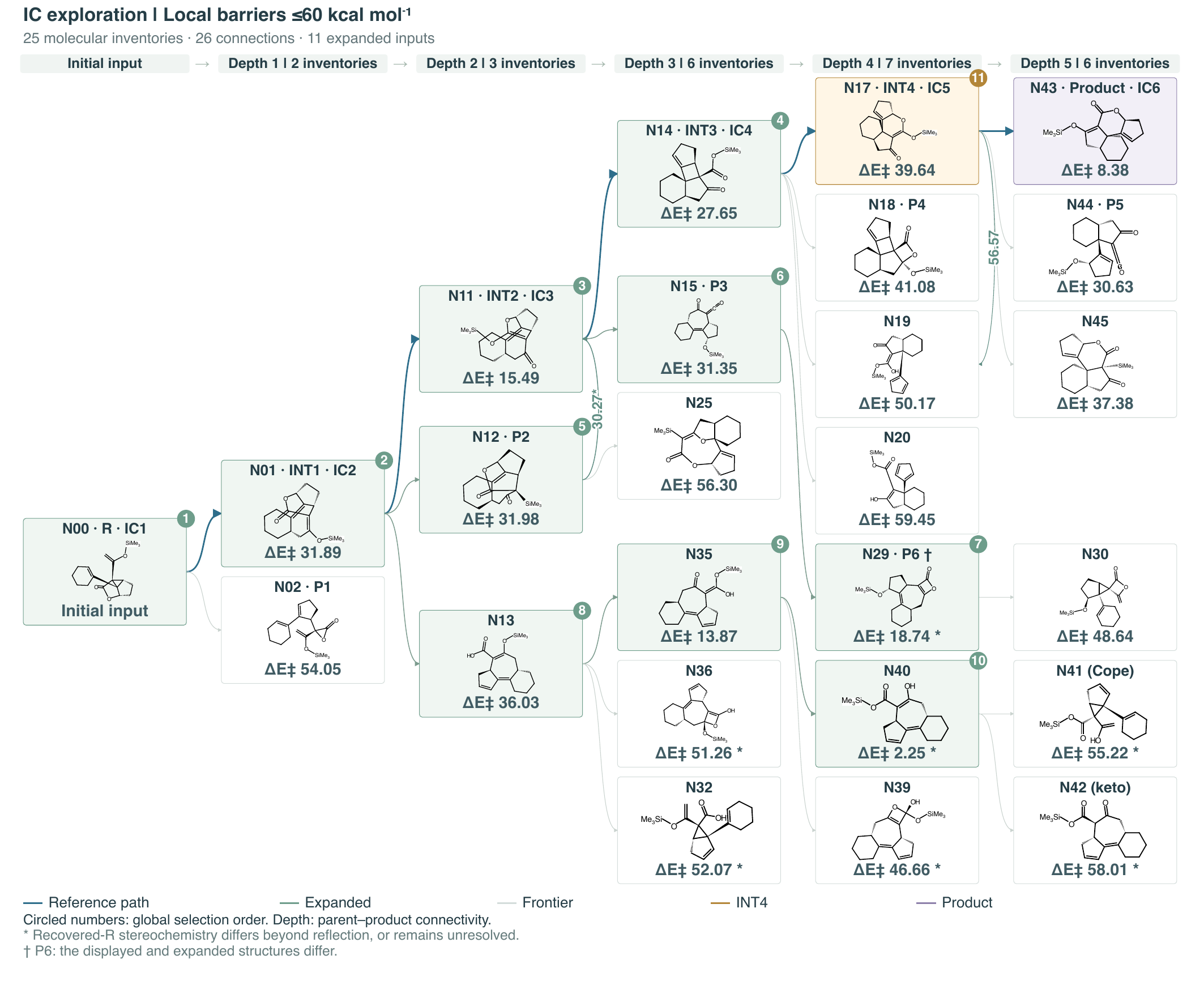}
\caption{\textbf{Fig. \thefigure\ \textbar{} Finite IC exploration and global selection.} The local-barrier display cutoff of 60~kcal/mol retains 25 nodes and 26 connections, including all 11 expanded inputs. Circle numbers indicate global-pool selection order, not reaction-step numbers. Values are local Carina-computed electronic barriers; the five reference connections use the adopted Fig.~\Mainref{fig:r1} TSs. The 48.64~kcal/mol barrier was evaluated from a separately optimized P6 structure. Asterisks mark connections for which equivalence between the optimized reactant and the displayed source structure was not established.}
\label{fig:si_ic_network}
\end{figure}

\FloatBarrier

\section{Carbocation network scope and rule comparison}\label{si:cation_comparison}
Figure~\Mainref{fig:cation}a groups stereoisomers by constitution. Its four-step cutoff counts fenchol-derived entries as step 1. Neutral products were assigned using HopCat quenching rules.\autocite{klucznik2024cationic}

Each directed connection was classified using the lowest-barrier record for its source--product pair, before redundant reverse arrows were suppressed in the drawing. Bond changes were assigned from mapped IRC endpoints or from optimized endpoints when relaxation was needed to reach the network nodes. A change is composable if its mapped bond changes can be ordered into two or more HopCat rules. This graph test does not check the cationic reactive site at each step, so the composable count is an upper bound under the tested rules. Figure~\Mainref{fig:cation}e labels connections outside any single rule as \emph{new}.
\begin{center}
\small
\begin{tabular}{lrrrr}
\toprule
Network & Single rule & Composable & Non-composable & Total \\
\midrule
Fenchol-derived & 46 & 48 & 0 & 94 \\
Linalool-derived & 67 & 109 & 6 & 182 \\
\bottomrule
\end{tabular}
\end{center}
The six non-composable connections have local electronic barriers of 24.2--51.9~kcal/mol and lie outside the identified reference-product routes. A sequence of rules describes net bond changes. To assess whether the sequence contains an intervening minimum, we inspect the calculated path and any subsequent endpoint relaxation.

\FloatBarrier

\section{Energy profiles for the multisite reaction}\label{si:multisite_pes}

\begin{figure}[H]
\centering
\includegraphics[width=\textwidth]{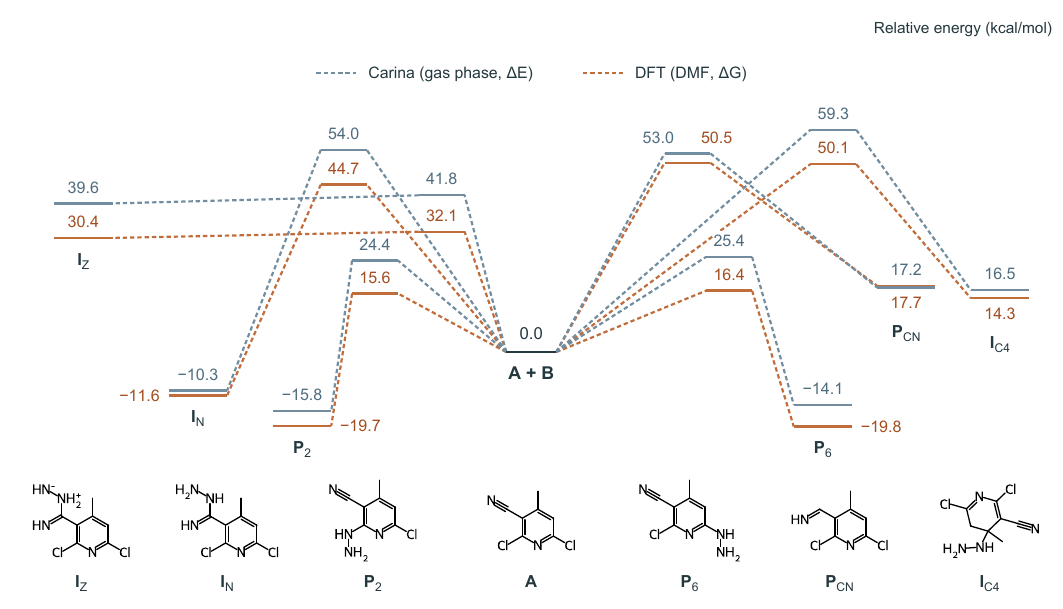}\captionsetup{justification=raggedright,singlelinecheck=false}
\captionsetup{font=small}
\caption{\color{black}\textbf{Fig.~\thefigure{} \textbar{} First-layer reaction energy profiles.} Carina gas-phase electronic energies (blue-grey, $\Delta E$) and DFT Gibbs free energies in DMF (orange, $\Delta G$). DFT calculations used the \mbox{$\omega$B97X-D4}/\allowbreak\mbox{def2-TZVPD}/\allowbreak\mbox{SMD(DMF)}//\allowbreak\mbox{$\omega$B97X-D4}/\allowbreak\mbox{def2-SVPD}/\allowbreak\mbox{SMD(DMF)} level. Energies are relative to \mbox{A + B}. Species labels follow Fig.~5a.}
\label{fig:si_multisite_l1_pes}
\end{figure}

The relative ordering of the first-substitution pathways was also assessed using fixed-geometry $\omega$B97X-D4/def2-TZVPD/SMD(DMF) single-point calculations on complete Carina reactant and TS complexes. The TS--reactant electronic energy differences were 14.25 and 12.93~kcal/mol for C2 (P$_2$) and C6 (P$_6$) substitution, respectively, referenced to the same complete A + B reactant complex. These values favor C6, whereas the DFT free energies of activation at 298.15~K in Fig.~\ref*{fig:si_multisite_l1_pes} are 15.6 and 16.4~kcal/mol, respectively, and slightly favor C2. The two comparisons differ in both geometry and energetic treatment: the fixed-geometry single points include no thermal, entropic, or standard-state corrections and do not establish DFT stationary points, whereas the plotted free-energy profiles use DFT-optimized geometries and thermal corrections.

The network also identifies candidate cyclization pathways (main-text Fig.~\Mainref{fig:applications}a). Closure from P$_2$ and routes through the nitrile-addition intermediate I$_Z$, either directly or through I$_N$, reach the charge-separated cyclic candidate $P_{\mathrm{cyc}}^{\pm}$. Reported CuI-promoted formation of pyrazolo[3,4-b]pyridines from 2-chloro-3-cyanopyridine and hydrazines provides a precedent in a related system for testing catalytic conditions that might favor this proposed cyclization.\autocite{lavecchia2004pyrazolopyridines} The cyclic product remains a computational candidate without experimental confirmation in this system.

\clearpage\section{LC--MS analysis of the multisite reaction}\label{si:lcms_experiment}
The PDA traces at 254~nm in main-text Fig.~5c are individually normalized, whereas the LC--MS records below retain the original absorbance and MS intensity scales. Ionization polarity and retention-time intervals are indicated for the selected averaged mass spectra. No LC--MS signal attributable to M178 was observed at 70~$^\circ$C. At 100~$^\circ$C, signals consistent with M178 were weak after 1--2~h and more pronounced after 4~h.

\begingroup
\setlength{\parindent}{0pt}
\setlength{\parskip}{0pt}
\subsection{LC--MS analysis at 70~$^\circ$C after 1~h}
\noindent\begin{minipage}{\textwidth}
\centering
\includegraphics[page=1,trim=15bp 531.322bp 15bp 0.000bp,clip,width=\textwidth]{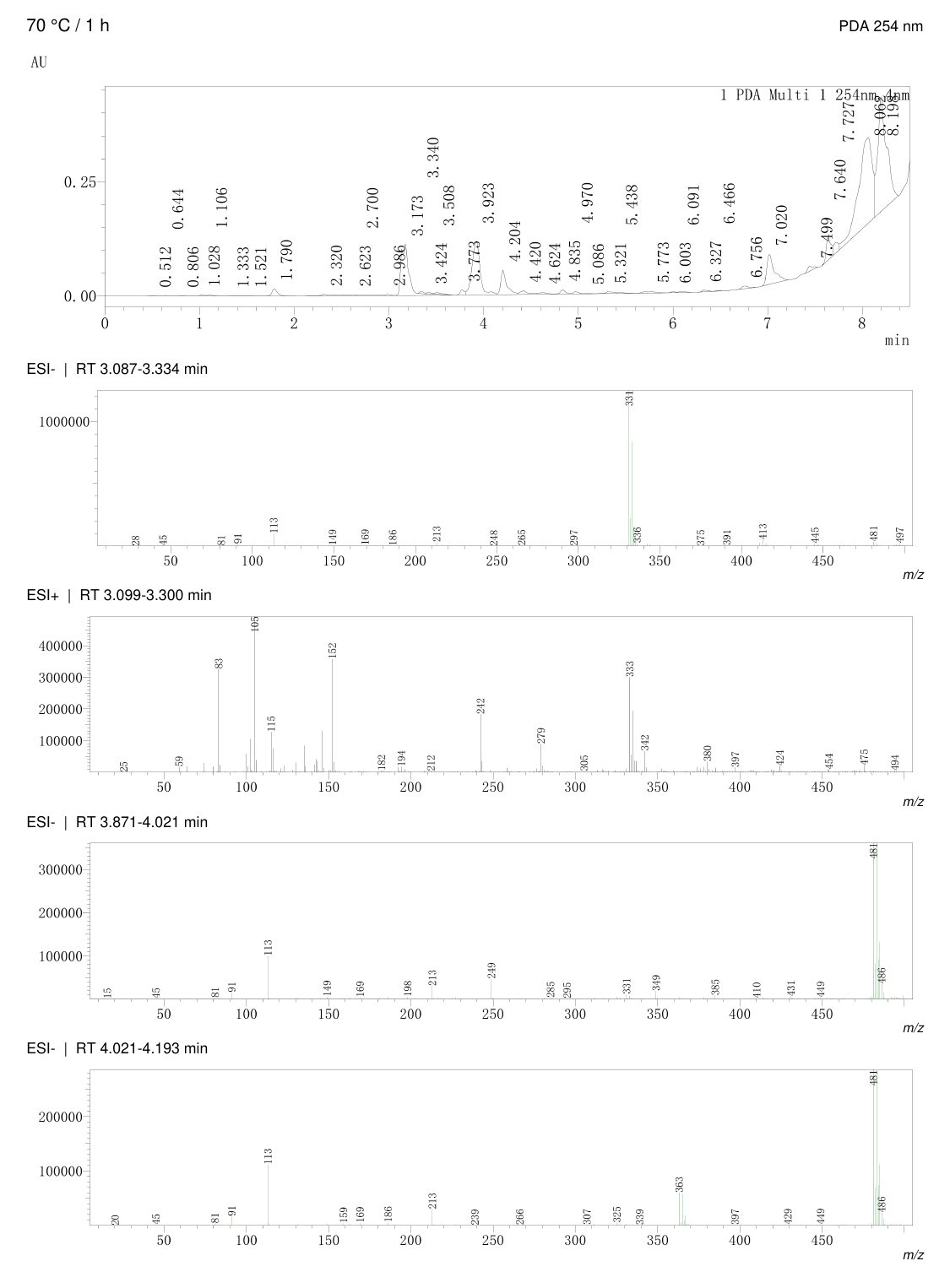}
\end{minipage}\par
\vspace{3pt}
\noindent\begin{minipage}{\textwidth}
\centering
\includegraphics[page=1,trim=15bp 399.506bp 15bp 203.940bp,clip,width=\textwidth]{figures/SI_exp_1_original_crops.pdf}
\end{minipage}\par
\vspace{3pt}
\noindent\begin{minipage}{\textwidth}
\centering
\includegraphics[page=1,trim=15bp 270.691bp 15bp 332.756bp,clip,width=\textwidth]{figures/SI_exp_1_original_crops.pdf}
\end{minipage}\par
\vspace{3pt}
\noindent\begin{minipage}{\textwidth}
\centering
\includegraphics[page=1,trim=15bp 141.875bp 15bp 461.572bp,clip,width=\textwidth]{figures/SI_exp_1_original_crops.pdf}
\end{minipage}\par
\vspace{3pt}
\noindent\begin{minipage}{\textwidth}
\centering
\includegraphics[page=1,trim=15bp 13.060bp 15bp 590.387bp,clip,width=\textwidth]{figures/SI_exp_1_original_crops.pdf}
\end{minipage}\par
\vspace{3pt}
\subsection{LC--MS analysis at 70~$^\circ$C after 2~h}
\noindent\begin{minipage}{\textwidth}
\centering
\includegraphics[page=1,trim=15bp 273.815bp 15bp 0.000bp,clip,width=\textwidth]{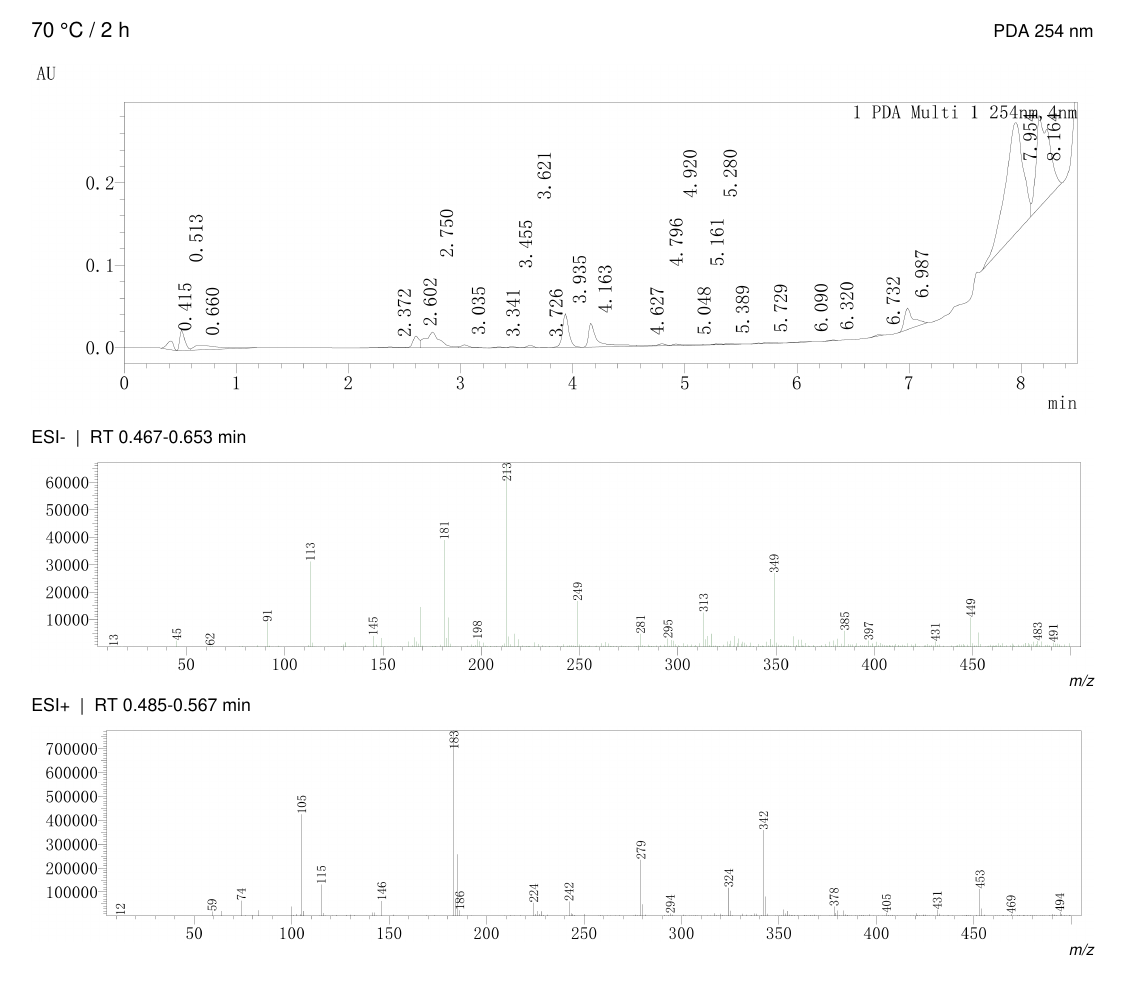}
\end{minipage}\par
\vspace{3pt}
\noindent\begin{minipage}{\textwidth}
\centering
\includegraphics[page=1,trim=15bp 141.999bp 15bp 203.940bp,clip,width=\textwidth]{figures/SI_exp_2_original_crops.pdf}
\end{minipage}\par
\vspace{3pt}
\noindent\begin{minipage}{\textwidth}
\centering
\includegraphics[page=1,trim=15bp 13.060bp 15bp 332.756bp,clip,width=\textwidth]{figures/SI_exp_2_original_crops.pdf}
\end{minipage}\par
\vspace{3pt}
\noindent\begin{minipage}{\textwidth}
\centering
\includegraphics[page=2,trim=15bp 270.631bp 15bp 22.000bp,clip,width=\textwidth]{figures/SI_exp_2_original_crops.pdf}
\end{minipage}\par
\vspace{3pt}
\noindent\begin{minipage}{\textwidth}
\centering
\includegraphics[page=2,trim=15bp 141.816bp 15bp 150.816bp,clip,width=\textwidth]{figures/SI_exp_2_original_crops.pdf}
\end{minipage}\par
\vspace{3pt}
\noindent\begin{minipage}{\textwidth}
\centering
\includegraphics[page=2,trim=15bp 13.000bp 15bp 279.631bp,clip,width=\textwidth]{figures/SI_exp_2_original_crops.pdf}
\end{minipage}\par
\vspace{3pt}
\subsection{LC--MS analysis at 70~$^\circ$C after 4~h}
\noindent\begin{minipage}{\textwidth}
\centering
\includegraphics[page=1,trim=15bp 531.446bp 15bp 0.000bp,clip,width=\textwidth]{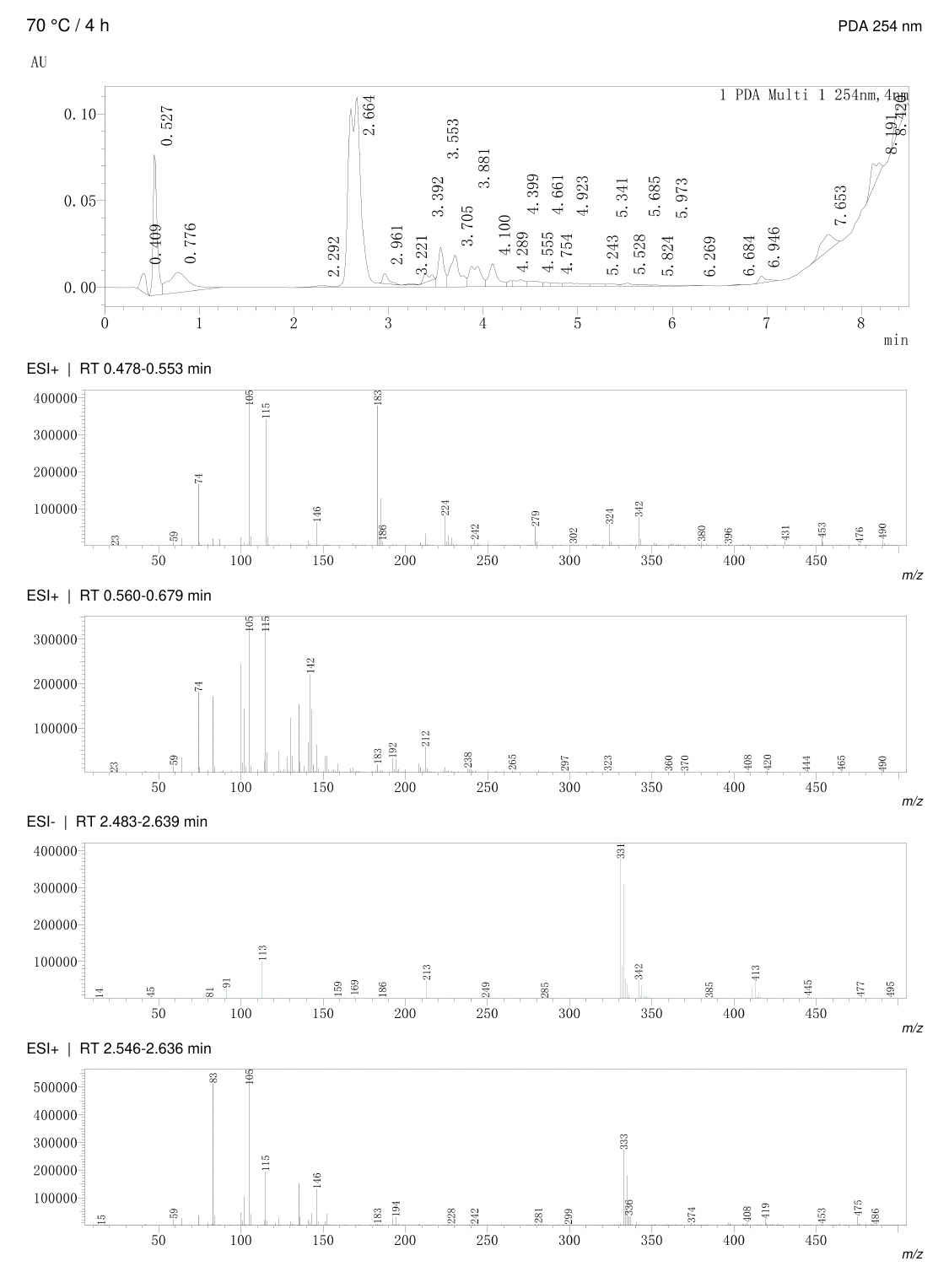}
\end{minipage}\par
\vspace{3pt}
\noindent\begin{minipage}{\textwidth}
\centering
\includegraphics[page=1,trim=15bp 399.630bp 15bp 203.940bp,clip,width=\textwidth]{figures/SI_exp_3_original_crops.pdf}
\end{minipage}\par
\vspace{3pt}
\noindent\begin{minipage}{\textwidth}
\centering
\includegraphics[page=1,trim=15bp 270.691bp 15bp 332.880bp,clip,width=\textwidth]{figures/SI_exp_3_original_crops.pdf}
\end{minipage}\par
\vspace{3pt}
\noindent\begin{minipage}{\textwidth}
\centering
\includegraphics[page=1,trim=15bp 141.875bp 15bp 461.819bp,clip,width=\textwidth]{figures/SI_exp_3_original_crops.pdf}
\end{minipage}\par
\vspace{3pt}
\noindent\begin{minipage}{\textwidth}
\centering
\includegraphics[page=1,trim=15bp 13.060bp 15bp 590.634bp,clip,width=\textwidth]{figures/SI_exp_3_original_crops.pdf}
\end{minipage}\par
\vspace{3pt}
\noindent\begin{minipage}{\textwidth}
\centering
\includegraphics[page=2,trim=15bp 528.262bp 15bp 22.000bp,clip,width=\textwidth]{figures/SI_exp_3_original_crops.pdf}
\end{minipage}\par
\vspace{3pt}
\noindent\begin{minipage}{\textwidth}
\centering
\includegraphics[page=2,trim=15bp 399.447bp 15bp 150.816bp,clip,width=\textwidth]{figures/SI_exp_3_original_crops.pdf}
\end{minipage}\par
\vspace{3pt}
\noindent\begin{minipage}{\textwidth}
\centering
\includegraphics[page=2,trim=15bp 270.631bp 15bp 279.631bp,clip,width=\textwidth]{figures/SI_exp_3_original_crops.pdf}
\end{minipage}\par
\vspace{3pt}
\noindent\begin{minipage}{\textwidth}
\centering
\includegraphics[page=2,trim=15bp 141.816bp 15bp 408.447bp,clip,width=\textwidth]{figures/SI_exp_3_original_crops.pdf}
\end{minipage}\par
\vspace{3pt}
\noindent\begin{minipage}{\textwidth}
\centering
\includegraphics[page=2,trim=15bp 13.000bp 15bp 537.262bp,clip,width=\textwidth]{figures/SI_exp_3_original_crops.pdf}
\end{minipage}\par
\vspace{3pt}
\subsection{LC--MS analysis at 100~$^\circ$C after 1~h}
\noindent\begin{minipage}{\textwidth}
\centering
\includegraphics[page=1,trim=15bp 273.815bp 15bp 0.000bp,clip,width=\textwidth]{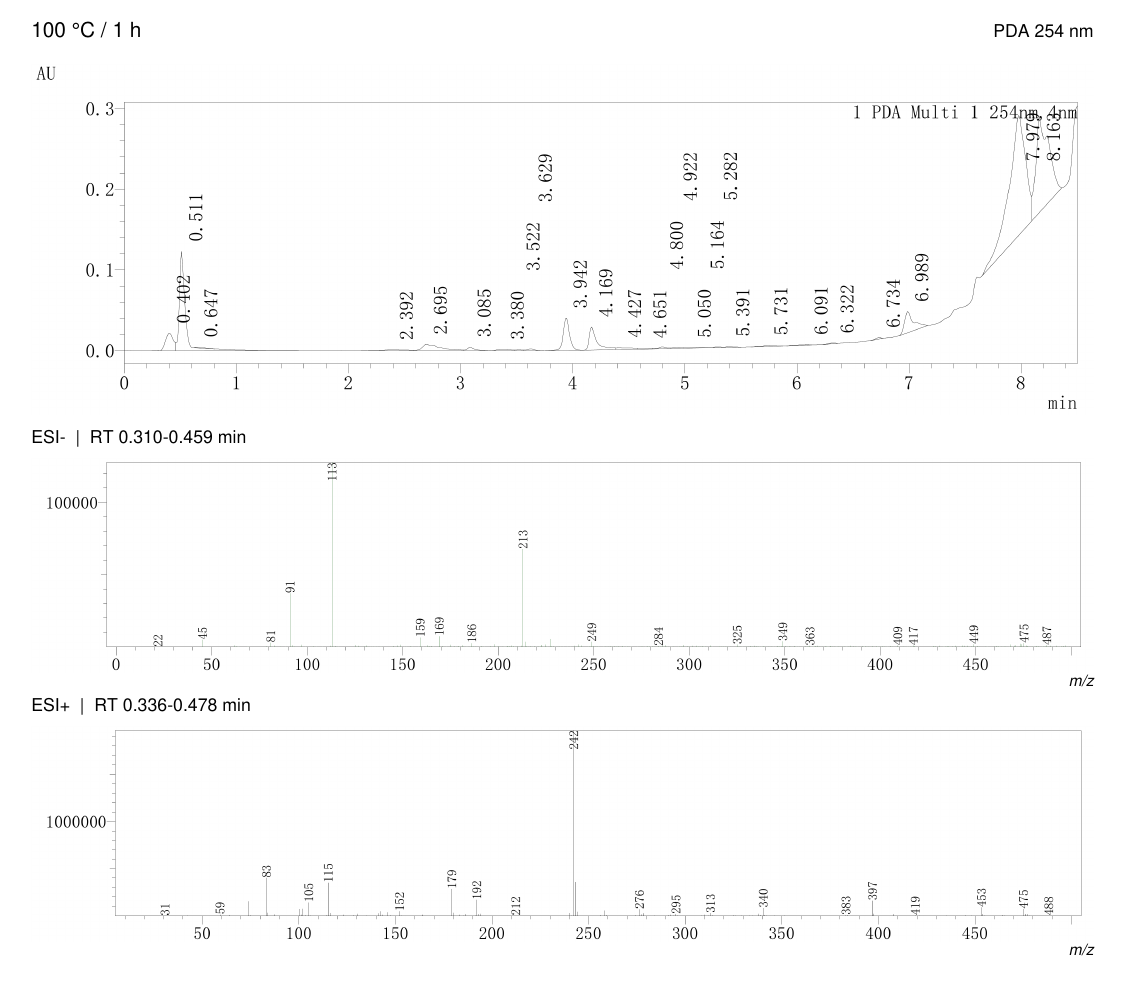}
\end{minipage}\par
\vspace{3pt}
\noindent\begin{minipage}{\textwidth}
\centering
\includegraphics[page=1,trim=15bp 141.999bp 15bp 203.940bp,clip,width=\textwidth]{figures/SI_exp_4_original_crops.pdf}
\end{minipage}\par
\vspace{3pt}
\noindent\begin{minipage}{\textwidth}
\centering
\includegraphics[page=1,trim=15bp 13.060bp 15bp 332.756bp,clip,width=\textwidth]{figures/SI_exp_4_original_crops.pdf}
\end{minipage}\par
\vspace{3pt}
\noindent\begin{minipage}{\textwidth}
\centering
\includegraphics[page=2,trim=15bp 270.755bp 15bp 22.000bp,clip,width=\textwidth]{figures/SI_exp_4_original_crops.pdf}
\end{minipage}\par
\vspace{3pt}
\noindent\begin{minipage}{\textwidth}
\centering
\includegraphics[page=2,trim=15bp 141.816bp 15bp 150.816bp,clip,width=\textwidth]{figures/SI_exp_4_original_crops.pdf}
\end{minipage}\par
\vspace{3pt}
\noindent\begin{minipage}{\textwidth}
\centering
\includegraphics[width=\textwidth]{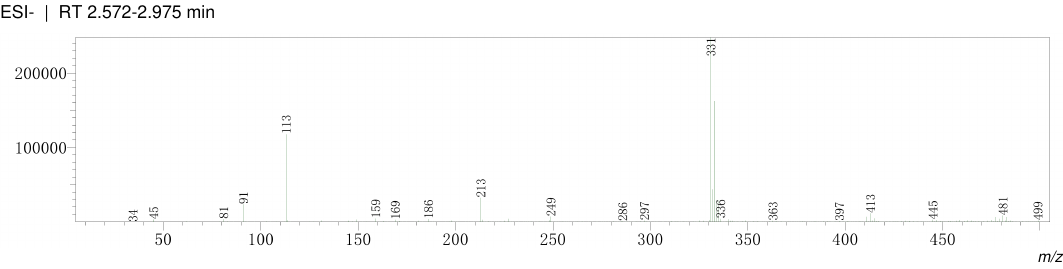}
\end{minipage}\par
\vspace{3pt}
\noindent\begin{minipage}{\textwidth}
\centering
\includegraphics[page=2,trim=15bp 13.000bp 15bp 279.755bp,clip,width=\textwidth]{figures/SI_exp_4_original_crops.pdf}
\end{minipage}\par
\vspace{3pt}
\subsection{LC--MS analysis at 100~$^\circ$C after 2~h}
\noindent\begin{minipage}{\textwidth}
\centering
\includegraphics[page=1,trim=15bp 273.815bp 15bp 0.000bp,clip,width=\textwidth]{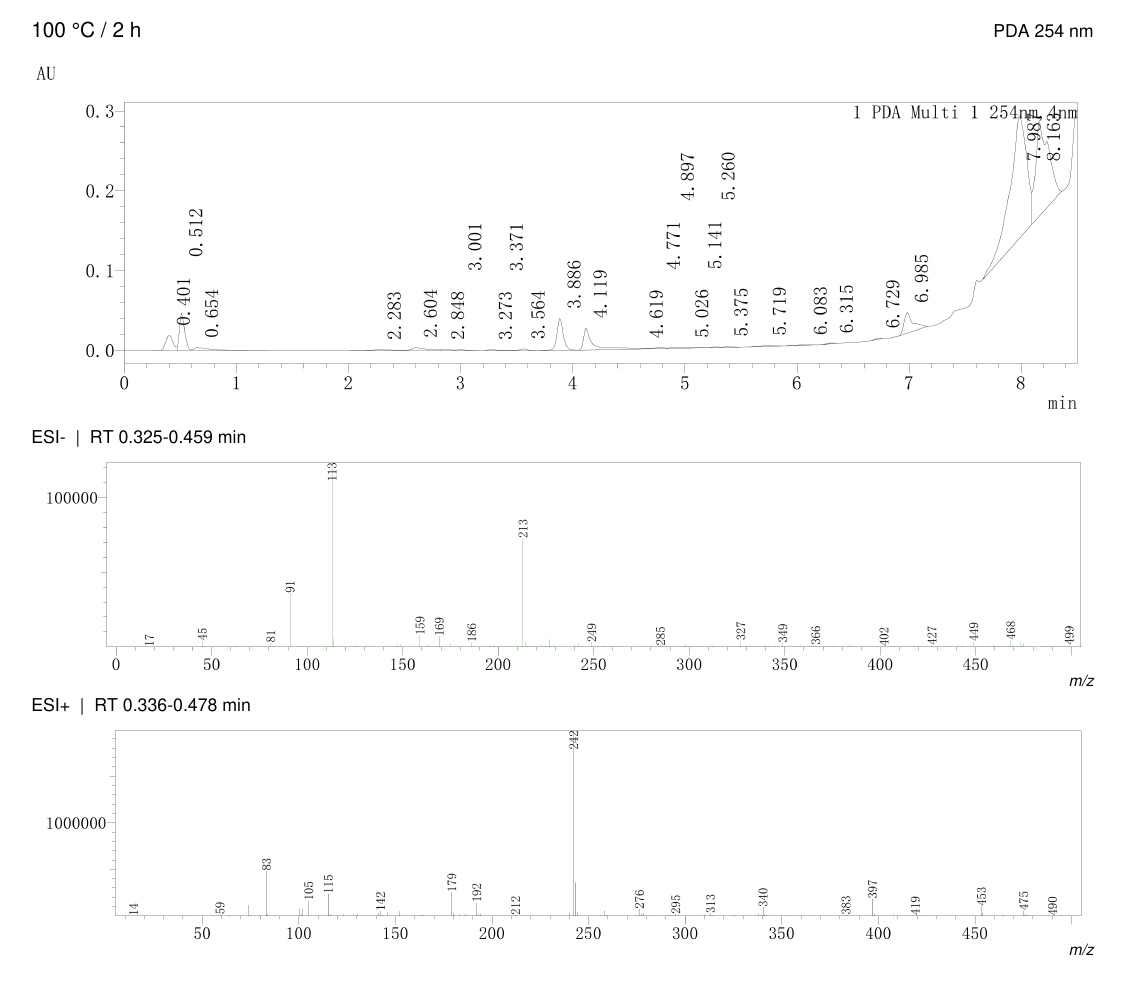}
\end{minipage}\par
\vspace{3pt}
\noindent\begin{minipage}{\textwidth}
\centering
\includegraphics[page=1,trim=15bp 141.999bp 15bp 203.940bp,clip,width=\textwidth]{figures/SI_exp_5_original_crops.pdf}
\end{minipage}\par
\vspace{3pt}
\noindent\begin{minipage}{\textwidth}
\centering
\includegraphics[page=1,trim=15bp 13.060bp 15bp 332.756bp,clip,width=\textwidth]{figures/SI_exp_5_original_crops.pdf}
\end{minipage}\par
\vspace{3pt}
\noindent\begin{minipage}{\textwidth}
\centering
\includegraphics[page=2,trim=15bp 270.755bp 15bp 22.000bp,clip,width=\textwidth]{figures/SI_exp_5_original_crops.pdf}
\end{minipage}\par
\vspace{3pt}
\noindent\begin{minipage}{\textwidth}
\centering
\includegraphics[page=2,trim=15bp 141.816bp 15bp 150.816bp,clip,width=\textwidth]{figures/SI_exp_5_original_crops.pdf}
\end{minipage}\par
\vspace{3pt}
\noindent\begin{minipage}{\textwidth}
\centering
\includegraphics[width=\textwidth]{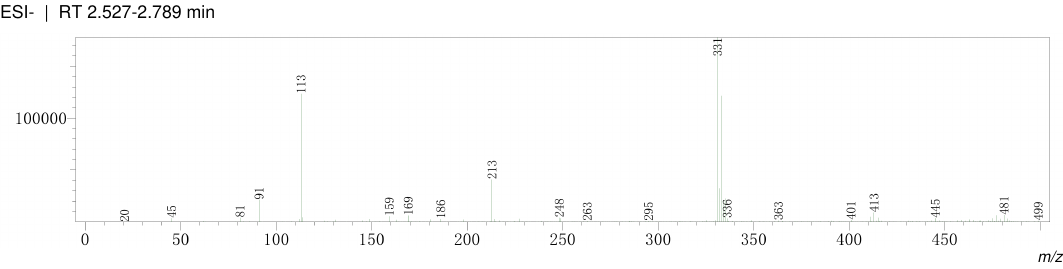}
\end{minipage}\par
\vspace{3pt}
\noindent\begin{minipage}{\textwidth}
\centering
\includegraphics[page=2,trim=15bp 13.000bp 15bp 279.755bp,clip,width=\textwidth]{figures/SI_exp_5_original_crops.pdf}
\end{minipage}\par
\vspace{3pt}
\subsection{LC--MS analysis at 100~$^\circ$C after 4~h}
\noindent\begin{minipage}{\textwidth}
\centering
\includegraphics[page=1,trim=15bp 273.815bp 15bp 0.000bp,clip,width=\textwidth]{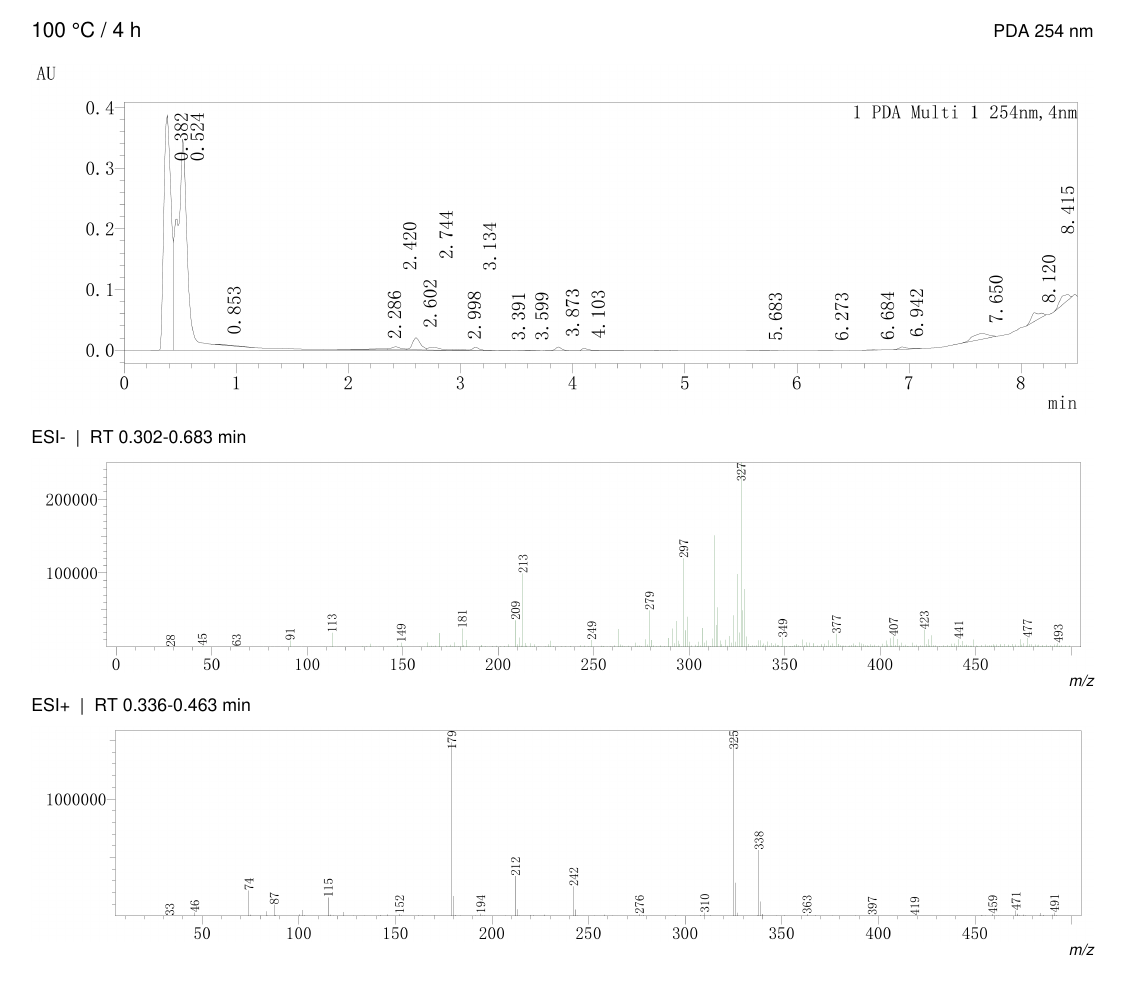}
\end{minipage}\par
\vspace{3pt}
\noindent\begin{minipage}{\textwidth}
\centering
\includegraphics[page=1,trim=15bp 141.999bp 15bp 203.940bp,clip,width=\textwidth]{figures/SI_exp_6_original_crops.pdf}
\end{minipage}\par
\vspace{3pt}
\noindent\begin{minipage}{\textwidth}
\centering
\includegraphics[page=1,trim=15bp 13.060bp 15bp 332.756bp,clip,width=\textwidth]{figures/SI_exp_6_original_crops.pdf}
\end{minipage}\par
\vspace{3pt}
\noindent\begin{minipage}{\textwidth}
\centering
\includegraphics[page=2,trim=15bp 270.755bp 15bp 22.000bp,clip,width=\textwidth]{figures/SI_exp_6_original_crops.pdf}
\end{minipage}\par
\vspace{3pt}
\noindent\begin{minipage}{\textwidth}
\centering
\includegraphics[page=2,trim=15bp 141.816bp 15bp 150.939bp,clip,width=\textwidth]{figures/SI_exp_6_original_crops.pdf}
\end{minipage}\par
\vspace{3pt}
\noindent\begin{minipage}{\textwidth}
\centering
\includegraphics[page=2,trim=15bp 13.000bp 15bp 279.878bp,clip,width=\textwidth]{figures/SI_exp_6_original_crops.pdf}
\end{minipage}\par
\vspace{3pt}
\endgroup

\clearpage
\section{NMR characterization of the multisite reaction products}\label{si:snar_nmr}
Four products isolated from the reaction at 70~$^\circ$C were characterized by $^1$H (500~MHz) and $^{13}$C (126~MHz) NMR spectroscopy in DMSO-$d_6$. The assignments correspond to P$_6$ and the 1,2-N-diarylation products P$_{332a}$ (C2/C2), P$_{332b}$ (C2/C6) and P$_{332c}$ (C6/C6) in main-text Fig.~5a. Chemical shifts ($\delta$) are reported in ppm; s denotes a singlet.

\vspace{8pt}
\noindent\begin{minipage}{\textwidth}
\raggedright
\setlength{\parskip}{3pt}
\includegraphics[page=1,trim=91.000bp 698.200bp 402.252bp 93.300bp,clip]{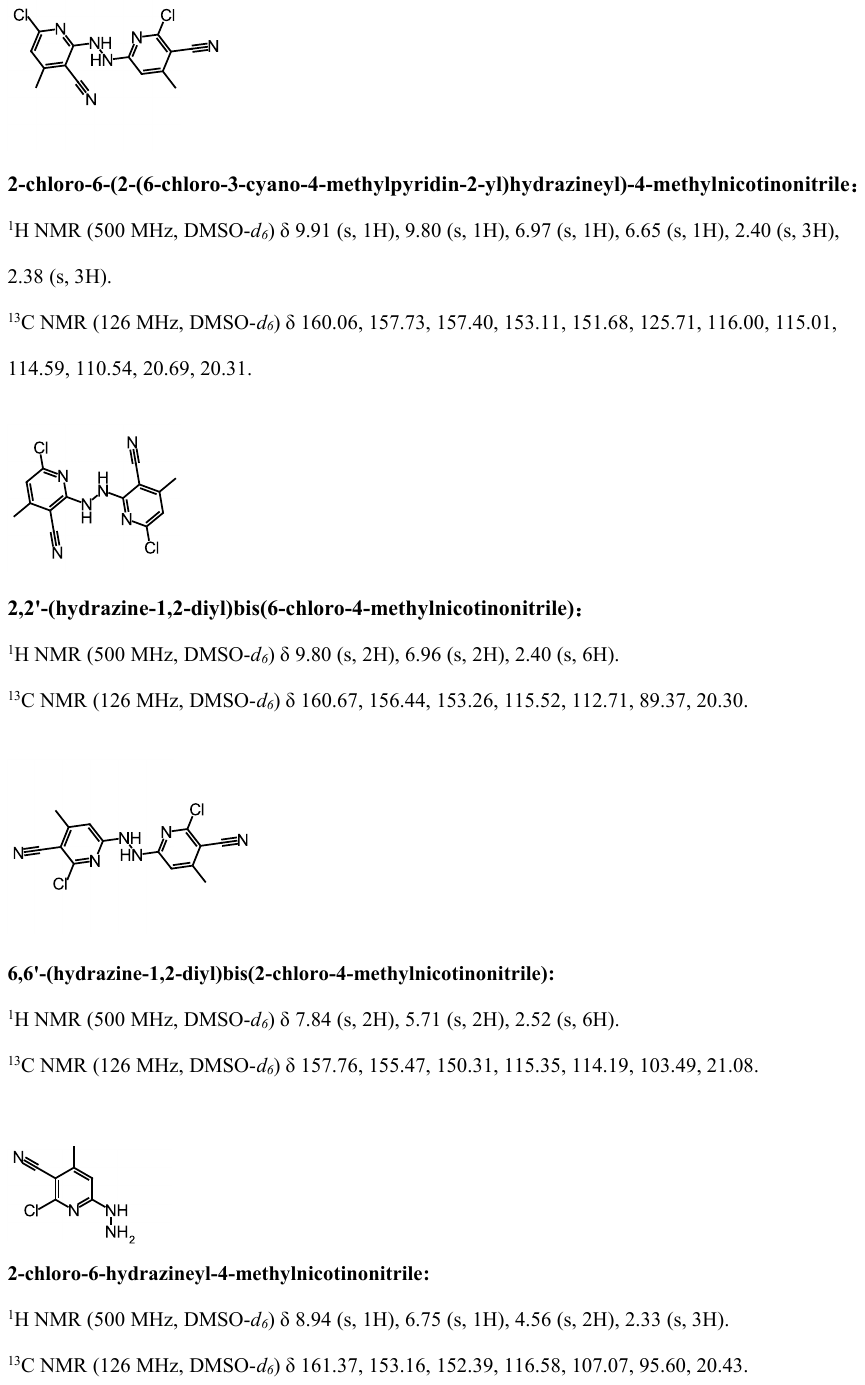}\par
\textbf{P$_{332b}$. 2-Chloro-6-(2-(6-chloro-3-cyano-4-methylpyridin-2-yl)hydrazineyl)-4-methylnicotinonitrile.}\par
$^1$H NMR (500~MHz, DMSO-$d_6$) $\delta$ 9.91 (s, 1H), 9.80 (s, 1H), 6.97 (s, 1H), 6.65 (s, 1H), 2.40 (s, 3H), 2.38 (s, 3H).\par
$^{13}$C NMR (126~MHz, DMSO-$d_6$) $\delta$ 160.06, 157.73, 157.40, 153.11, 151.68, 125.71, 116.00, 115.01, 114.59, 110.54, 20.69, 20.31.
\end{minipage}\par\vspace{12pt}

\noindent\begin{minipage}{\textwidth}
\raggedright
\setlength{\parskip}{3pt}
\includegraphics[page=1,trim=91.000bp 480.588bp 422.476bp 298.688bp,clip]{figures/SNAr_NMR_original_layout_v2.pdf}\par
\textbf{P$_{332a}$. 2,2'-(Hydrazine-1,2-diyl)bis(6-chloro-4-methylnicotinonitrile).}\par
$^1$H NMR (500~MHz, DMSO-$d_6$) $\delta$ 9.80 (s, 2H), 6.96 (s, 2H), 2.40 (s, 6H).\par
$^{13}$C NMR (126~MHz, DMSO-$d_6$) $\delta$ 160.67, 156.44, 153.26, 115.52, 112.71, 89.37, 20.30.
\end{minipage}\par\vspace{12pt}

\noindent\begin{minipage}{\textwidth}
\raggedright
\setlength{\parskip}{3pt}
\includegraphics[page=1,trim=91.000bp 321.580bp 388.108bp 474.720bp,clip]{figures/SNAr_NMR_original_layout_v2.pdf}\par
\textbf{P$_{332c}$. 6,6'-(Hydrazine-1,2-diyl)bis(2-chloro-4-methylnicotinonitrile).}\par
$^1$H NMR (500~MHz, DMSO-$d_6$) $\delta$ 7.84 (s, 2H), 5.71 (s, 2H), 2.52 (s, 6H).\par
$^{13}$C NMR (126~MHz, DMSO-$d_6$) $\delta$ 157.76, 155.47, 150.31, 115.35, 114.19, 103.49, 21.08.
\end{minipage}\par\vspace{12pt}

\noindent\begin{minipage}{\textwidth}
\raggedright
\setlength{\parskip}{3pt}
\includegraphics[page=1,trim=91.000bp 151.832bp 442.316bp 639.412bp,clip]{figures/SNAr_NMR_original_layout_v2.pdf}\par
\textbf{P$_{6}$. 2-Chloro-6-hydrazineyl-4-methylnicotinonitrile.}\par
$^1$H NMR (500~MHz, DMSO-$d_6$) $\delta$ 8.94 (s, 1H), 6.75 (s, 1H), 4.56 (s, 2H), 2.33 (s, 3H).\par
$^{13}$C NMR (126~MHz, DMSO-$d_6$) $\delta$ 161.37, 153.16, 152.39, 116.58, 107.07, 95.60, 20.43.
\end{minipage}\par\vspace{12pt}

\clearpage
\begin{figure}[p]
\centering
\includegraphics[page=2,trim=80bp 135bp 80bp 70bp,clip,width=\textwidth,height=0.89\textheight,keepaspectratio]{figures/SNAr_NMR_original_layout_v2.pdf}
\captionsetup{font=small}
\caption*{\textbf{NMR spectra of P$_{332b}$.} $^1$H NMR (500~MHz, top) and $^{13}$C NMR (126~MHz, bottom) in DMSO-$d_6$ for the C2/C6 1,2-N-diarylation product shown in main-text Fig.~5a.}
\end{figure}

\clearpage
\begin{figure}[p]
\centering
\includegraphics[page=3,trim=80bp 135bp 80bp 70bp,clip,width=\textwidth,height=0.89\textheight,keepaspectratio]{figures/SNAr_NMR_original_layout_v2.pdf}
\captionsetup{font=small}
\caption*{\textbf{NMR spectra of P$_{332a}$.} $^1$H NMR (500~MHz, top) and $^{13}$C NMR (126~MHz, bottom) in DMSO-$d_6$ for the C2/C2 1,2-N-diarylation product shown in main-text Fig.~5a.}
\end{figure}

\clearpage
\begin{figure}[p]
\centering
\includegraphics[page=4,trim=80bp 135bp 80bp 70bp,clip,width=\textwidth,height=0.89\textheight,keepaspectratio]{figures/SNAr_NMR_original_layout_v2.pdf}
\captionsetup{font=small}
\caption*{\textbf{NMR spectra of P$_{332c}$.} $^1$H NMR (500~MHz, top) and $^{13}$C NMR (126~MHz, bottom) in DMSO-$d_6$ for the C6/C6 1,2-N-diarylation product shown in main-text Fig.~5a.}
\end{figure}

\clearpage
\begin{figure}[p]
\centering
\includegraphics[page=5,trim=80bp 135bp 80bp 70bp,clip,width=\textwidth,height=0.89\textheight,keepaspectratio]{figures/SNAr_NMR_original_layout_v2.pdf}
\captionsetup{font=small}
\caption*{\textbf{NMR spectra of P$_6$.} $^1$H NMR (500~MHz, top) and $^{13}$C NMR (126~MHz, bottom) in DMSO-$d_6$ for the C6 monosubstitution product shown in main-text Fig.~5a.}
\end{figure}

\clearpage
\printbibliography[resetnumbers=1]